\documentclass[11pt,preprint]{aastex}
\pdfoutput=1
\slugcomment{{\sc Accepted to ApJ:} August 15, 2011} 

\usepackage{natbib}
\usepackage{epstopdf}
\usepackage{rotating}
\usepackage{epsfig}
\usepackage{subfig}
\usepackage{color}
\newcommand{\arcs}{\mbox{\ensuremath{^{\prime\prime}}}} 

\shorttitle{\emph{HST} Imaging of PSQs}
\shortauthors{Cales et al.}

\begin{document}

\title{Hubble Space Telescope Imaging of Post-Starburst Quasars}


\author{S. L. Cales\altaffilmark{1}, M. S. Brotherton\altaffilmark{1}, Zhaohui Shang\altaffilmark{2,1}, Vardha Nicola Bennert\altaffilmark{3}, G. Canalizo\altaffilmark{4,5}, R. Stoll\altaffilmark{6}, R. Ganguly\altaffilmark{7}, D. Vanden Berk\altaffilmark{8}, C. Paul\altaffilmark{9}, and A. Diamond-Stanic\altaffilmark{10}}


\altaffiltext{1}{Department of Physics and Astronomy, University of Wyoming, Laramie, WY 82071; scales@uwyo.edu, (307)766-3162}
\altaffiltext{1}{Department of Physics and Astronomy, University of Wyoming, Laramie, WY 82071; mbrother@uwyo.edu, (307) 766-5402}
\altaffiltext{1}{Department of Physics and Astronomy, University of Wyoming, Laramie, WY 82071; shang@uwyo.edu}
\altaffiltext{2}{Tianjin Normal University, Tianjin 300074, China, (+86) 22 23766527}
\altaffiltext{3}{Department of Physics, University of California, Santa Barbara, CA 93106;  bennert@physics.ucsb.edu, (805)893-3768}
\altaffiltext{4}{Department of Physics and Astronomy, University of California, Riverside, CA 92521; gabriela.canalizo@ucr.edu, (951)827-5310}
\altaffiltext{5}{Institute of Geophysics and Planetary Physics, University of California, Riverside, CA 92521} 
\altaffiltext{6}{Department of Astronomy, The Ohio State University, Columbus, OH 43210; stoll@astronomy.ohio-state.edu, (614) 292-1773}
\altaffiltext{7}{Department of Computer Science, Engineering, \& Physics, University of Michigan-Flint, Flint, MI 48502; ganguly@umflint.edu, (810)762-0787} 
\altaffiltext{8}{Department of Physics, Saint Vincent College, Latrobe, PA 15650; daniel.vandenberk@email.stvincent.edu, (724)532-6600}
\altaffiltext{9}{University of California, Davis, CA 95616; capaul@ucdavis.edu, 	(530)754-9779}
\altaffiltext{10}{Center for Astrophysics and Space Sciences, University of California, San Diego; La Jolla, CA 92093; aleks@ucsd.edu; (858)534-2230}



\begin{abstract}

   We present images of 29 post-starburst quasars (PSQs) from a Hubble Space Telescope (\emph{HST}) Advanced Camera for Surveys (ACS) Wide Field Channel Snapshot program. These broad-lined active galactic nuclei (AGN) possess the spectral signatures of massive ($M_{burst} \sim 10^{10} M_{\odot}$), moderate-aged stellar populations (hundreds of Myrs). Thus, their composite nature provides insight into the AGN-starburst connection. We measure quasar-to-host galaxy light contributions via semi-automated two-dimensional light profile fits of PSF-subtracted images. We examine the host morphologies, as well as, model the separate bulge and disk components. The \emph{HST}/ACS-F606W images reveal an equal number of spiral (13/29) and early-type (13/29) hosts, with the remaining three hosts having indeterminate classifications. AGNs hosted by early-type galaxies have on average greater luminosity than those hosted by spiral galaxies. Disturbances, such as tidal tails, shells, star-forming knots, and asymmetries are seen as signposts of interaction/merger activity. Disturbances such as these were found in 17 of the 29 objects and are evenly distributed among early-type and spiral galaxies. Two of these systems are clearly merging with their companions. Compared to other AGN of similar luminosity and redshift these PSQs have a higher fraction of early-type hosts and disturbances. Our most luminous objects with disturbed early-type host galaxies appear to be consistent with merger products. Thus, these luminous disturbed galaxies may represent a phase in an evolutionary scenario for merger driven activity and of hierarchical galaxy evolution. Our less luminous objects appear to be consistent with Seyfert galaxies not requiring triggering by major mergers. Many of these Seyferts are barred spiral galaxies. 

\end{abstract}
\clearpage



\section{Introduction}
\label{sec:Intro}
   
    The correlation between the masses of central supermassive black holes in local galaxies and their stellar bulge \citep[$M_{BH} \sim 0.15\% M_{bulge}$;][]{merrittferrarese01, magorrian98, gebhardt00b} suggests that the growth of black holes is intimately linked to the formation and evolution of their host galaxies. There is a tighter correlation between black hole mass and stellar velocity dispersion \citep[$M - \sigma_*$ relation;][]{gebhardt00a, ferraresemerritt00, tremaine02}. These relations have been extended by more recent studies of galaxies and quasars at high redshift \citep[e.g., ][]{shields03, treu04, treu07, peng06a, peng06b, woo06, woo08, bennert10}. The link between the evolution of  supermassive black holes and  their host galaxies hints at a common origin via a process  synchronizing the growth of both \citep{kauffmannhaehnelt00, granato04, hopkins08}. 
              
    Observational evidence and theoretical modeling supports a `merger hypothesis', a mechanism responsible for the formation of large (i.e., elliptical) galaxies via mergers of smaller (e.g., spiral) systems \citep{toomre77}. Ultraluminous infrared galaxies (ULIRGs) are interacting systems powered by both starburst and active galactic nuclei (AGN). ULIRGs have been hypothesized to evolve into  ``normal'' optically luminous quasars after the central engine clears away the dust associated with the massive star formation \citep{sanders88, sandersmirabel96, veilleux06} and are possibly the first step in the evolution of large elliptical galaxies. Post-starburst galaxies, objects with the spectral signatures of massive, moderate-aged stellar populations, show signs of fossil AGN activity and are thought to be systems a few 100 Myr after the ULIRG phase \citep*{tremonti07}. Furthermore, deep imaging reveals that the host galaxies of normal optically luminous quasars are disturbed \citep[e.g.,][]{canalizostockton01, kauffmann03, canalizo07, bennert08}, a telltale sign of interaction, and reveal the presence of young stellar populations \citep[e.g.,][]{canalizostockton01, kauffmann03, jahnke04, vandenberk06}. It has been suggested that these distinct phenomenological phases are part of an evolutionary merger sequence \citep{hopkins08}.
    
    Numerical simulations have successfully reproduced the physical properties of elliptical galaxies and bulges through major mergers of gas-rich disk galaxies \citep{granato04, dimatteo05, hopkins06}. These simulations follow star formation and black hole growth simultaneously during gas-rich galaxy-galaxy collisions and find that mergers lead to strong gaseous inflows that feed quasars and intense starbursts (i.e., a ULIRG phase). In due course, feedback from the quasar quenches both star formation and further black hole growth \citep{dimatteo05}. Without further star formation the blue remnant evolves onto the red sequence in $\sim$1 Gyr \citep{springel05}. Thus, AGN feedback is deemed responsible for the coupling of the black hole mass with the host bulge mass as well as the bimodality of the color distribution of galaxies. 
    
    Galaxy bimodality is observed in galaxy colors, morphology, star formation rates and galaxy stellar masses \citep[e.g.,][]{kauffmann03, blanton09}. A color-magnitude diagram of galaxies reveals this bimodality as two strong peaks in the red and blue. The blue cloud consists of a population of galaxies that are actively star-forming, gas-rich and morphologically disk-dominated. The red sequence population is typically quiescent, gas-poor and morphologically spheroidal. Since $z = 1$, the stellar mass contained within the red sequence doubled while that of the blue cloud remained more or less constant \citep{bell04}. Some blue cloud galaxies quench their star formation and passively evolve to the red sequence while other blue galaxies continue with ongoing star formation. These both happen at such a rate that the stellar mass generated via star formation is balanced by the mass of galaxies moving off of the blue cloud \citep{martin07}. Galaxies that lie intermediate between the blue cloud and red sequence (i.e., in the ``green valley'') are likely transition galaxies. Local AGN hosts lie in this green valley of the color distribution of galaxies. Their intermediate colors suggest that they are now, without significant star formation within the past $\sim$100 Myr, passively evolving from the blue cloud to the red sequence \citep{schawinski09}. 
  
  Some of the research into the bimodality of the galaxy population has centered on post-starburst galaxies. Traditionally, post-starburst galaxies are defined by strong Balmer absorption lines, indicating intense star formation in the past $\sim$1 Gyr, as well as, a lack of ongoing star formation which is observed as an absence of nebular emission lines \citep{dresslergunn83}. Their spectra are best modeled as a superposition of an elliptical spectrum and an A-star spectrum, hence they are often called $E + A$ or $k + a$ and sometimes called $H\delta$-strong galaxies. These galaxies are thought to be in transition between actively star-forming late-type galaxies and passive early-type galaxies observable on a timescale of 100s of Myr ($< 1$ Gyr). It is important to note that placing limits against nebular emission lines biases the sample selection against AGN \citep{yan06, wild09}. Additionally, it has been found that the traditional post-starburst galaxy definition is too narrow to encompass a full range of post-starburst galaxies \citep{falkenberg09}. 
    
  A post-starburst phase can be induced via starbursts and/or halting of star formation \citep{falkenberg09}. There are a number of mechanisms attributed to triggering the post-starburst phase (e.g., merger, harassment, gas stripping, strangulation, etc.). For example, \citet{falkenberg09} model star formation activity of harassment as a starburst with gradual termination of star formation as the galaxy gradually consumes its fuel. However, a gas-rich major merger triggers a strong burst of star formation followed by a rapid truncation on the order of $\sim$ 0.1-0.4 Gyr. This rapid truncation may be due to the rapid consumption of fuel or its expulsion from the galaxy via AGN and/or supernova feedback. Observationally, post-starburst galaxies have been linked to mergers and AGN \citep[e.g.,][and references therein]{brown09, falkenberg09, wild09}. The environments and morphologies of post-starburst galaxies are heterogeneous, which is suggestive of multiple triggering mechanisms; with merger triggered AGN possibly being responsible for the most luminous post-starburst galaxies \citep{brown09}.  
  
    A key evolutionary phase in merger driven evolutionary scenarios is the ignition of AGN activity that, through outflows, can inhibit both star formation and its own fueling. Such objects would be expected to have luminous quasar activity, starburst or post-starburst signatures, along with indications of a recent merger (e.g., companion, tidal tails, star forming knots, asymmetries, etc.). Indeed, the post-starburst quasar (PSQ) prototype, UN J1025-0040, possesses many of these features; it is hosted by a galaxy with a $\sim$400 Myr old strong starburst \citep{brotherton99}, has a companion galaxy in a post-starburst phase \citep{canalizo00}, and is morphologically classified as a merger remnant \citep{brotherton02}. A younger UN J1025-0040 (tens of Myrs after the starburst) would have a more luminous stellar population and would likely be dust-enshrouded, placing it in the ULIRG class. These observations suggest that UN J1025-0040 is a plausible transition object between ULIRGs and quasars. 
    
  We investigate a sample of PSQs in this paper at somewhat lower redshifts and luminosities than the prototype, thus it is necessary to consider the premise of \citet{hasinger08}. They suggest that major merger-driven evolution dominates in the early universe, producing the bulk of the brightest quasars at $z = 2-3$. Later, around $z\sim1$, secular evolution and minor interactions take the lead, becoming the main fueling mechanisms. This is seen observationally as we move to lower redshift less massive systems show activity \citep{heckman04}. A recent study by \citet{cisternas11} of lower luminosity AGN at $z < 1$, find that there is no difference in the disturbance fraction between AGN and inactive galaxies. Thus, as we move from quasar to Seyfert luminosities, fueling rates and triggering mechanisms may differ \citep{hopkinshernquist09}. For example, bars in spiral galaxies may be sufficient to fuel such nuclear activity. For this reason, and since post-starburst galaxies tend to be a heterogeneous population, our sample selection may be expected to include objects of both types.   
  
  We test the idea that PSQs are a phase in the life of galaxies triggered by external events (e.g., mergers, tidal interactions) or whether they are a more heterogenous population in which multiple mechanisms can contribute to the class (i.e., external events and internal processes). This paper is the first in a series devoted to understanding the properties of PSQs. Our aim for this paper is to characterize the morphology and other host galaxy parameters of a sample of 29 PSQs via Hubble Space Telescope (\emph{HST}) Advanced Camera for Surveys (ACS) F606W imaging. The sample selection and data used in this paper are discussed in \S~\ref{sec:Data}. We visually characterize the morphologies in \S~\ref{sec:Morph}. The PSQ hosts are characterized via semi-automated two-dimensional light profile fits in \S~\ref{sec:Decomp}. Host galaxy correlations between morphology and degree of disturbance are given in \S~\ref{sec:Analysis}. In \S~\ref{sec:Disc} we  compare this sample's properties with other types of galaxies (e.g., post-starburst galaxies, AGN/quasars host galaxies). We present a summary of our conclusions in \S~\ref{sec:Summ} along with a brief discussion of future directions of PSQ study. SDSS DR7 spectra of the targets are given in Appendix~A. Descriptions of the individual targets are given in Appendix~B. We adopt a Hubble constant of  $H_o = 73$ km s$^{-1}$Mpc$^{-1}$ and a flat universe where $\Omega_{M}$ is equal to 0.27 with $\Omega_{\Lambda}$ of 0.73, which is consistent with Wilkinson Microwave Anisotropy Probe (WMAP) three-year cosmology \citep{spergel07}.

\section{Data}
\label{sec:Data}

\subsection{SDSS Sample}
\label{sec:Data.SDSSSample}

  Of the 16,067 objects with $z < 1$ classified as broad-lined (FWHM $> 1000$ km s$^{-1}$) by Sloan Digital Sky Survey data release 3 \citep[SDSS DR3;][]{abazajian05} 609 were spectroscopically selected as PSQs and cataloged by \citet[hereafter B11]{brotherton11}. To select PSQs, we modified a post-starburst galaxy selection algorithm used by \citet{zabludoff96}. The selection criteria required that the spectra display both the broad emission lines of luminous AGNs (Type 1 Seyfert galaxies or quasars) and the Balmer absorption lines characteristic of massive stellar populations with ages $\sim$100s Myr. We note that B11 choose not to place limits on nebular emission (most notably $H_{\alpha}$ and [\ion{O}{2}]) since it may result from AGN emission. Thus, while a limit on nebular emission restricts the amount of ongoing star formation, it introduce an unfair bias against AGN. For this reason B11 cannot ensure that these objects lack current star formation, however the additional use of a Balmer break criterion can help select galaxies with older and possibly weaker post-starburst features.
  
  The B11 catalog was built with conservative selection criteria as an exploratory exercise to develop a more thorough understanding into the physical nature of PSQs. The aim of the selection criteria is to find broad-lined objects showing Balmer absorption. Through extensive trial and error B11 settled on the following criteria:

\begin{enumerate}
\item{$S/N > 8$}
\item{$H_{total} > 2$}
\item{$H_{sig}> 6$}
\item{Balmer Break $> 0.9$}
\end{enumerate}
The continuum signal-to-noise ratio ($S/N$) was calculated between the rest-wavelengths of 4150 and 4250 \AA. $H_{total}$ is the summation of the Balmer lines $H\delta$, $H\zeta$, and $H\eta$. The $H_{sig}$ parameter is the significance level of the detection of the $H_{total}$ parameter. The Balmer Break parameter is defined as a strength based on the ratio of the fluxes at two 100 \AA\ wide regions starting at rest-wavelengths of 3985 and 3740 \AA. After removing false positives by visual inspection of the spectra the final catalog of 609 objects was assembled. 
  
   Commonly, post-starburst galaxies are selected on the basis of only the strength of $H\delta$ \citep[e.g.,][]{zabludoff96}. This is problematic for selection of PSQs since the AGN component can dilute the absorption to the point of non-detection, so a stringent limit on $H\delta$ is not advisable. Estimating the dilution of the AGN component is non-trivial and beyond the scope of B11. Thus, the B11 selection criteria, although originally based on the algorithm of \citet{zabludoff96}, is more reminiscent with that of \citet{wild09}; they use a combination of Balmer lines and Balmer jump strength without a nebular emission criterion to select their sample of post-starburst galaxies without biasing against AGN.

\subsection{\emph{HST} Sample}
\label{sec:Data.HSTSample}
      
    Focusing on the most luminous ($M_r  \sim -22.9$) examples of PSQs with $H\delta$ absorption equivalent widths $> 1$\AA\ and in the redshift range $0.25 < z < 0.45$, a pool of 80 PSQs were selected as candidate targets for Hubble Space Telescope (\emph{HST}) Advanced Camera for Surveys (ACS) Snapshot program. The most luminous objects also possessing $H\delta > 1$ \AA\ are more likely to have dominant post-starburst stellar populations, rather than just being systems with weaker AGNs or significant line-of-sight dust reddening. This redshift range ensures similar size scales, resolving structures about a half kpc across. We received 29 \emph{HST}/ACS-F606W images that are listed in Table~\ref{tab:sample}. SDSS DR7 spectra of the targets are given in Appendix~A. This sample will hereafter be referred to as the PSQ sample.
    
    The F606W filter corresponds to the rest-frame \emph{B} and probes wavelengths near the peak output for $\sim$100 Myr stellar populations enhancing the stellar/AGN contrast. At a redshift of $z  \sim 0.319$ one kpc subtends 0\farcs23. 
         
   It is important to note that since the PSQs were selected using an apparent magnitude cut, the total luminosity given in Table~\ref{tab:sample} reflects contributions from both stellar and AGN components. In order to detect an object as a PSQ, the post-starburst population and AGN luminosity must be of comparable strength. Objects in this sample are all classified as active galaxies in the \citet{veroncettyveron06} catalog.

\subsection{Observations and Data Reductions}
\label{sec:Data.Reduc}

   ACS  Wide Field Channel (WFC) observations (proposal ID 10588: PI M. Brotherton) were carried out between July 2005 and November 2006. For each field, two 360s 202\arcs $\times$202\arcs\ F606W images, offset by a pixel in each direction, were acquired in order to facilitate the removal of cosmic rays and hot pixels. The images were bias-subtracted and flat-fielded with the standard ACS calibration pipeline. The Multidrizzle implementation provides fully geometrically corrected, cosmic-ray cleaned, drizzle-combined images. The offsets and rotations were determined between the exposures by cross-correlatingthe bright field stars and then drizzling the individual frames onto geometrically corrected output frames. Cosmic-ray masks were created for each frame. The original input images, together with the final cosmic ray masks were then all drizzled onto a single output image. We use an output pixel scale of 0.05\arcs$/pixel$ and the square interpolation kernel for the final drizzle image. 
      
   Absolute magnitudes were calculated in the following manner. We obtained SDSS DR7 spectra for all targets. We scaled the spectra to match the reported SDSS photometric $r$-band magnitudes (modelMag\_r). We then correct for Galactic extinction using the \citet{schlegel98} maps and the IRAF\footnotemark\ task \textit{deredden} which utilizes the \citet*{cardelliclaytonmathis89} extinction curves. We used the resulting spectra to make K-corrections between the observed and rest-frame $r$-band. We elaborate on the approach for calculating F606W magnitudes of the host galaxy and PSF in \S~\ref{sec:Decomp.AbsMag}.
   
  Our method of performing K-corrections a more robust method since it does not assume a spectral shape but utilizes the data to make the calculation. We must note however that there is a possibility of aperture bias when calculating K-corrections in this fashion. The SDSS fiber diameter is 3\arcs. Thus, some of our objects are larger than the aperture and SDSS magnitudes are missing some of the light from the galaxy. Missing light from the galaxy may not be important unless there is a color gradient with radial distance from the central source. Therefore, capturing that missing light would change the shape of the spectra and by extension the K-correction. We estimate that on average $\sim$60\% of the light falls in the fiber.
  
 \footnotetext{IRAF (Image Reduction and Analysis Facility) is distributed by the National Optical Astronomy Observatories, which are operated by AURA, Inc., under cooperative agreement with the National Science Foundation.}

\begin{deluxetable}{clrrrcc}
\tabletypesize{\scriptsize}
\tablecolumns{7}
\tablewidth{0pc}
\tablecaption{\emph{HST} Sample Observations}
\tablehead{
\colhead{Obj} & \colhead{SDSS} & \colhead{z} & \colhead{$r$\tablenotemark{a}} & \colhead{M$_r$\tablenotemark{b}} & \colhead{Scale} & \colhead{\emph{HST}} \\
\colhead{ID} & \colhead{Name} & \colhead{} & \colhead{} & \colhead{} & \colhead{(kpc/\arcs)} & \colhead{Obs Date} 
}
\startdata
1 & J003043$-$103517 & 0.296 & 18.36 & $-$22.80 & 0.21 & 05 Nov 2006 \\
2 & J005739+010044 & 0.253 & 17.68 & $-$23.08 & 0.19 & 11 Jun 2006 \\
3 & J020258$-$002807 & 0.339 & 18.26 & $-$23.35 & 0.23 & 16 Oct 2005 \\
4 & J021447$-$003250 & 0.349 & 18.66 & $-$22.90 & 0.24 & 03 Dec 2005 \\
5 & J023700$-$010130 & 0.344 & 18.67 & $-$22.90 & 0.24 & 11 Oct 2005 \\
6 & J040210$-$054630 & 0.270 & 18.96 & $-$22.11 & 0.20 & 07 Aug 2005 \\
7 & J074621+335040 & 0.284 & 18.11 & $-$22.90 & 0.21 & 03 Jan 2006 \\
8 & J075045+212546 & 0.408 & 18.19 & $-$23.94 & 0.26 & 15 Dec 2005 \\
9 & J075521+295039 & 0.334 & 18.83 & $-$22.61 & 0.23 & 08 Dec 2005 \\
10 & J075549+321704 & 0.420 & 18.99 & $-$23.00 & 0.27 & 27 Oct 2005 \\
11 & J081018+250921 & 0.263 & 17.55 & $-$22.99 & 0.20 & 10 Dec 2005 \\
12 & J105816+102414 & 0.275 & 18.37 & $-$22.48 & 0.20 & 22 Jan 2006 \\
13 & J115159+673604 & 0.274 & 18.47 & $-$22.52 & 0.20 & 13 Aug 2005 \\
14 & J115355+582442 & 0.319 & 18.33 & $-$22.76 & 0.22 & 19 Aug 2005 \\
15 & J123043+614821 & 0.324 & 18.66 & $-$22.67 & 0.23 & 17 Mar 2006 \\
16 & J124833+563507 & 0.266 & 17.47 & $-$23.35 & 0.20 & 31 Jul 2005 \\
17 & J145640+524727 & 0.277 & 18.18 & $-$22.73 & 0.20 & 15 Aug 2005 \\
18 & J145658+593202 & 0.326 & 18.60 & $-$22.76 & 0.23 & 12 Jul 2006 \\
19 & J154534+573625 & 0.268 & 18.11 & $-$22.77 & 0.20 & 28 Oct 2005 \\
20 & J164444+423304 & 0.317 & 19.02 & $-$22.05 & 0.22 & 10 Mar 2006 \\
21 & J170046+622056 & 0.276 & 18.65 & $-$22.37 & 0.20 & 12 Feb 2006 \\
22 & J210200+000501 & 0.329 & 18.36 & $-$23.12 & 0.23 & 24 May 2006 \\
23 & J211343$-$075017 & 0.420 & 18.77 & $-$23.55 & 0.27 & 06 Jun 2006 \\
24 & J211838+005640 & 0.384 & 18.76 & $-$23.38 & 0.25 & 03 Jun 2006 \\
25 & J212843+002435 & 0.346 & 18.92 & $-$22.70 & 0.24 & 17 Jul 2005 \\
26 & J230614$-$010024 & 0.267 & 17.91 & $-$23.00 & 0.20 & 06 Jun 2006 \\
27 & J231055$-$090107 & 0.364 & 18.60 & $-$23.23 & 0.24 & 19 Aug 2005 \\
28 & J233430+140649 & 0.363 & 18.81 & $-$22.95 & 0.24 & 22 Jul 2006 \\
29 & J234403+154214 & 0.288 & 18.41 & $-$22.77 & 0.21 & 10 Jun 2006 \\\enddata
\tablenotetext{a}{SDSS DR7 $r$ AB magnitudes (modelMag\_r).}
\tablenotetext{b}{SDSS DR7 dereddened K-corrected $r$ absolute AB magnitudes. \label{tab:sample}}\end{deluxetable}

\clearpage

\begin{figure}[tbhp]
\centering
\figurenum{1} 
\includegraphics[width=6in]{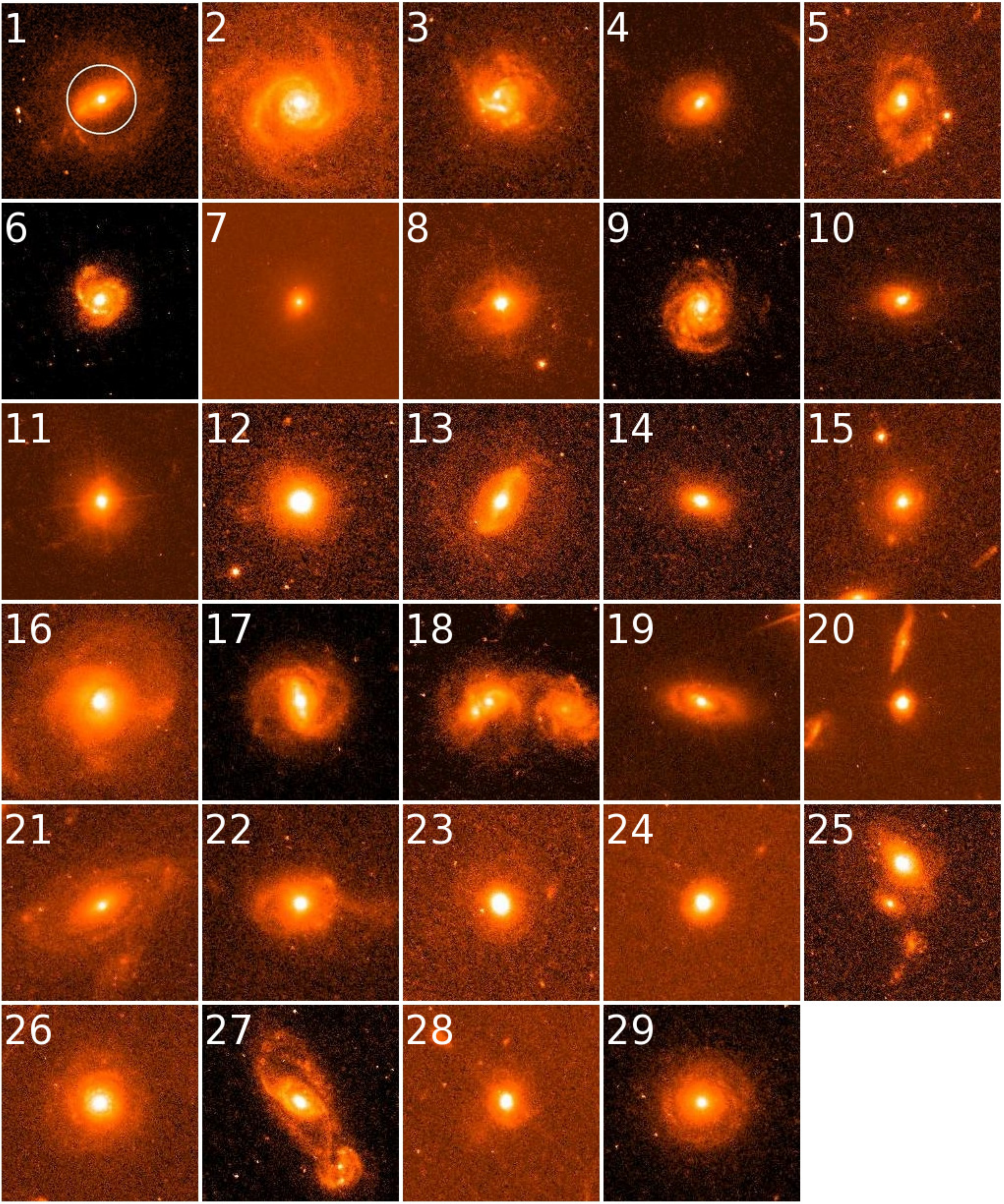}
\caption{{\small A mosaic of the 29 \emph{HST}/ACS images. Each image is 10\arcs $\times$ 10\arcs. NE is towards the upper left. At an average redshift of $\langle z \rangle \sim 0.319$, these 200 $\times$ 200 pixel images correspond to about 44 kpc across. For reference, we show the 3\arcs\ SDSS fiber aperture overlaid on the first image. See also \texttt{http://physics.uwyo.edu/agn/psq/index.html} for a finer detailed view. \label{fig:mosaic}}}
\end{figure}

\section{Morphologies}
\label{sec:Morph}

   The clearly resolved images reveal morphologies and companions without the need for PSF subtraction. Visual classifications were performed by S. L. Cales, M. S. Brotherton and Z. Shang. In cases where the classifiers disagree, the majority classification was adopted as the final classification. Galaxies were morphologically classified into 4 main categories:  
\begin{itemize}
\item \textbf{Spiral:} Host galaxies containing a disk, arms and/or bars. Grand design spiral galaxies have two arms which can be traced from the center to the outskirts of the galaxy. Flocculent spiral galaxies  have arms which are fragmented.
\item \textbf{Probable spiral:} Host galaxies with smooth distributions on the large scale but disk-like appearance upon careful inspection (i.e.,  by varying the size and dynamic scale of the image).
\item \textbf{Early-type:} Host galaxies with smooth, somewhat featureless (lacking arms/bars although sometimes shows tidal features) light distributions.
\item \textbf{Indeterminate:} Host galaxies for which it is difficult to give a classification because of the degree of disturbance.
\end{itemize}   
   
   The majority of systems are disturbed. In two of the images we see obvious merging. Galaxies were classified by degree of disturbance into 4 main categories:
\begin{itemize}
\item \textbf{Disturbed:} Galaxies showing signs of interaction/merger activity, such as, tidal tails, shells, star-forming knots and asymmetries.
\item \textbf{Highly Disturbed:} Host galaxies which appear to be the result or in the process of a major merger, such as, overall lack of structure with star-forming arcs, rings and obvious multiple shell structure. These signatures are brighter with a galaxy-wide appearance indicating a stronger disturbance when compared to the Disturbed classification. Note this is highly subjective.
\item \textbf{Undisturbed:} Galaxies lacking asymmetries or signs of interaction/merger activity.
\item \textbf{Indeterminate:} Galaxies where competing possibilities, such as, companion versus line of sight neighbor, asymmetry or star-forming knots, make classification difficult.
\end{itemize}

   Interactions, harassment, and outright mergers make classification difficult, thus some objects were originally categorized as indeterminate and final classification was performed after two-dimensional light profile fitting (\S~\ref{sec:Decomp}). We determine whether close neighbors are indeed companions with Keck spectroscopy in an upcoming paper \citep[\emph{in preparation}]{stoll11}. Figure~\ref{fig:mosaic} gives postage stamp images of our sample. Table \ref{tab:morph} summarizes the host galaxy morphologies, along with a choice of model fit which we will discuss in \S~\ref{sec:Decomp}. Descriptions of the individual targets are given in Appendix~B.

\begin{deluxetable}{clcccccl}
\tabletypesize{\scriptsize}
\tablecolumns{8}
\tablewidth{0pc}
\tablecaption{PSQ Host Morphologies}
\tablehead{
\colhead{Obj} & \colhead{SDSS} & \multicolumn{4}{c}{Visual Classification} &  \colhead{Adopted} & \colhead{Notes} \\
\cline{3-6}
\colhead{ID} & \colhead{Name} & \colhead{Morphology} & \colhead{Bar} & \colhead{Arm(s)} & \colhead{Disturbance} & \colhead{Model} & \colhead{} 
}
\startdata
1 & J003043$-$103517 & \begin{sideways} {\normalsize \S} \end{sideways} & \checkmark & \checkmark & & B+D & Grand-design \\
2 & J005739+010044 & \begin{sideways} {\normalsize \S} \end{sideways} & & \checkmark & & B+D & Grand-design, Asymmetric \\
3 & J020258$-$002807 & ? & & & ! & B+D & Asymmetric, SF arc \\
4 & J021447$-$003250 & \begin{sideways} {\normalsize 0} \end{sideways} & & & & S\'{e}rsic & \\
5 & J023700$-$010130 & \begin{sideways} {\normalsize \S} \end{sideways} & & \checkmark & ! & B+D & Ring, SF knot \\
6 & J040210$-$054630 & ? & & & \checkmark & S\'{e}rsic & Tidal tail/Intersecting companion? \\
7 & J074621+335040 & \begin{sideways} {\normalsize 0} \end{sideways} & & & & S\'{e}rsic & \\
8 & J075045+212546 & \begin{sideways} {\normalsize 0} \end{sideways} & & & \checkmark & S\'{e}rsic & Shells \\
9 & J075521+295039 & \begin{sideways} {\normalsize \S} \end{sideways} & & \checkmark & & B+D & Flocculent, Asymmetric \\
10 & J075549+321704 & \begin{sideways} {\normalsize \S} \end{sideways} & \checkmark & \checkmark & & S\'{e}rsic & \\
11 & J081018+250921 & \begin{sideways} {\normalsize 0} \end{sideways} & & & \checkmark & S\'{e}rsic & \\
12 & J105816+102414 & \begin{sideways} {\normalsize 0} \end{sideways} & & & & S\'{e}rsic & \\
13 & J115159+673604 & \begin{sideways} {\normalsize \S} \end{sideways} & \checkmark & \checkmark & & B+D & Grand-design \\
14 & J115355+582442 & \begin{sideways} {\normalsize 0} \end{sideways} & & & & S\'{e}rsic & \\
15 & J123043+614821 & ? & & & ? & S\'{e}rsic & Flocculent \\
16 & J124833+563507 & \begin{sideways} {\normalsize 0} \end{sideways} & & & ! & S\'{e}rsic & Shells \\
17 & J145640+524727 & \begin{sideways} {\normalsize \S} \end{sideways} & \checkmark & \checkmark & & B+D & Grand-design, Asymmetric, Fine structure \\
18 & J145658+593202 & \begin{sideways} {\normalsize \S} \end{sideways}? & & & ! & B+D & 1 Grand-design?, Merger of 2-3 \begin{sideways} {\normalsize \S} \end{sideways} \\
19 & J154534+573625 & \begin{sideways} {\normalsize \S} \end{sideways}? & & & & B+D & \\
20 & J164444+423304 & \begin{sideways} {\normalsize 0} \end{sideways} & & & & S\'{e}rsic & \\
21 & J170046+622056 & \begin{sideways} {\normalsize \S} \end{sideways} & & & \checkmark & B+D & Ring, Dust lane, Merging companion \\
22 & J210200+000501 & \begin{sideways} {\normalsize 0} \end{sideways} & & & \checkmark & S\'{e}rsic & Tidal tail \\
23 & J211343$-$075017 & \begin{sideways} {\normalsize 0} \end{sideways} & & & & S\'{e}rsic & Companion \\
24 & J211838+005640 & \begin{sideways} {\normalsize 0} \end{sideways} & & & & S\'{e}rsic & \\
25 & J212843+002435 & \begin{sideways} {\normalsize 0} \end{sideways} & & & & S\'{e}rsic & Companion \\
26 & J230614$-$010024 & \begin{sideways} {\normalsize \S} \end{sideways} & & \checkmark ? & & B+D & Flocculent, 1 Arm, Tidal features, Companion \\
27 & J231055$-$090107 & \begin{sideways} {\normalsize \S} \end{sideways} & \checkmark & \checkmark & \checkmark & B+D & Grand-design?, Merging companion \\
28 & J233430+140649 & \begin{sideways} {\normalsize 0} \end{sideways} & & & \checkmark & S\'{e}rsic & Tidal feature \\
29 & J234403+154214 & \begin{sideways} {\normalsize \S} \end{sideways} & & & & B+D & Flocculent \\
\enddata
\tablecomments{ \begin{sideways} {\normalsize 0} \end{sideways}, \begin{sideways} {\normalsize \S} \end{sideways},  \begin{sideways} {\normalsize \S} \end{sideways}?, and ? represent early-type, spiral, `probable' spiral, and indeterminate host morphology, respectively. A \checkmark\ indicates the presence of a bar, spiral arm(s), and/or disturbance. We denote highly disturbed objects with a !. A ? denotes that there is evidence suggesting the property exists but resolution limits a definite classification. We discuss our adopted model in Table~\ref{sec:Decomp.Picks}. \label{tab:morph}}
\end{deluxetable}
\clearpage

\section{Quasar-Host Decomposition}
\label{sec:Decomp}

   We decompose the quasar/host of the PSQ to determine whether there is more fine structure in the nuclear regions as well as to determine the quasar-to-host galaxy light contributions which will be necessary to constrain spectral modeling of the sample \citep[\emph{in preparation}]{cales11}. We use a point-spread function (PSF) to model the AGN/quasar core of a PSQ. We then fit a variety of 2-dimensional light profiles to the host and prescribe a method for choosing the best fitting model.


\subsection{Modeling the Point Spread Function}
\label{sec:Decomp.Real}
   
   We followed the prescription set forth by \citet{canalizo07} to generate an empirical PSF model which we describe briefly here. The star was chosen from the \emph{HST} archive (GO-9433) to be within 30 pixels of the target position of ACS/WFC and processed in a standard manner. The PSF was then adaptively smoothed by comparing the data to the standard deviation $s$ of the sky, according to the following: i) for data of high signal ($> 7s$), the PSF was unmodified; ii) for values between 3$s$ and 7$s$, a Gaussian kernel of $\sigma = 0.5$ pixel was used to smooth the image, iii) for data of  $< 3s$ we used a Gaussian kernel of $\sigma = 2.0$ pixels and finally, iv) after this last step if the data values were less than $1s$ the value was replaced by zero. The unaltered PSF was retained to test the effects of the adaptive smoothing.
   
   We also generate Tiny Tim \citep{kristhook04} synthetic HST PSFs. Through extensive testing of the three types of PSFs we found the empirical models (adaptively smoothed and unaltered versions) yielded superior models and thus we preferred these PSFs to the Tiny Tim PSFs.

\subsection{Two-Dimensional Image Analysis}
\label{sec:Decomp.Galfit}

   GALFIT \citep{peng02} is a 2-dimensional light profile fitting program utilizing chi-squared minimization to simultaneously fit components in an image having different light distributions (e.g., sky, PSF, de Vaucouleurs 1948, S\'{e}rsic 1968). The S\'{e}rsic power law is a generalized power law defined by
   \begin{equation}
   \Sigma(r) = \Sigma _{eff} \exp \left[-\kappa\left(\left(\frac{r}{r_{eff}}\right)^{1/n} - 1\right)\right].
   \end{equation}
The pixel surface brightness at the effective radius $r_{eff}$ is $\Sigma _{eff}$. The effective radius is defined such that half of the flux is within $r_{eff}$ and constrains $\kappa$ to be coupled to the S\'{e}rsic index, $n$. The S\'{e}rsic index, also called the concentration index, describes the shape or concentration of the brightness profile. A large index gives a steeply sloping profile towards small radii with extended wings. Conversely, small values of $n$ have shallow inner profiles with steep truncation at large radius. Special cases of the S\'{e}rsic power law are the exponential ($n = 1$), de Vaucouleurs ($n = 4$), and  Gaussian profile ($n = 0.5$) used to fit the disk, bulge, and PSF of galaxies, respectively. 

   We decompose the image to determine the quasar-to-host light fraction, host morphology, and other host parameters. The method is as follows. First, we created a mask to exclude the surrounding objects. Then, as suggested by \citet{peng02}, we begin by fitting a PSF simultaneously with the sky. In GALFIT the fitting functions/light distributions are convolved with the PSF to simulate blurring caused by the telescope optics. We then introduce a number of different models: i) S\'{e}rsic, ii)  de Vaucouleurs, iii) Bulge-plus-Disk ($n = 4$ and $n = 1$), and the fit is then recalculated. The modeled components are all initially centered on the position of the quasar, however, in subsequent iterations the centroid is free to move.  
   
   For each light component the parameters of the fit include: centroid of the component, magnitude, effective radius, S\'{e}rsic index, axis ratio, and position angle. All of the components are free to vary. The exception to this rule is when the fitting routine settles on large values of the S\'{e}rsic index. Large S\'{e}rsic values are generally due to a mismatch between the data and the galaxy component (or PSF, or sky) model. While there is no defined number for the physical/unphysical boundary, C. Peng (\emph{private communication}) mentions that above $n = 5$ Sersic values are generally unreliable/unconvincing. Hence, the S\'{e}rsic values above $n = 5$ are reset to the de Vaucouleurs ($n = 4$) profile with little effect on the reduced-chi squared ($\chi_{\nu}^2$) value.
      
   We fit the host galaxy component with simple light distributions. We note that it is not the goal of this study to make detailed model fits to every feature of the targets. Subtracted simple light distribution models from the data allows us to see asymmetries and fine structure such as tidal tails, shells, star-forming knots, rings, bars, and spiral arms.

\subsection{Model Choice}
\label{sec:Decomp.Picks}

  There are many permutations between the two types of PSFs (adaptively smoothed and unaltered versions) and the three options for host components (S\'{e}rsic,  de Vaucouleurs, and Bulge-plus-Disk). Our visual morphological classifications can help discriminate in cases where $\chi_{\nu}^2$ values of model choices are similar. We outline a method for choosing the model that best describes the host and quasar features of the PSQ sample. 
  
\begin{enumerate}
\item We fit the PSQ images by varying the type of PSF and host galaxy component. This generates a library of fits from which we choose the lowest reduced-chi squared ($\chi_{\nu}^2$) value. 
\item PSF: The chosen PSFs for the models were either the empirical unaltered or adaptively smoothed PSF based on the best $\chi_{\nu}^2$ model. Generally, the adaptively smoothed PSF generated the best fit (S\'{e}rsic $\sim$27/29 and Bulge-plus-Disk $\sim$24/29), however is sometimes made a difference in the $\chi_{\nu}^2$ if the unaltered PSF was used. Thus, on rare occasions we chose the unaltered PSF.
\item Host component: We compare the model with the best  $\chi_{\nu}^2$ to the original morphological classifications. In our subsequent analysis we adopt a S\'{e}rsic model for early-types and a Bulge-plus-Disk model for spirals (spiral galaxies) and probable spirals. In the case of Object 10 we were unable to run a successful Bulge-plus-Disk model, thus, we were forced to choose the simpler S\'{e}rsic model. For the rest of the objects, the model adopted based upon our visual classification also yields the best $\chi_{\nu}^2$.
\end{enumerate}

\begin{figure}[tbhp]
\centering
\figurenum{3} 
\includegraphics[width=4in]{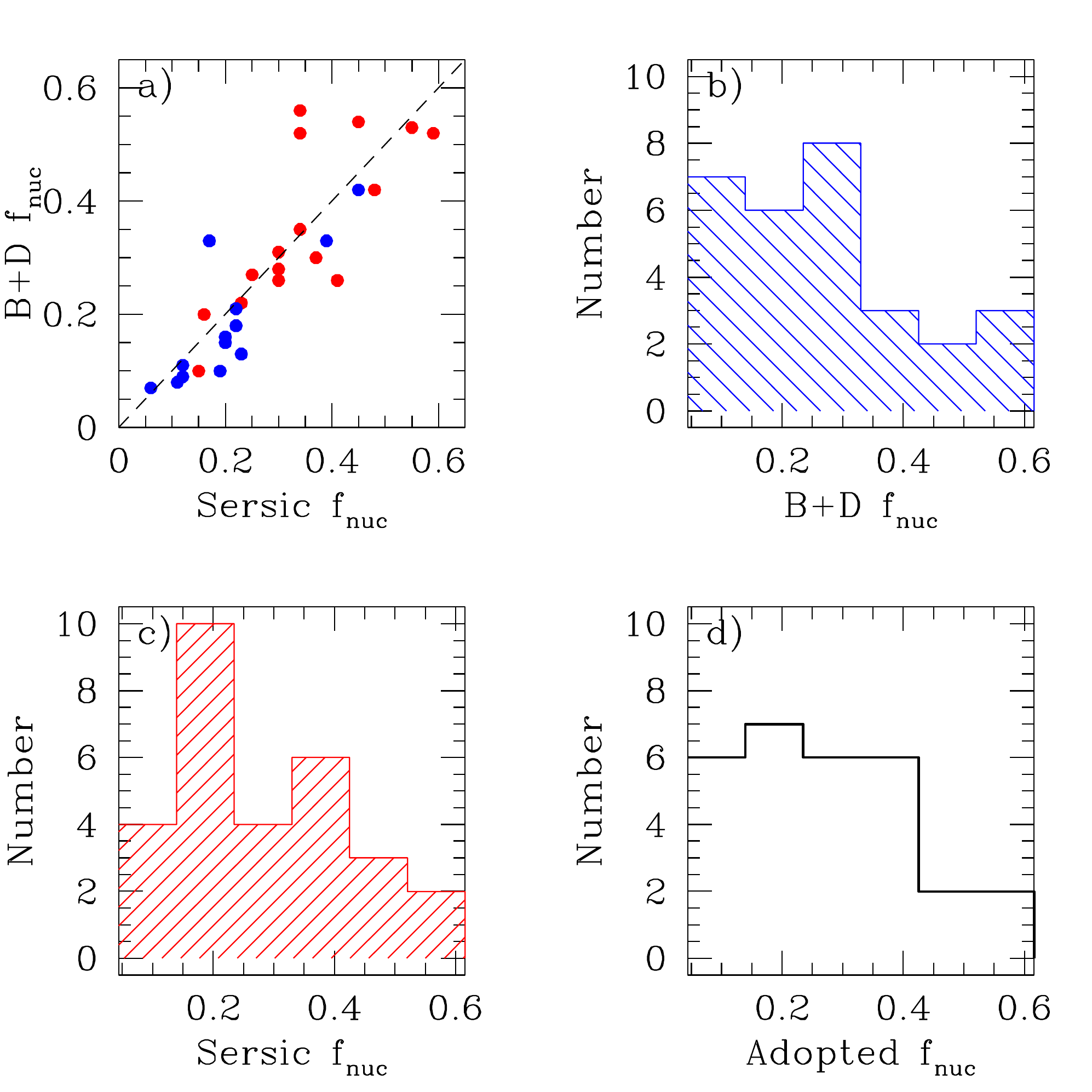}
\caption{a) Comparison of the nuclear light fraction determined by the S\'{e}rsic and Bulge-plus-Disk model runs. Red points are PSQs with the S\'{e}rsic model adopted as the best fit. Blue points are PSQs with the Bulge-plus-Disk model adopted as the best fit. The dashed line has slope of unity. b) The distribution of Bulge-plus-Disk model nuclear light fractions. c) The distribution of  S\'{e}rsic model nuclear light fractions. d) The distribution of adopted model nuclear light fractions. \label{fig:hist_FnucVmodel}}
\end{figure}

  Tables~\ref{tab:sresults} and~\ref{tab:bdresults} give the AGN plus host galaxy modeling parameter results for the S\'{e}rsic and Bulge-plus-Disk model runs, respectively. The fitted parameters for each adopted model are denoted with an asterisk in the model parameter results tables. The residual images of the adopted models that best describe AGN plus host galaxy components are given in Figure~\ref{fig:Sersic}. Figure~\ref{fig:hist_FnucVmodel} compares the nuclear light fraction for the two models as well as the adopted models which are formed from a combination of them. We show that in most cases the nuclear light fraction is not overly sensitive to the adopted model.  Thus, the Quasar-Host decomposition is fairly robust.

\label{sec:Decomp.AbsMag}
  
   As in \S~\ref{sec:Data.Reduc}, we use measurements of continuum slope of the Galactic dereddened SDSS DR7 target spectra to make the K-corrections to the PSF and host component magnitudes. In order to determine K-corrected host galaxy absolute magnitudes, we need to estimate the spectral shape of AGN and host component separately. We assume the AGN component to have the spectral shape of the SDSS quasar composite of \citet{vandenberk01}, scaled it by $f_{nuc}$ and then subtracted from our Galactic dereddened SDSS DR7 spectra. For S\'{e}rsic or de Vaucouleurs fits, we use the resulting spectrum to determine the K-correction. For hosts with Bulge-plus-Disk components, we assume spectral shapes consistent with the \citet{kinney96} bulge and Sc disk templates for the PSQ host bulge and disk components, respectively.


\begin{figure}[tbhp]
   \centering
   \figurenum{2} 
      \subfloat[][Fig. 2$a$]{\includegraphics[width=6in]{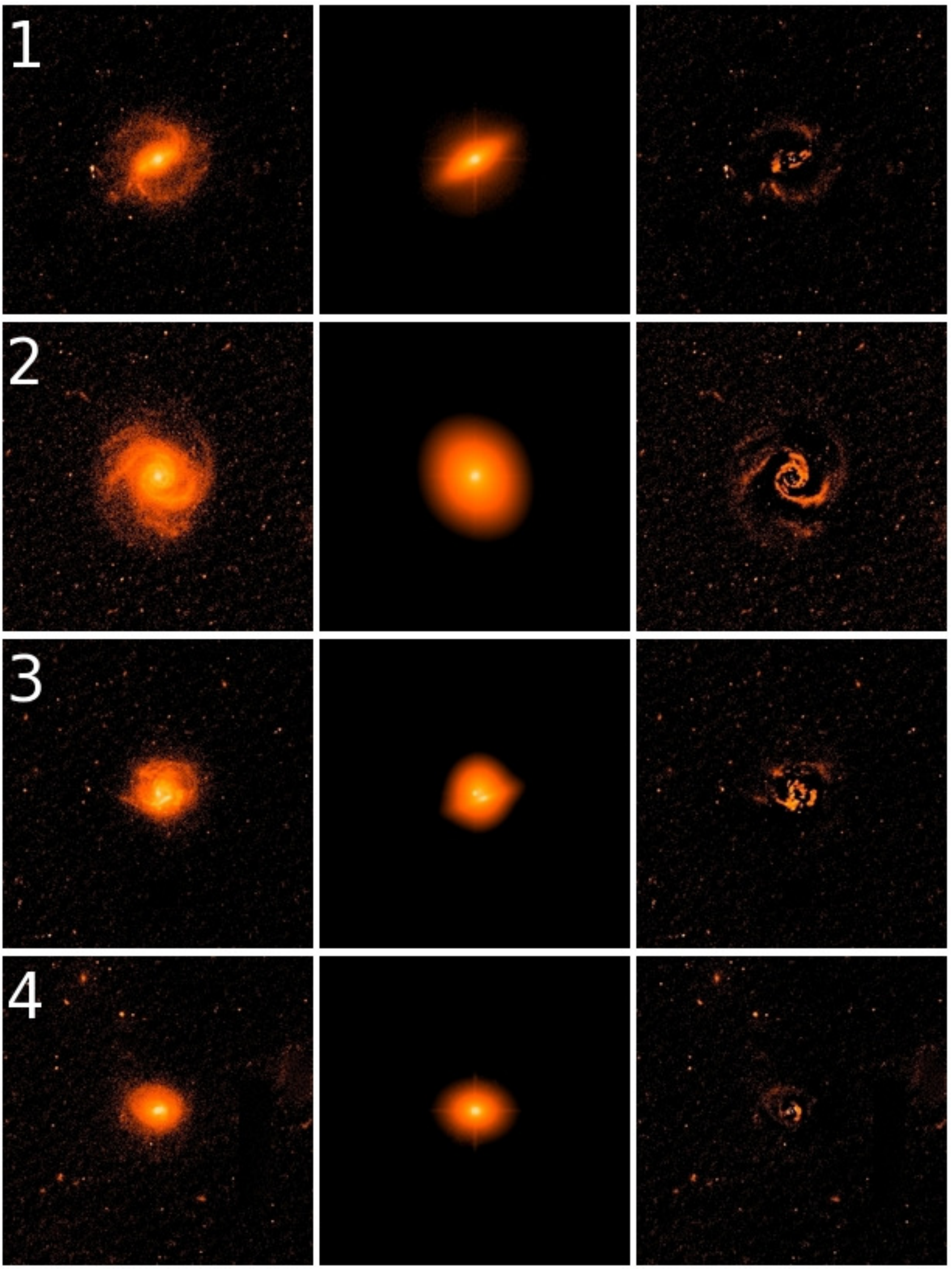}}
      \caption{Raw (\emph{left}), GALFIT model (\emph{middle}), and residual (\emph{right}) images of the PSQs. The images are 20\arcs $\times$ 20\arcs. We note here that in order to retain the orientation of the diffraction spikes the images are not North aligned as in Figure~\ref{fig:mosaic} but oriented as the targets were originally imaged on the ACS/WFC. Companion galaxies are masked prior to fitting. \label{fig:Sersic}}
\end{figure}
\begin{figure}
   \ContinuedFloat
   \centering
      \subfloat[][Fig. 2$b$]{\includegraphics[width=6in]{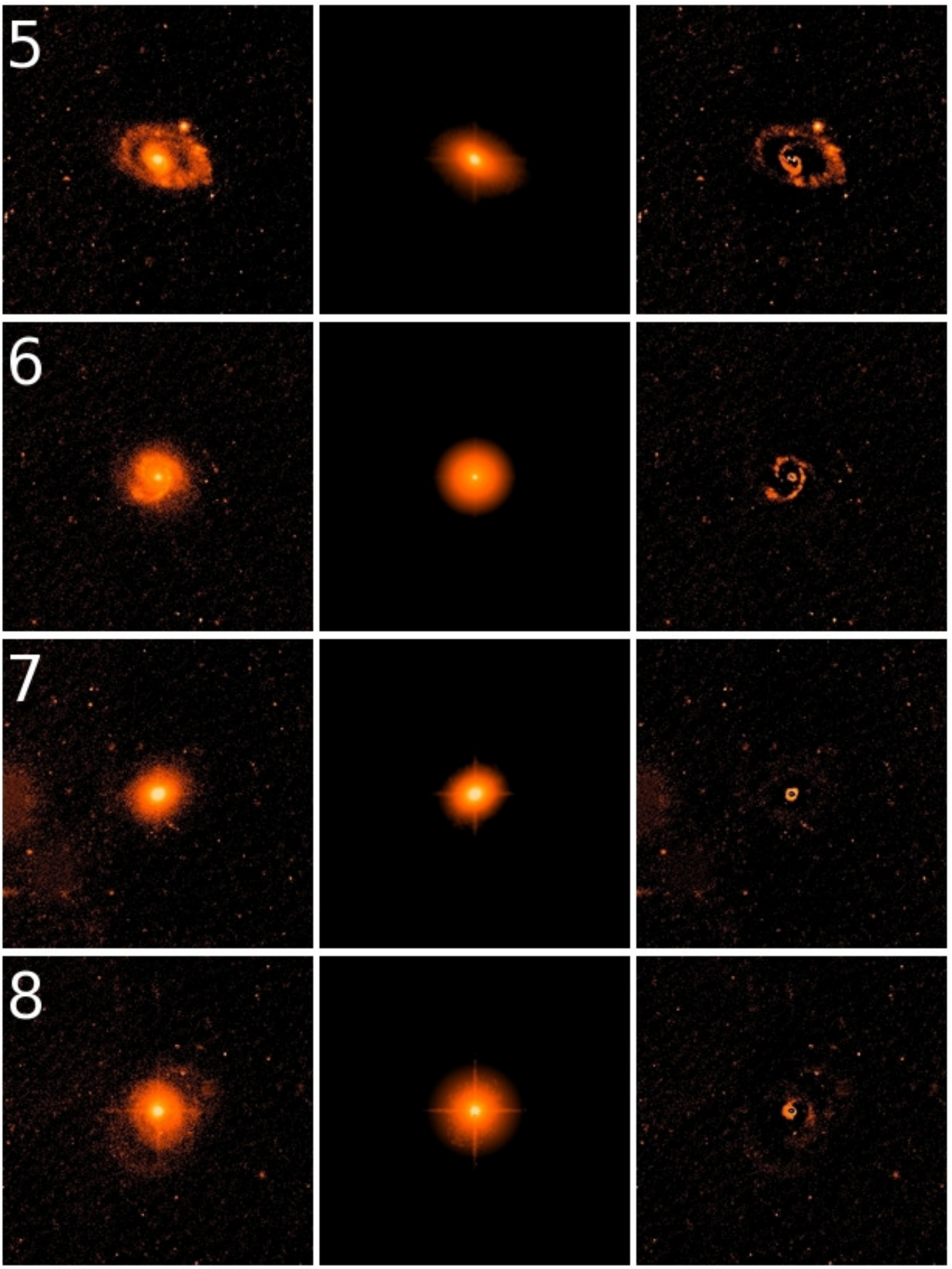}}
      \end{figure}
\begin{figure}
   \ContinuedFloat
   \centering
      \subfloat[][Fig. 2$c$]{\includegraphics[width=6in]{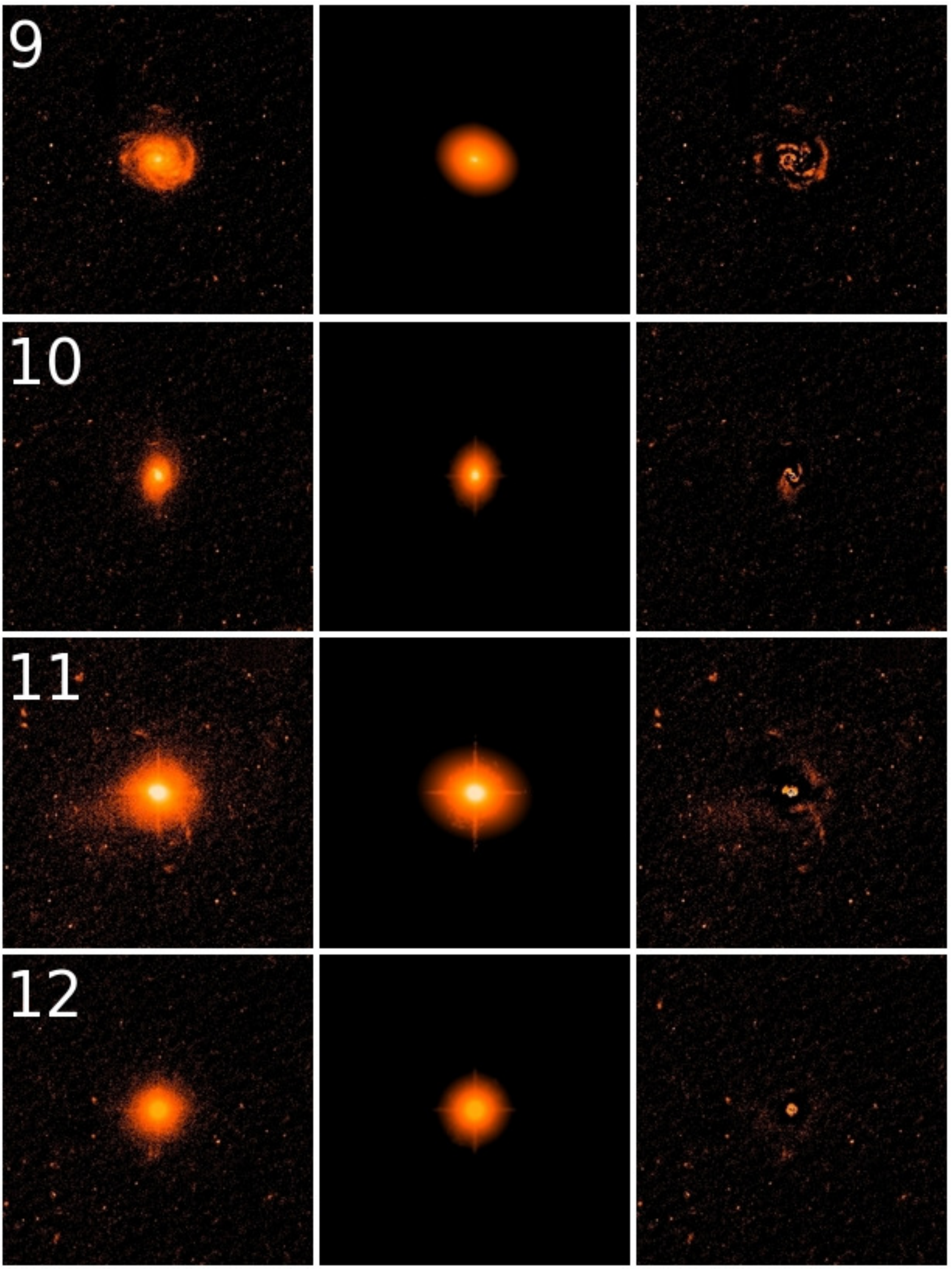}}
      \end{figure}
\begin{figure}
   \ContinuedFloat
   \centering
      \subfloat[][Fig. 2$d$]{\includegraphics[width=6in]{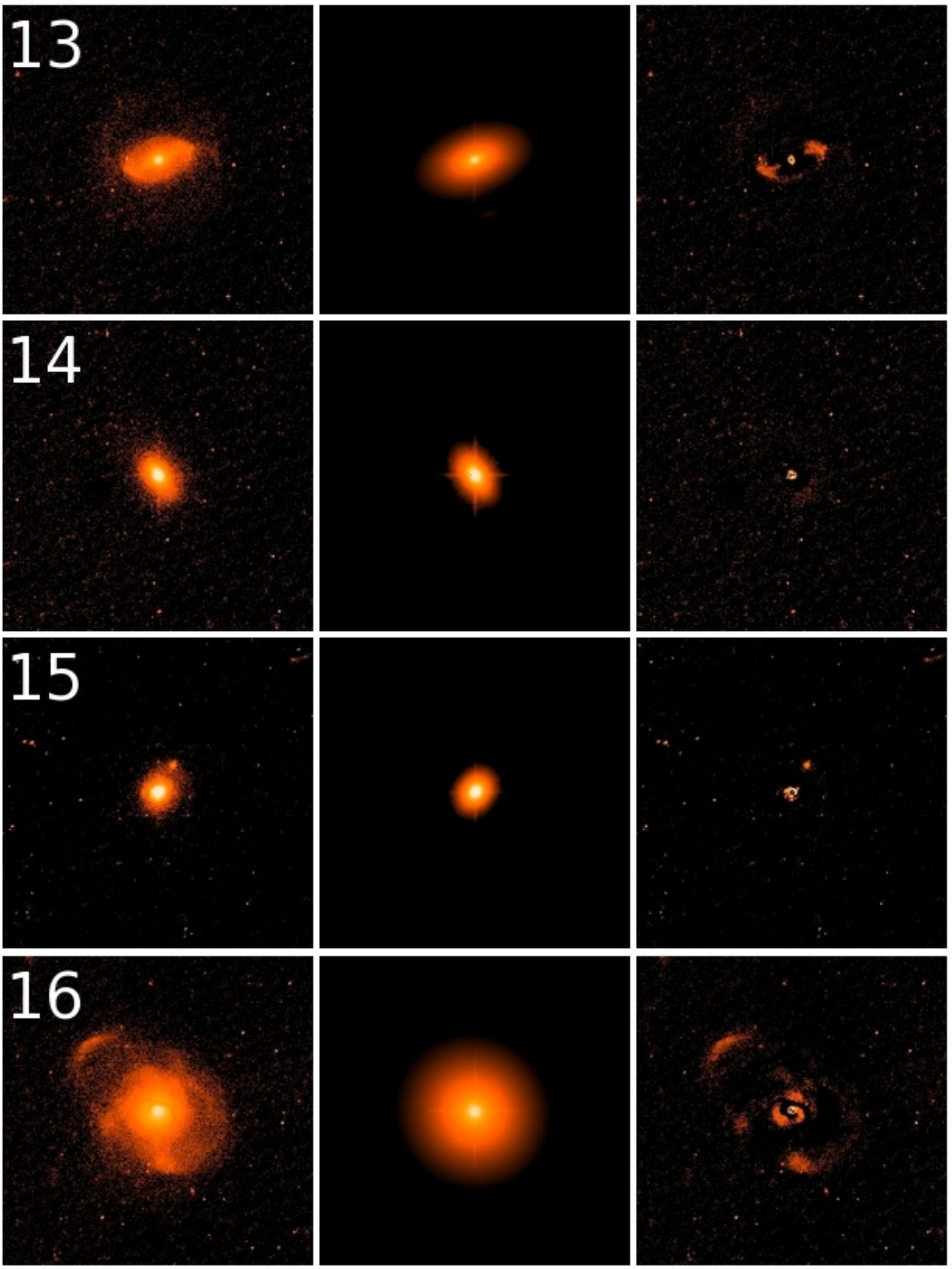}}
      \end{figure}
\begin{figure}
   \ContinuedFloat
   \centering
      \subfloat[][Fig. 2$e$]{\includegraphics[width=6in]{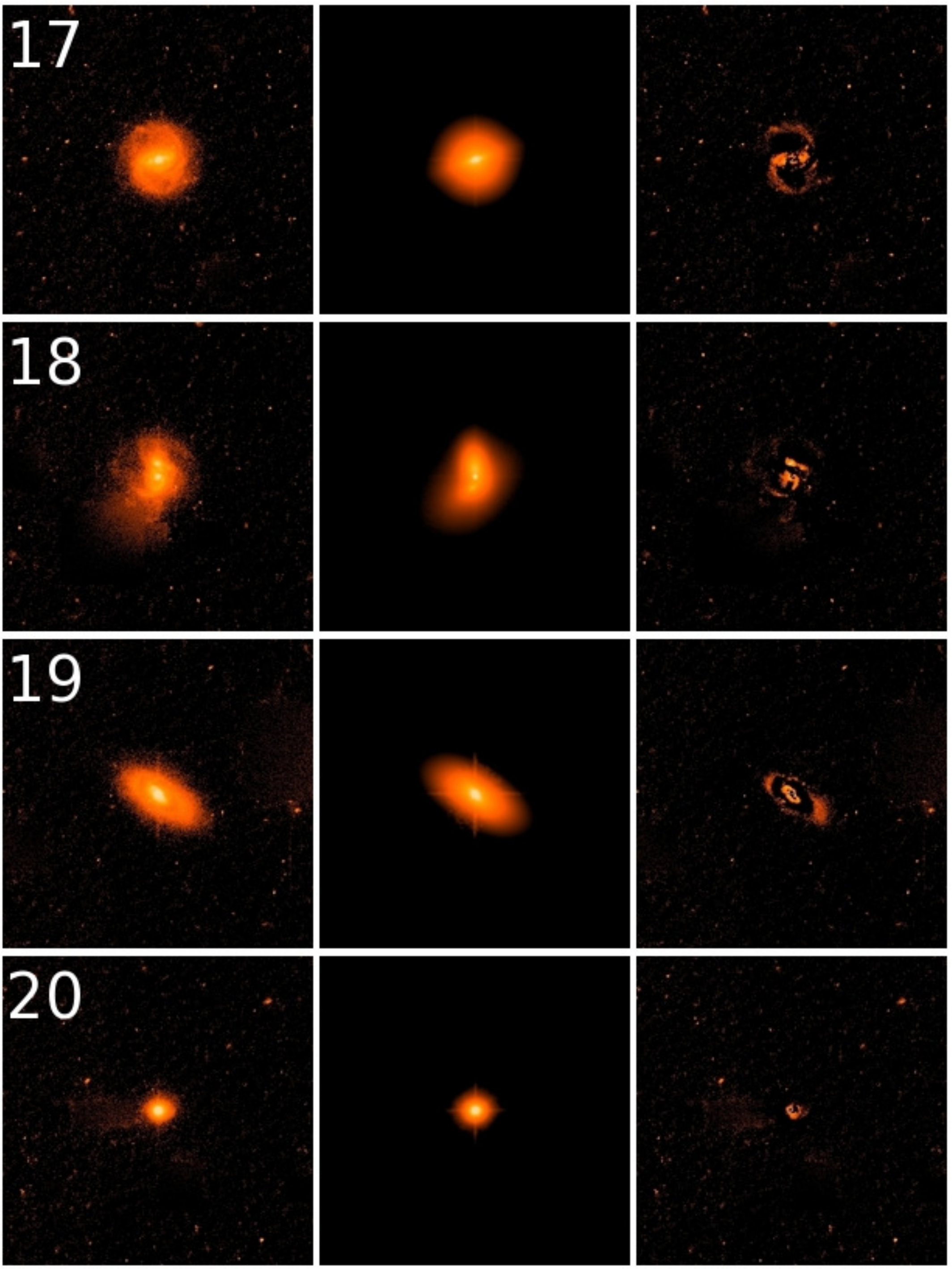}}
      \end{figure}
\begin{figure}
   \ContinuedFloat
   \centering
      \subfloat[][Fig. 2$f$]{\includegraphics[width=6in]{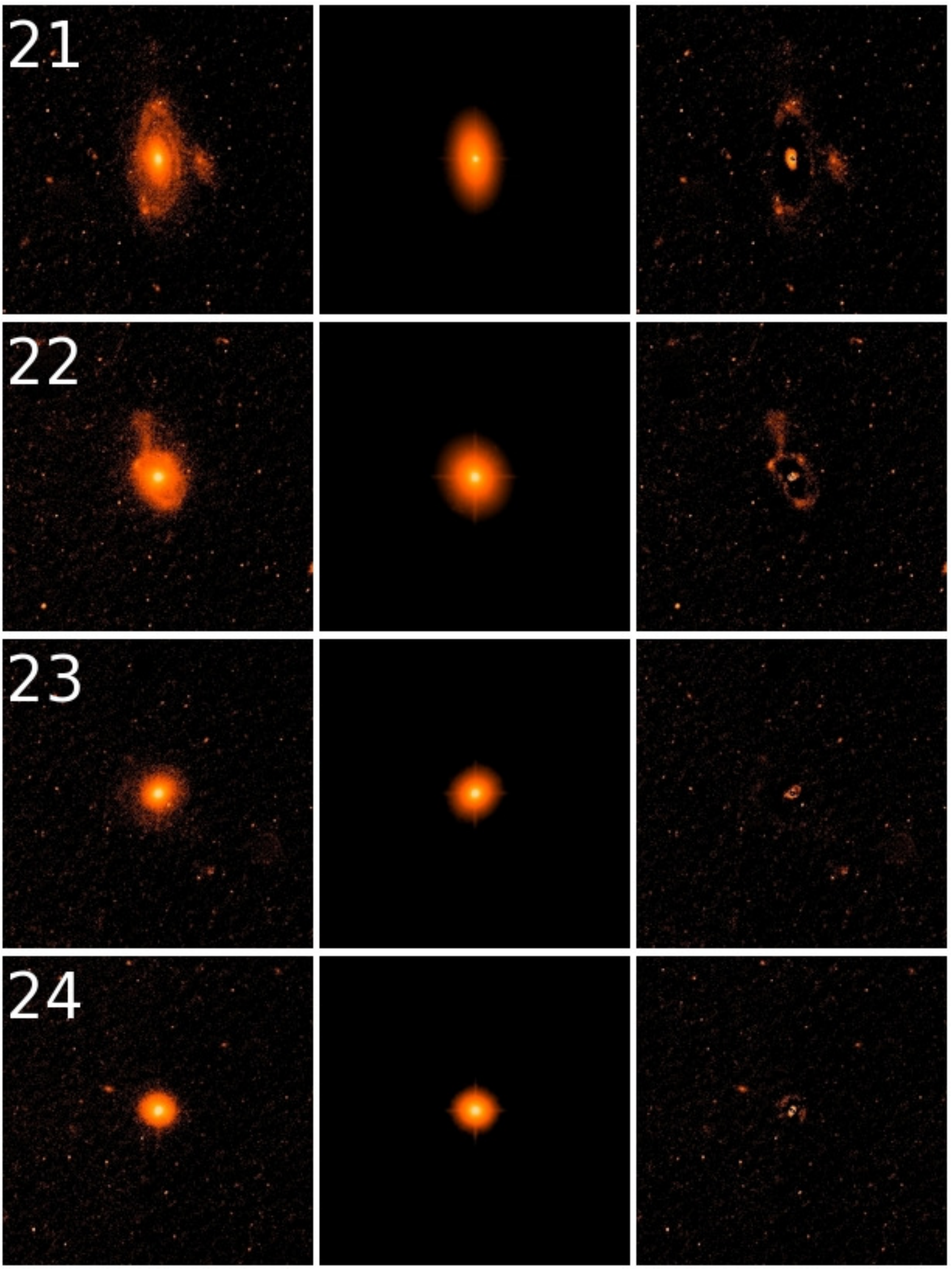}}
      \end{figure}
\begin{figure}
   \ContinuedFloat
   \centering
      \subfloat[][Fig. 2$g$]{\includegraphics[width=6in]{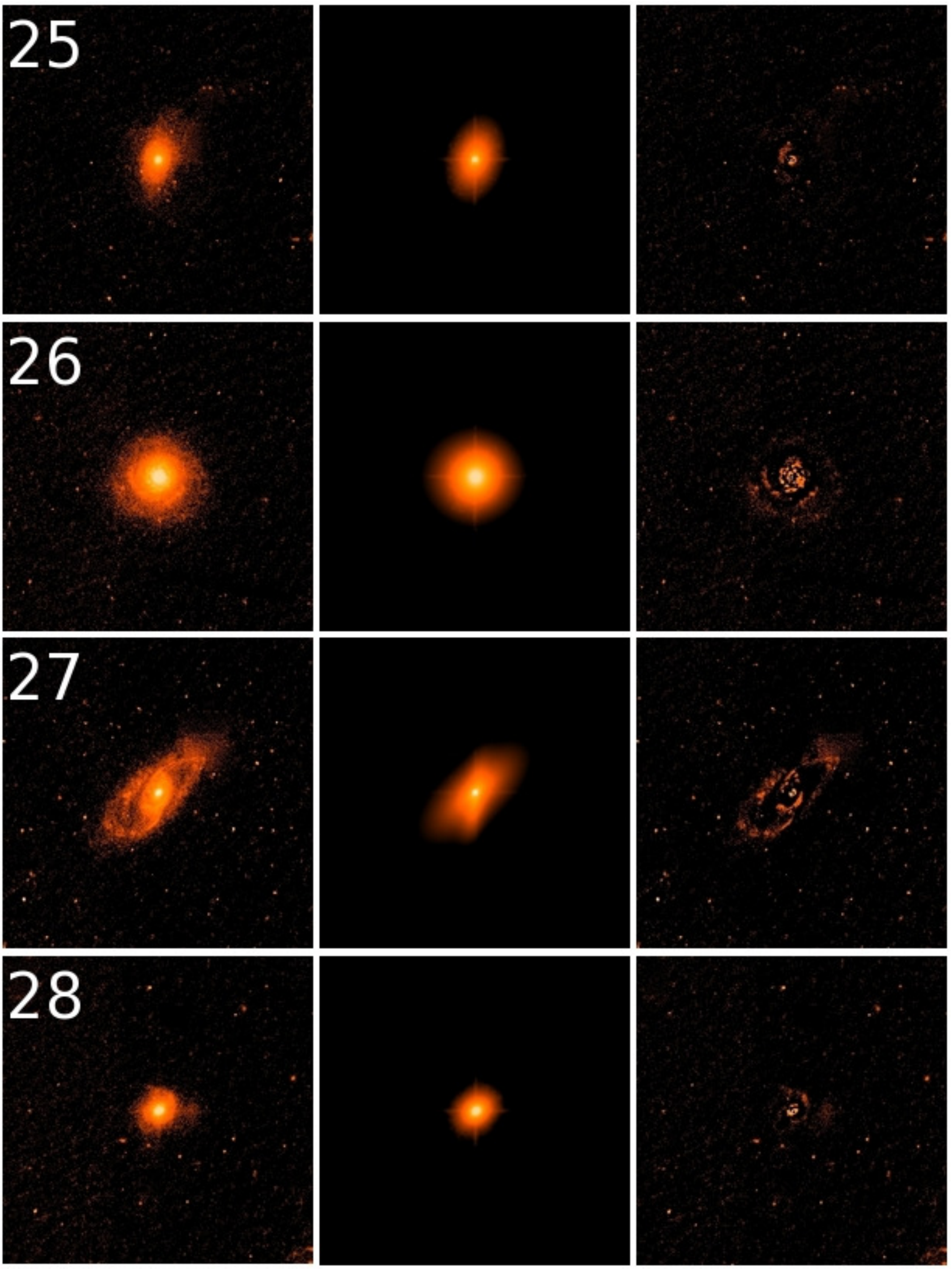}}
      \end{figure}
\begin{figure}
   \ContinuedFloat
   \centering
      \subfloat[][Fig. 2$h$]{\includegraphics[width=6in]{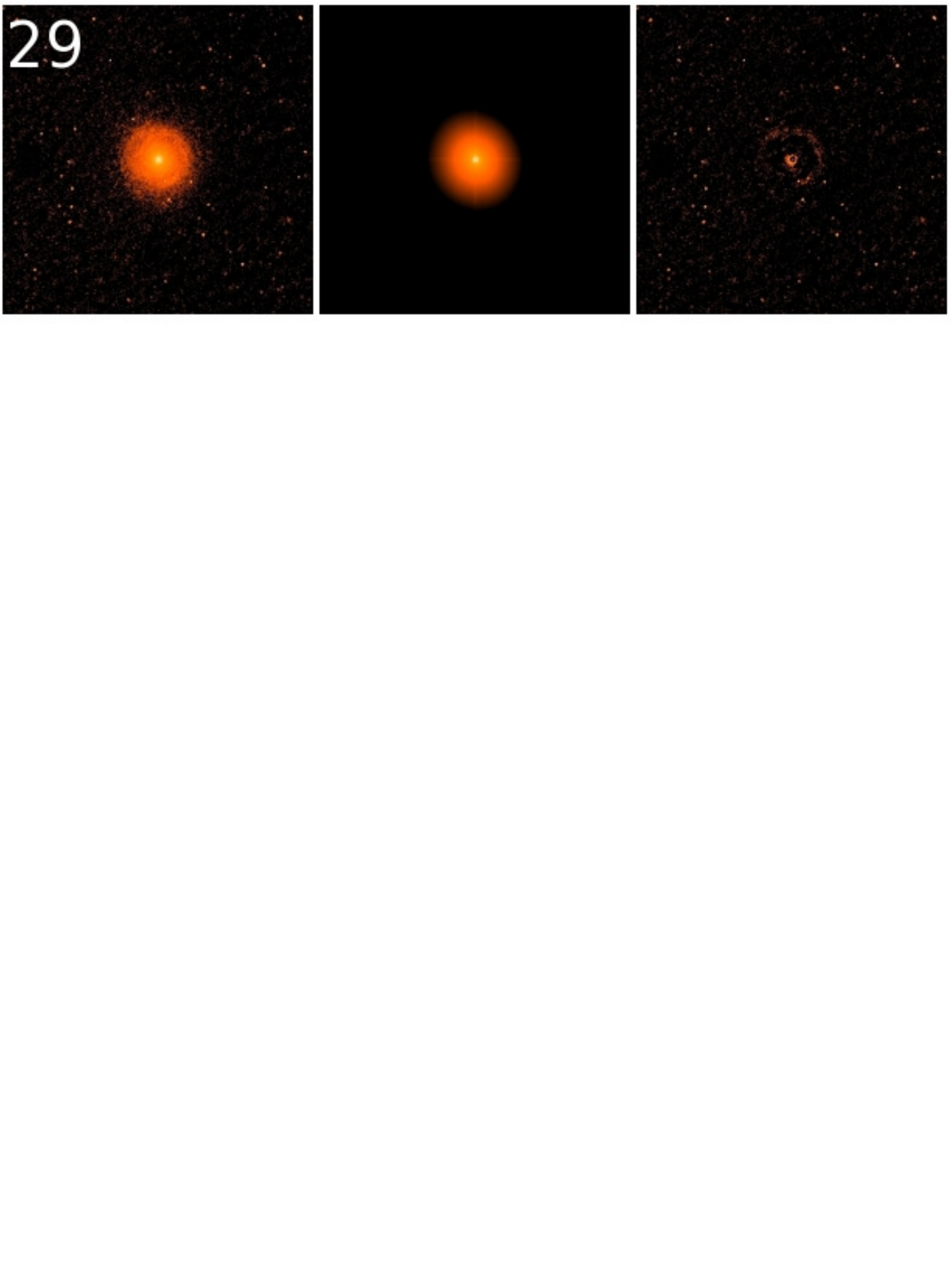}}
\end{figure}

\begin{deluxetable}{clccccrrrr}
\tabletypesize{\tiny}
\tablecolumns{10}
\tablewidth{0pc}
\tablecaption{PSQ Host Galaxy S\'{e}rsic Modeling Results}
\tablehead{
\colhead{} & \colhead{} & \multicolumn{1}{c}{PSF} & \colhead{} & \multicolumn{6}{c}{Host/S\'{e}rsic Component} \\
\cline{3-3} \cline{5-10}
\colhead{Obj} & \colhead{SDSS} & \colhead{$M_{\textrm{\tiny F606W}}$} & \colhead{} &  \colhead{$M_{\textrm{\tiny F606W}}$} & \colhead{$f_{nuc}$} & \colhead{$r_{eff}$} & \colhead{$n$} & \colhead{$b/a$} & \colhead{P.A.} \\
\colhead{ID} & \colhead{Name} & \colhead{(mag)} & \colhead{} & \colhead{(mag)} & \colhead{} &  \colhead{(kpc)} & \colhead{} & \colhead{} & \colhead{(\arcdeg)} \\
\colhead{(1)} & \colhead{(2)} & \colhead{(3)} & \colhead{} & \colhead{(4)} & \colhead{(5)} & \colhead{(6)} & \colhead{(7)} & \colhead{(8)} & \colhead{(9)}
}
\startdata
1 & J003043$-$103517 & $-$21.3 & & $-$22.1 & 0.39 & 4.51 & 1.54 & 0.4786 & $-$58.67  \\
2 & J005739+010044 & $-$20.1 & & $-$22.5 & 0.12 & 6.21 & 1.51 & 0.8620 & 27.28  \\
3 & J020258$-$002807 & $-$20.0 & & $-$22.6 & 0.11 & 3.52 & 0.78 & 0.9514 & $-$61.59  \\
4 & J021447$-$003250* & $-$21.2 & & $-$22.2 & 0.34 & 4.24 & 2.23 & 0.7771 & 87.44  \\
5 & J023700$-$010130 & $-$20.6 & & $-$22.3 & 0.20 & 3.97 & 4.00 & 0.6811 & 52.61  \\
6 & J040210$-$054630* & $-$19.7 & & $-$21.7 & 0.16 & 4.45 & 0.88 & 0.9672 & $-$80.98  \\
7 & J074621+335040* & $-$20.9 & & $-$22.2 & 0.25 & 1.38 & 3.12 & 0.7904 & $-$63.84  \\
8 & J075045+212546* & $-$23.1 & & $-$23.3 & 0.59 & 8.96 & 2.80 & 0.9835 & $-$80.21  \\
9 & J075521+295039 & $-$19.2 & & $-$22.3 & 0.06 & 4.89 & 1.05 & 0.8127 & 65.76  \\
10 & J075549+321704* & $-$21.7 & & $-$22.4 & 0.41 & 3.85 & 3.83 & 0.7334 & $-$4.55  \\
11 & J081018+250921* & $-$21.9 & & $-$22.4 & 0.37 & 2.26 & 4.00 & 0.8061 & 79.94  \\
12 & J105816+102414* & $-$20.1 & & $-$22.2 & 0.15 & 1.46 & 4.00 & 0.9797 & $-$41.72  \\
13 & J115159+673604 & $-$20.3 & & $-$21.9 & 0.23 & 5.96 & 2.22 & 0.5121 & $-$71.77  \\
14 & J115355+582442* & $-$21.5 & & $-$21.9 & 0.45 & 2.22 & 4.00 & 0.5999 & 32.56  \\
15 & J123043+614821* & $-$21.0 & & $-$22.1 & 0.30 & 3.17 & 4.00 & 0.7866 & $-$34.83  \\
16 & J124833+563507* & $-$21.4 & & $-$23.0 & 0.23 & 8.72 & 2.70 & 0.9568 & 46.34  \\
17 & J145640+524727 & $-$20.3 & & $-$22.0 & 0.20 & 4.65 & 1.40 & 0.8184 & $-$71.79  \\
18 & J145658+593202 & $-$19.8 & & $-$22.2 & 0.12 & 5.14 & 1.10 & 0.5596 & $-$5.69  \\
19 & J154534+573625 & $-$21.5 & & $-$22.0 & 0.45 & 4.97 & 1.90 & 0.4994 & 55.49  \\
20 & J164444+423304* & $-$20.7 & & $-$21.6 & 0.34 & 0.83 & 4.13 & 0.8653 & $-$80.95  \\
21 & J170046+622056 & $-$20.0 & & $-$21.9 & 0.17 & 9.18 & 4.00 & 0.5299 & 4.81  \\
22 & J210200+000501* & $-$21.3 & & $-$22.3 & 0.30 & 4.63 & 3.75 & 0.8607 & 22.74  \\
23 & J211343$-$075017* & $-$21.5 & & $-$22.7 & 0.30 & 2.63 & 3.43 & 0.8661 & $-$47.48 \\
24 & J211838+005640* & $-$22.0 & & $-$22.5 & 0.55 & 2.28 & 1.30 & 0.9070 & 76.10  \\
25 & J212843+002435* & $-$21.5 & & $-$21.9 & 0.48 & 7.31 & 1.89 & 0.6474 & $-$15.48  \\
26 & J230614$-$010024 & $-$20.9 & & $-$22.5 & 0.22 & 3.13 & 2.00 & 0.9611 & $-$29.18 \\
27 & J231055$-$090107 & $-$20.8 & & $-$23.8 & 0.19 & 13.51 & 1.85 & 0.4974 & $-$40.57 \\
28 & J233430+140649* & $-$21.4 & & $-$22.9 & 0.34 & 1.65 & 3.84 & 0.7597 & $-$49.09  \\
29 & J234403+154214 & $-$20.3 & & $-$22.1 & 0.22 & 6.03 & 1.27 & 0.8930 & 28.25  \\
\enddata
\tablenotetext{*}{Denotes the best/adopted model of the S\'{e}rsic, de Vaucouleurs and Bulge-plus-Disk model runs.}
\tablecomments{Col. (1): Object ID number. Col. (2): SDSS name. Col. (3): Integrated Galactic dereddened K-corrected absolute ST magnitude of the PSF in the F606W filter. Col. (4): Integrated Galactic dereddened K-corrected absolute ST magnitude of the host in the F606W filter. Col. (5): PSF to total light fraction. Col. (6): Effective radius in kpc. Col. (7): S\'{e}rsic index. Col. (8): Axis ratio. Col. (9): Position angle (east of north). \label{tab:sresults}}
\end{deluxetable}
\clearpage

\begin{deluxetable}{clcccrccrlcrccr}
\tabletypesize{\tiny}
\tablecolumns{15}
\tablewidth{0pc}
\tablecaption{PSQ Host Galaxy Bulge$+$Disk Modeling Results}
\tablehead{
\colhead{} & \colhead{} & \multicolumn{1}{c}{PSF} & \colhead{} & \multicolumn{5}{c}{Bulge Component}  & \colhead{} & \multicolumn{5}{c}{Disk Component}  \\
\cline{3-3} \cline{5-9} \cline{11-15}
\colhead{Obj} & \colhead{SDSS} & \colhead{$M_{\textrm{\tiny F606W}}$} &  \colhead{$f_{nuc}$} & \colhead{$M_{\textrm{\tiny F606W}}$} & \colhead{$r_{eff}$} & \colhead{$n$} & \colhead{$b/a$} & \colhead{P.A.} & \colhead{} & \colhead{$M_{\textrm{\tiny F606W}}$} & \colhead{$r_{eff}$} & \colhead{$n$} & \colhead{$b/a$} & \colhead{P.A.}\\
\colhead{ID} & \colhead{Name} & \colhead{(mag)} & \colhead{Fraction} & \colhead{(mag)} & \colhead{(kpc)} & \colhead{} & \colhead{} & \colhead{(\arcdeg)}  & \colhead{} & \colhead{(mag)} & \colhead{(kpc)} & \colhead{} & \colhead{} & \colhead{(\arcdeg)} \\
\colhead{(1)} & \colhead{(2)} & \colhead{(3)} &\colhead{(4)} & \colhead{(5)} & \colhead{(6)} & \colhead{(7)} & \colhead{(8)}  & \colhead{(9)} & \colhead{} & \colhead{(10)} & \colhead{(11)} & \colhead{(12)} & \colhead{(13)} & \colhead{(14)}
}
\startdata
1 & J003043$-$103517* & $-$21.3 & 0.33 & $-$21.8 & 10.99 & 4.0 & 0.8833 & $-$79.15 & & $-$21.1 & 2.50 & 1.00 & 0.3653 & $-$58.34  \\
2 & J005739+010044* & $-$20.1 & 0.11 & $-$21.9 & 8.91 & 4.0 & 0.8096 & $-$40.41 & & $-$21.7 & 3.97 & 1.00 & 0.7433 & 34.29  \\
3 & J020258$-$002807* & $-$19.7 & 0.08 & $-$21.5 & 4.63 & 4.0 & 0.2954 & $-$71.22 & & $-$21.8 & 2.63 & 1.00 & 0.8807 & 5.56  \\
4 & J021447$-$003250 & $-$21.3 & 0.35 & $-$22.1 & 5.35 & 4.0 & 0.8555 & 78.57 & & $-$20.8 & 2.51 & 1.00 & 0.7734 & 71.02  \\
5 & J023700$-$010130* & $-$20.3 & 0.15 & $-$22.4 & 1.08 & 4.0 & 0.6607 & 51.18 & & $-$20.7 & 6.41 & 1.00 & 0.5091 & 72.26  \\
6 & J040210$-$054630 & $-$20.0 & 0.20 & $-$17.6 & 2.35 & 4.0 & 0.0019 & 64.36 & & $-$21.4 & 2.41 & 1.00 & 0.9577 & $-$86.98  \\
7 & J074621+335040 & $-$21.1 & 0.27 & $-$22.1 & 3.09 & 4.0 & 0.7379 & $-$66.99 & & $-$21.0 & 0.59 & 1.00 & 0.8951 & $-$64.36  \\
8 & J075045+212546 & $-$23.1 & 0.52 & $-$23.7 & 15.55 & 4.0 & 0.8540 & $-$80.54 & & $-$20.1 & 3.70 & 1.00 & 0.4676 & 10.96  \\
9 & J075521+295039* & $-$19.2 & 0.07 & $-$20.5 & 4.34 & 4.0 & 0.6554 & $-$12.95 & & $-$21.6 & 3.05 & 1.00 & 0.7855 & 66.39  \\
10 & J075549+321704 & $-$18.7 & 0.26 & $-$22.9 & 3.93 & 4.0 & 0.7326 & $-$4.43 & & $-$20.0 & 3.93 & 1.00 & 0.7326 & $-$4.43  \\
11 & J081018+250921 & $-$21.7 & 0.30 & $-$22.7 & 1.04 & 4.0 & 0.7732 & 73.29 & & $-$21.1 & 6.80 & 1.00 & 0.7182 & $-$67.89  \\
12 & J105816+102414 & $-$19.7 & 0.10 & $-$22.4 & 0.90 & 4.0 & 0.9710 & $-$19.69 & & $-$20.1 & 4.40 & 1.00 & 0.8824 & $-$78.55  \\
13 & J115159+673604* & $-$19.9 & 0.13 & $-$22.3 & 10.01 & 4.0 & 0.5729 & $-$71.66 & & $-$17.7 & 6.38 & 1.00 & 0.3030 & $-$74.06  \\
14 & J115355+582442 & $-$21.7 & 0.54 & $-$20.5 & 0.00 & 4.0 & 0.3096 & 90.00 & & $-$20.9 & 2.06 & 1.00 & 0.6195 & 35.47  \\
15 & J123043+614821 & $-$20.9 & 0.26 & $-$22.3 & 2.07 & 4.0 & 0.8307 & $-$37.03 & & $-$20.0 & 11.53 & 1.00 & 0.4339 & $-$21.75  \\
16 & J124833+563507 & $-$21.4 & 0.22 & $-$22.9 & 8.97 & 4.0 & 0.8427 & $-$64.65 & & $-$21.1 & 7.28 & 1.00 & 0.5611 & 30.98  \\
17 & J145640+524727* & $-$20.1 & 0.16 & $-$20.9 & 2.63 & 4.0 & 0.2985 & $-$71.64 & & $-$21.5 & 2.94 & 1.00 & 0.9796 & 15.05  \\
18 & J145658+593202* & $-$19.8 & 0.09 & $-$22.4 & 20.32 & 4.0 & 0.5579 & $-$51.99 & & $-$21.1 & 2.29 & 1.00 & 0.5595 & 6.83  \\
19 & J154534+573625* & $-$21.4 & 0.42 & $-$21.1 & 1.66 & 4.0 & 0.4092 & 38.86 & & $-$21.3 & 3.88 & 1.00 & 0.4748 & 61.27  \\
20 & J164444+423304 & $-$21.3 & 0.56 & $-$19.9 & 1.22 & 4.0 & 0.3733 & $-$74.44 & & $-$20.6 & 1.21 & 1.02 & 0.9369 & 74.39  \\
21 & J170046+622056* & $-$20.5 & 0.17 & $-$18.9 & 8.68 & 4.0 & 0.4979 & 4.78 & & $-$21.1 & 2.91 & 1.00 & 0.4618 & 34.94  \\
22 & J210200+000501 & $-$21.2 & 0.28 & $-$22.1 & 1.23 & 4.0 & 0.8617 & $-$68.40 & & $-$21.2 & 4.70 & 1.00 & 0.6419 & 21.07  \\
23 & J211343$-$075017 & $-$21.6 & 0.31 & $-$23.0 & 5.04 & 4.0 & 0.9558 & 61.48 & & $-$20.9 & 1.80 & 1.08 & 0.6476 & $-$48.37  \\
24 & J211838+005640 & $-$22.0 & 0.53 & $-$21.0 & 2.92 & 4.0 & 0.4756 & 55.96 & & $-$21.4 & 2.30 & 1.16 & 0.9141 & $-$84.44  \\
25 & J212843+002435 & $-$21.5 & 0.42 & $-$22.2 & 16.02 & 4.0 & 0.6240 & $-$31.18 & & $-$20.3 & 3.59 & 1.00 & 0.5233 & $-$2.62  \\
26 & J230614$-$010024* & $-$20.9 & 0.18 & $-$22.5 & 7.38 & 4.0 & 0.9262 & $-$86.60 & & $-$21.5 & 1.45 & 1.00 & 0.8762 & $-$16.58  \\
27 & J231055$-$090107* & $-$20.7 & 0.10 & $-$23.7 & 44.49 & 4.0 & 0.4459 & $-$48.88 & & $-$20.3 & 3.54 & 1.00 & 0.3051 & $-$10.08  \\
28 & J233430+140649 & $-$21.7 & 0.52 & $-$10.7 & 0.01 & 4.0 & 0.7895 & 71.41 & & $-$21.4 & 1.32 & 1.00 & 0.8094 & $-$46.05  \\
29 & J234403+154214* & $-$20.3 & 0.21 & $-$20.5 & 7.30 & 4.0 & 0.5708 & 29.59 & & $-$21.4 & 3.54 & 1.00 & 0.9575 & 28.93  \\
\enddata
\tablenotetext{*}{Denotes the best/adopted model of the S\'{e}rsic, de Vaucouleurs and Bulge-plus-Disk model runs.}
\tablecomments{Col. (1): Object ID number. Col. (2): SDSS name. Col. (3): Integrated Galactic dereddened K-corrected absolute ST magnitude of the PSF in the F606W filter. Col. (4): PSF to total light fraction. Bulge Component: Col. (5): Integrated Galactic dereddened K-corrected absolute ST magnitude in the F606W filter. Col. (6): Effective radius in kpc. Col. (7): S\'{e}rsic index. Col. (8): Axis ratio. Col. (9): Position angle (east of north). Disk Component: Col. (10): Integrated Galactic dereddened K-corrected absolute ST magnitude in the F606W filter. Col. (11): Effective radius in kpc. Col. (12): S\'{e}rsic index. Col. (13): Axis ratio. Col. (14): Position angle (east of north).\label{tab:bdresults}}
\end{deluxetable}
\clearpage

\section{Analysis} 
\label{sec:Analysis}

   If PSQs are indeed transitioning objects, we might expect to see correlations between host galaxy parameters involving morphology and degree of disturbance with AGN activity. For example, AGN activity may show a preference for a given merger stage. We investigate the relationships between morphology, degree of disturbance, and the fitted host galaxy parameters. We test differences between the population medians binned by morphology (\S~\ref{sec:Analysis.Morph}) and degree of disturbance (\S~\ref{sec:Analysis.DI}). We characterize the correlations between fitted host galaxy parameters for the; i) S\'{e}rsic model run of all PSQ hosts, ii) PSQ hosts that are best characterized by one component (S\'{e}rsic), and, iii) PSQ hosts that are best characterized by two components (Bulge-plus-Disk), in \S~\ref{sec:Analysis.Param}.
  
\subsection{Morphology}
\label{sec:Analysis.Morph}
 
   We classify 13 of the 29 objects as spirals, including two `probable' spirals. At least four of these objects appear to be flocculent, while six are grand design spirals. The frequencies of grand design, flocculent and multiple spiral arm structure in the local Universe are 9\%, 30\% and, 61\%, respectively, regardless of whether a bar is present \citep{elmegreen90}. Although this is a small sample, the high frequency of grand design spirals in the sample supports interaction/merger simulations of tidally induced grand design spiral patterns \citep{byrdhoward92}. Another 13 of the 29 hosts were classified as early-type galaxies. Although elliptical galaxies are sometimes thought of as old, red, dead objects, PSQ early-types are notably disturbed, revealing shells, dust lanes, and asymmetries, signposts of interactions that have yet to fade away. The three objects classified as indeterminate show disturbances or flocculent structure without revealing a disk or spiral arms characteristic of spiral galaxies.
    
\begin{figure}[!h]
\figurenum{4} 
  \centering
    \subfloat[][(a) \label{fig:hist_morph_PSFmag}]{\includegraphics[scale=0.3]{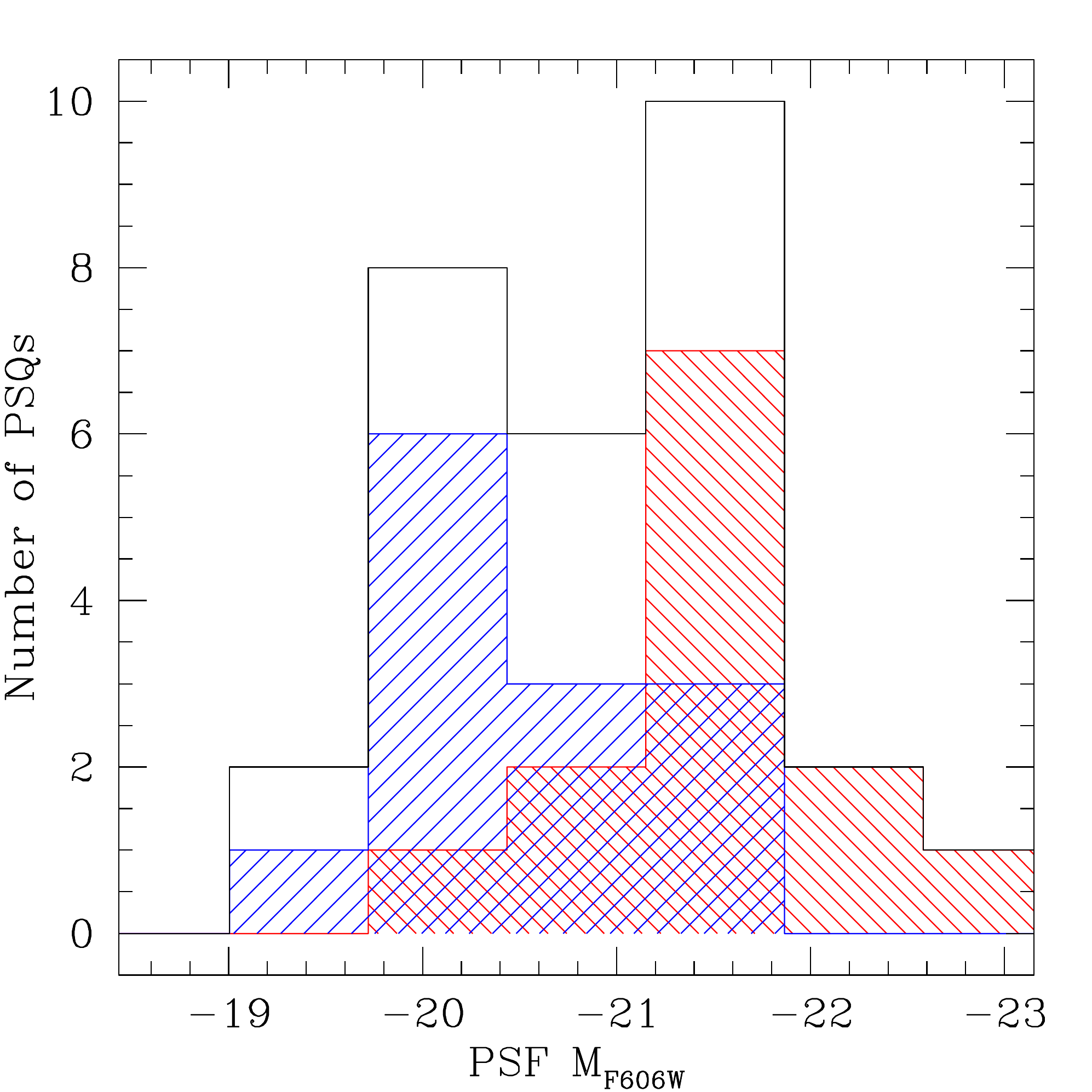}}
    \subfloat[][(b)\label{fig:hist_morph_CompMag}]{\includegraphics[scale=0.3]{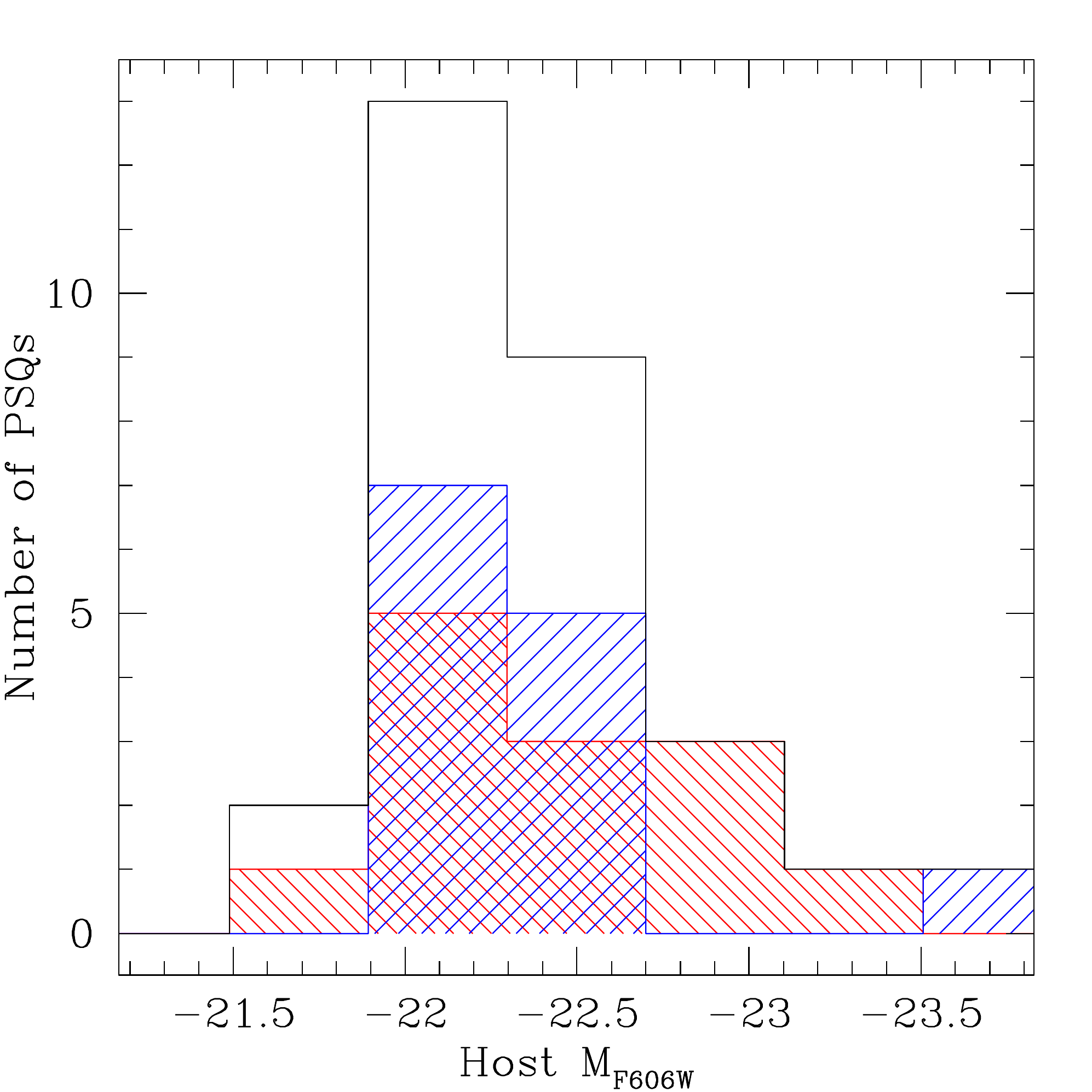}}
    \subfloat[][(c) \label{fig:hist_morph_LF}]{\includegraphics[scale=0.3]{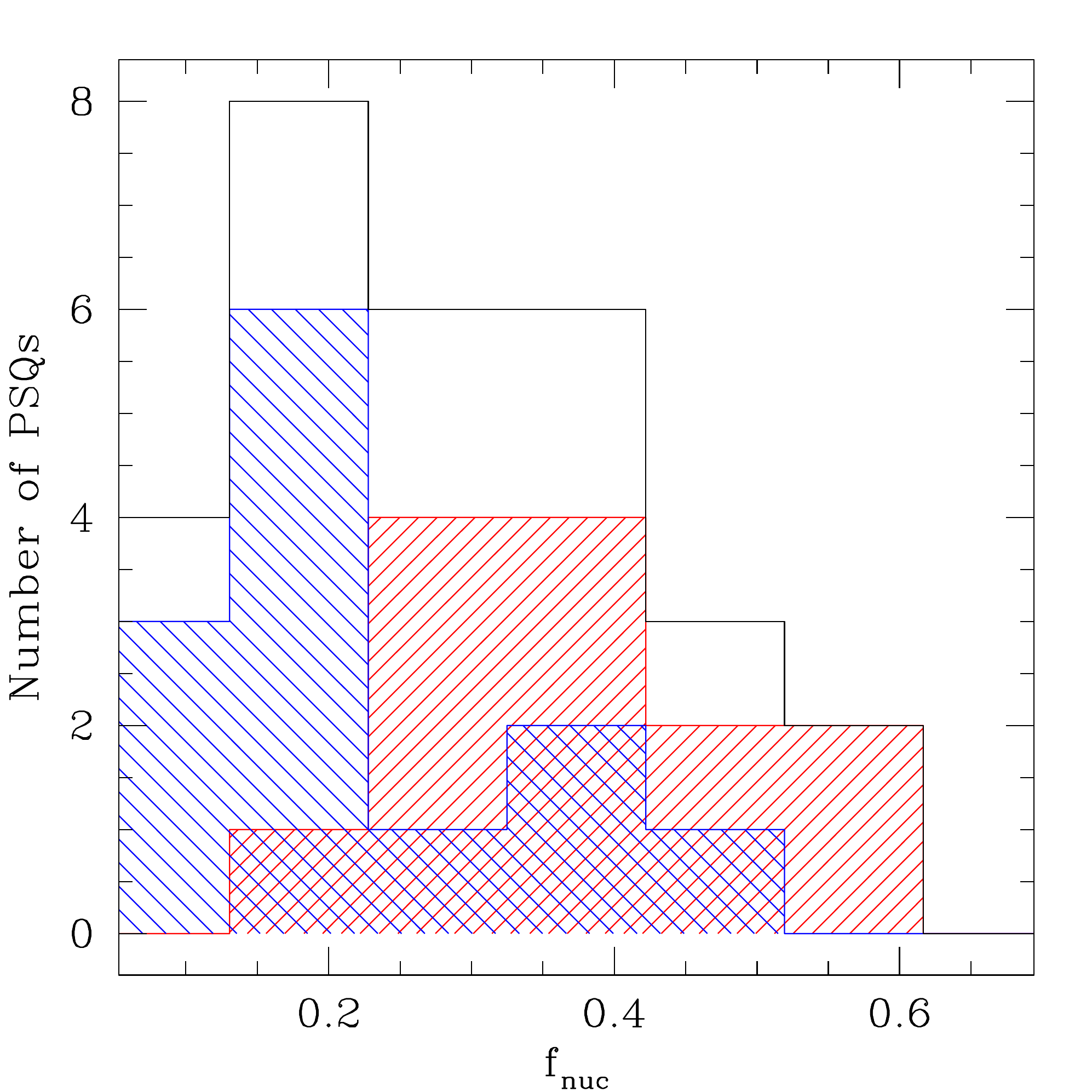}}\\
    \subfloat[][(d) \label{fig:hist_morph_Re}]{\includegraphics[scale=0.3]{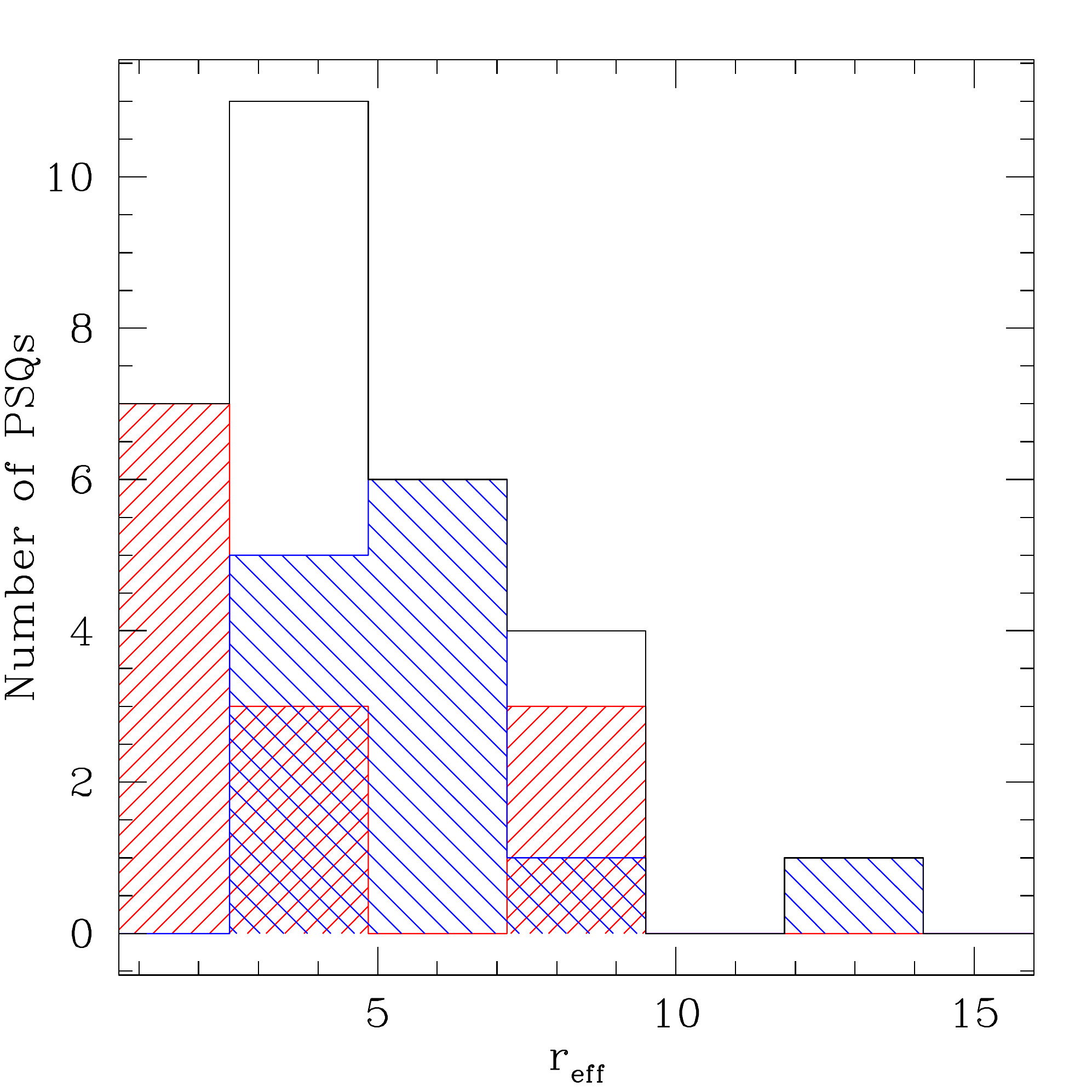}}       
    \subfloat[][(e) \label{fig:hist_morph_n}]{\includegraphics[scale=0.3]{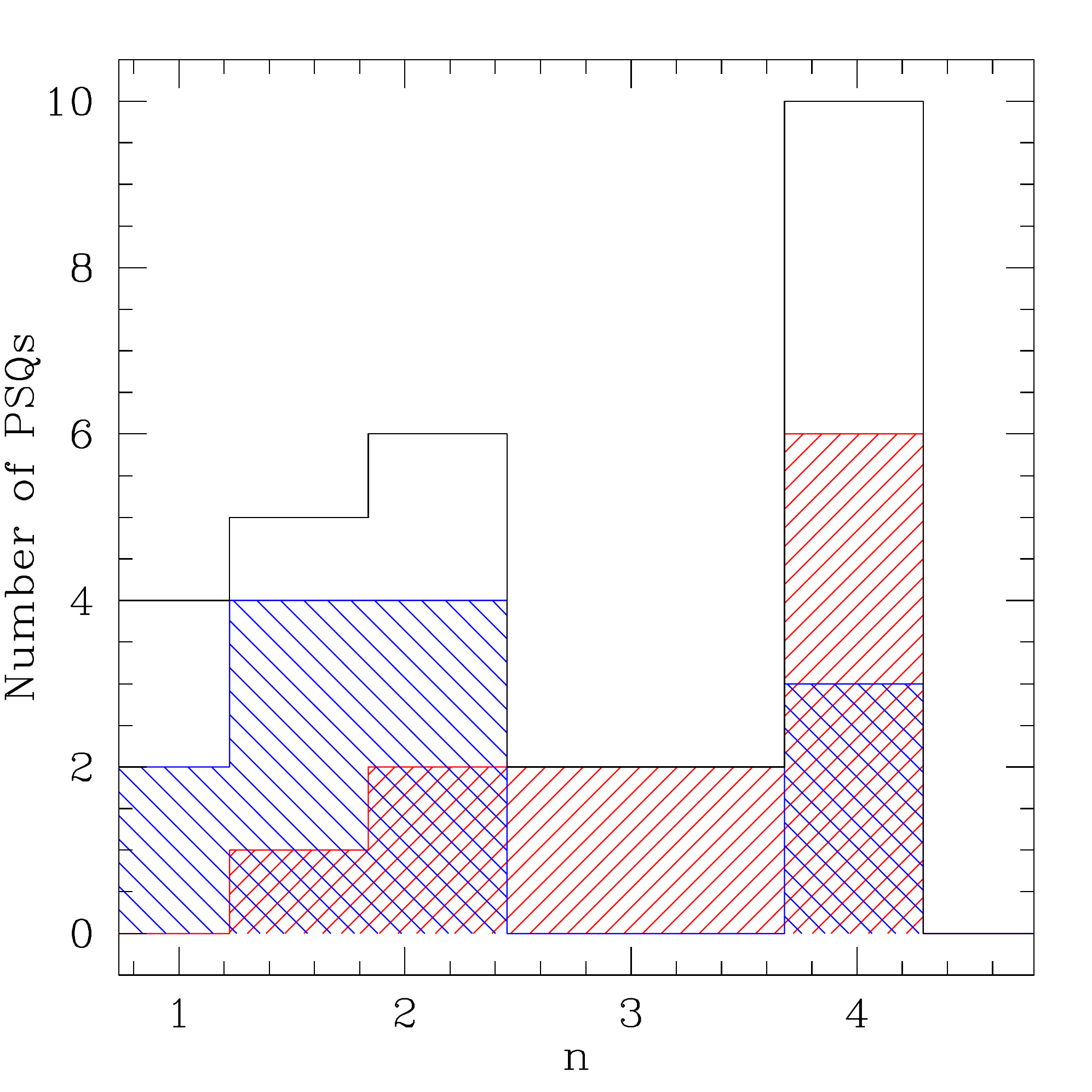}} \\    
    \caption{\small{a-e) The black histogram represents the total distribution while the red and blue histograms indicate the early-type and spiral plus `probable' spiral morphology distributions. \label{fig:morphhist}}}
  \centering
\end{figure}

  Based on the S\'{e}rsic model fits (Table~\ref{tab:sresults}), we investigate whether the early-type and spiral hosts (including `probable' spirals) are drawn from the same parent population using the nonparametric Mann-Whitney test. We give the median host parameter values for early-types and spirals and the two tailed probability that they are drawn from the same parent population in Table~\ref{tab:spop}. P-values above the 2$\sigma$ level are in listed in bold. At this level we see bimodal distributions between early-type and spiral host galaxies in PSF absolute magnitude, nuclear light fraction, effective radius and S\'{e}rsic index. Histograms of the tested parameters are given in Figure~\ref{fig:morphhist}a-e.  Although some of the tests may be dependent (e.g., PSF absolute magnitude, host absolute magnitude and nuclear light fraction) we expect less than 0.25 spurious results ($\sigma > 5\%$).
    
   Early-type and spiral host galaxies differ in several ways. Early-type galaxies have greater PSF absolute magnitudes (Figure~\ref{fig:hist_morph_PSFmag}). However, there is not a significant difference between host galaxy absolute magnitudes (Figure~\ref{fig:hist_morph_CompMag}). Consistent with these two results, early-type hosts have larger nuclear light fractions (Figure~\ref{fig:hist_morph_LF}). Since early-type and spiral galaxies have similar host magnitudes and, by selection, similar stellar population ages, this results in their having similar total masses. However, the black hole mass is expected to correlate with the mass of the spheroidal component only. Thus, early-type galaxies having greater AGN luminosities may simply reflect the fact that they have larger black holes but similar fueling rates (e.g., similar Eddington ratios). 
      
  A visual inspection of the images reveals that the spiral galaxies appear larger than the early-type galaxies (see Figure~\ref{fig:hist_morph_Re}). The Mann-Whitney test of the early-type and spiral host effective radii gives only a $\sim$ 2\% probability that the populations are the same. This is consistent with results from studies of SDSS galaxies \citep[e.g.,][]{shen03,blanton09}, which have shown that spirals tend to have larger effective radii than all but the most massive ellipticals (elliptical galaxies).
   
   We expect that our quantitative (S\'{e}rsic index) and qualitative (morphology) measures of galaxy type are correlated (see also Figure~\ref{fig:hist_morph_n}). However, within their respective distributions, early-type and spiral galaxies can cover a range of S\'{e}rsic indices, usually $n \sim1-2$ for spiral types and $n \sim1-5$ for elliptical hosts \citep{blanton09}. Our study yielded S\'{e}rsic indices medians of 1.85 and 3.43 for spiral hosts and early-type types, respectively.

\begin{deluxetable}{lccc}
\tablecolumns{4}
\tablewidth{0pc}
\tablecaption{PSQ Host Galaxy S\'{e}rsic Modeling Population Tests \label{tab:spop}}
\tablehead{
\colhead{} & \multicolumn{2}{c}{Morphology} \\
\cline{2-3}
\colhead{ } & \colhead{\begin{sideways} {\normalsize 0} \end{sideways}} & \colhead{\begin{sideways} {\normalsize \S} \end{sideways}} & \colhead{Prob(W)} 
}
\startdata
PSF $M_{\textrm{\tiny F606W}}$ & -21.4 & -20.3 & \textbf{0.0077} \\
Host $M_{\textrm{\tiny F606W}}$ & -22.3 & -22.2 & 0.5050 \\
$f_{nuc}$ & 0.34 & 0.20 & \textbf{0.0120} \\
$r_{eff}$  & 2.28 & 4.97 & \textbf{0.0240} \\
$n$  & 3.43 & 1.85 & \textbf{0.0210} \\
\hline
\colhead{} & \multicolumn{2}{c}{Degree of Disturbance} \\
\cline{2-3} 
\colhead{ } & \colhead{Undisturbed} & \colhead{Disturbed} & \colhead{Prob(W)}  \\
\hline
PSF $M_{\textrm{\tiny F606W}}$ & -20.9 & -20.8 & 0.6214 \\
Host $M_{\textrm{\tiny F606W}}$ & -22.2 & -22.4 & 0.0701 \\
$f_{nuc}$ & 0.30 & 0.20 & 0.2396 \\
$r_{eff}$  & 4.24 & 4.63 & 0.1878 \\
$n$  & 2.00 & 2.801 & 0.7419 \\
\enddata
\tablecomments{Median values for the S\'{e}rsic modeling parameters of objects binned by morphology (early-type and spiral, respectively). Prob(W) are the two-tailed probabilities that the medians of the early-type and spiral subsamples are as different as they are if their parent populations shared the same median, computed using the nonparametric Mann-Whitney test. P-values above the 2$\sigma$ level are in listed in bold.}
\end{deluxetable}
\clearpage

\clearpage

\begin{figure}[!h]
\figurenum{5} 
  \centering
    \subfloat[][(a) \label{fig:hist_DI_PSFmag}]{\includegraphics[scale=0.3]{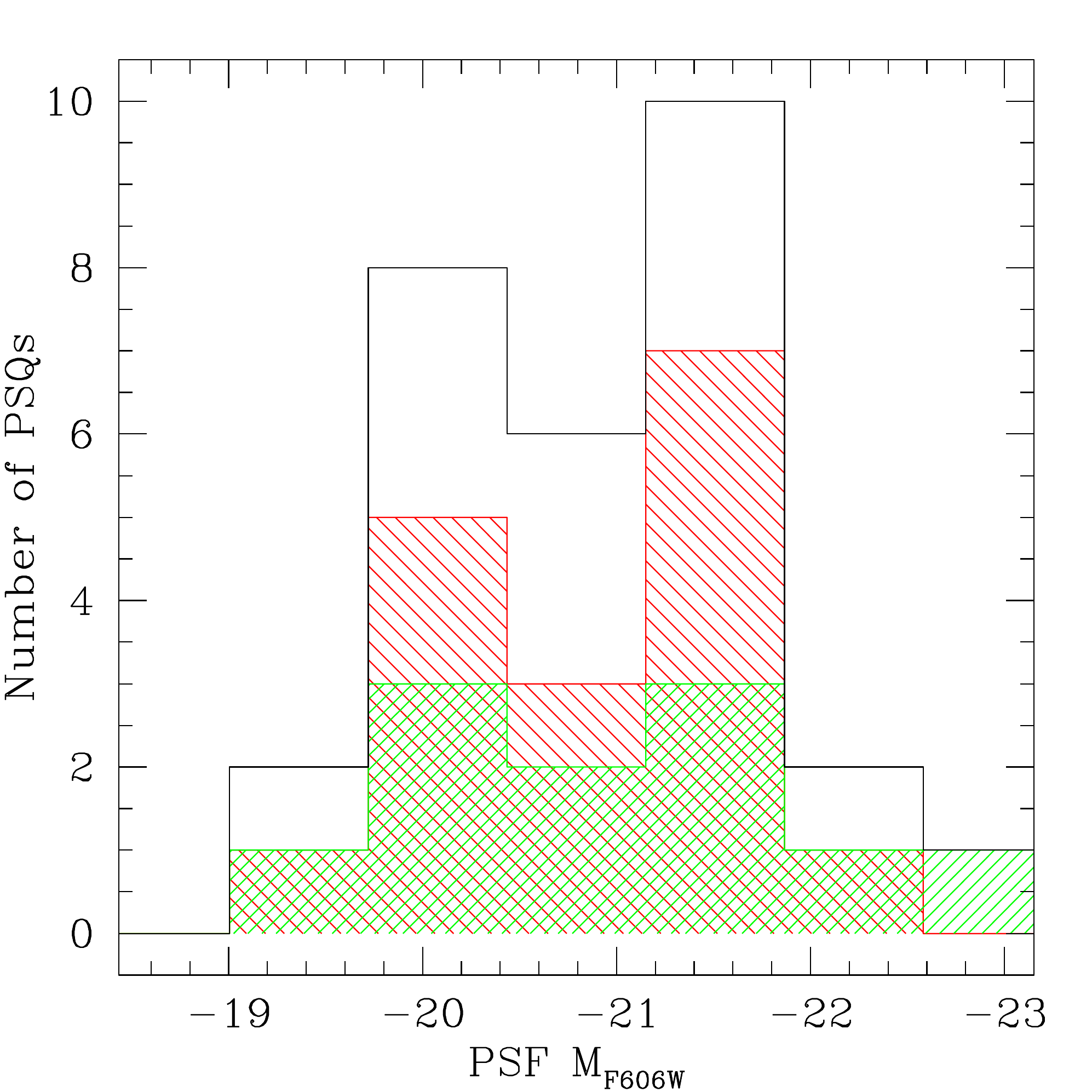}}
    \subfloat[][(b)\label{fig:hist_DI_CompMag}]{\includegraphics[scale=0.3]{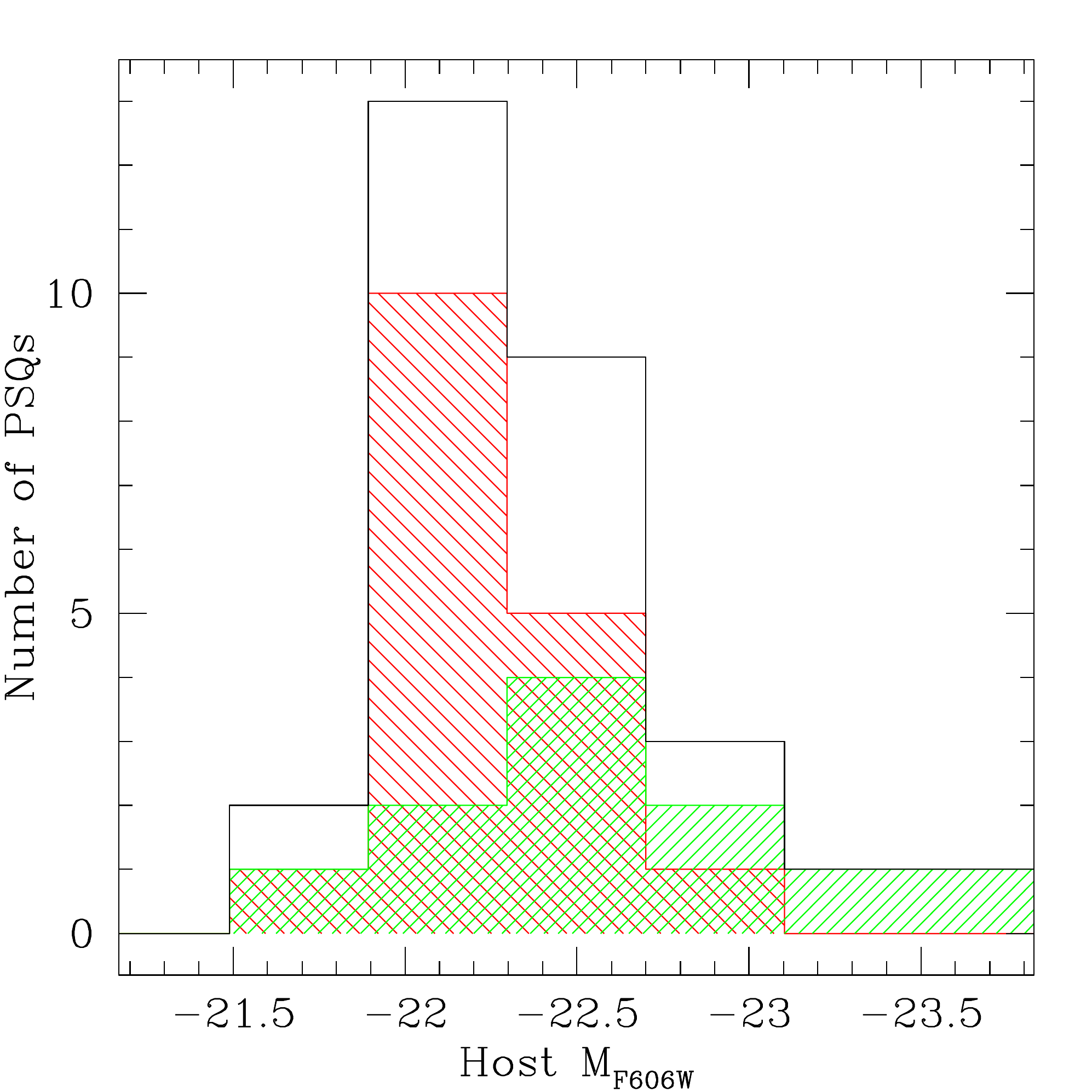}}
    \subfloat[][(c) \label{fig:hist_DI_LF}]{\includegraphics[scale=0.3]{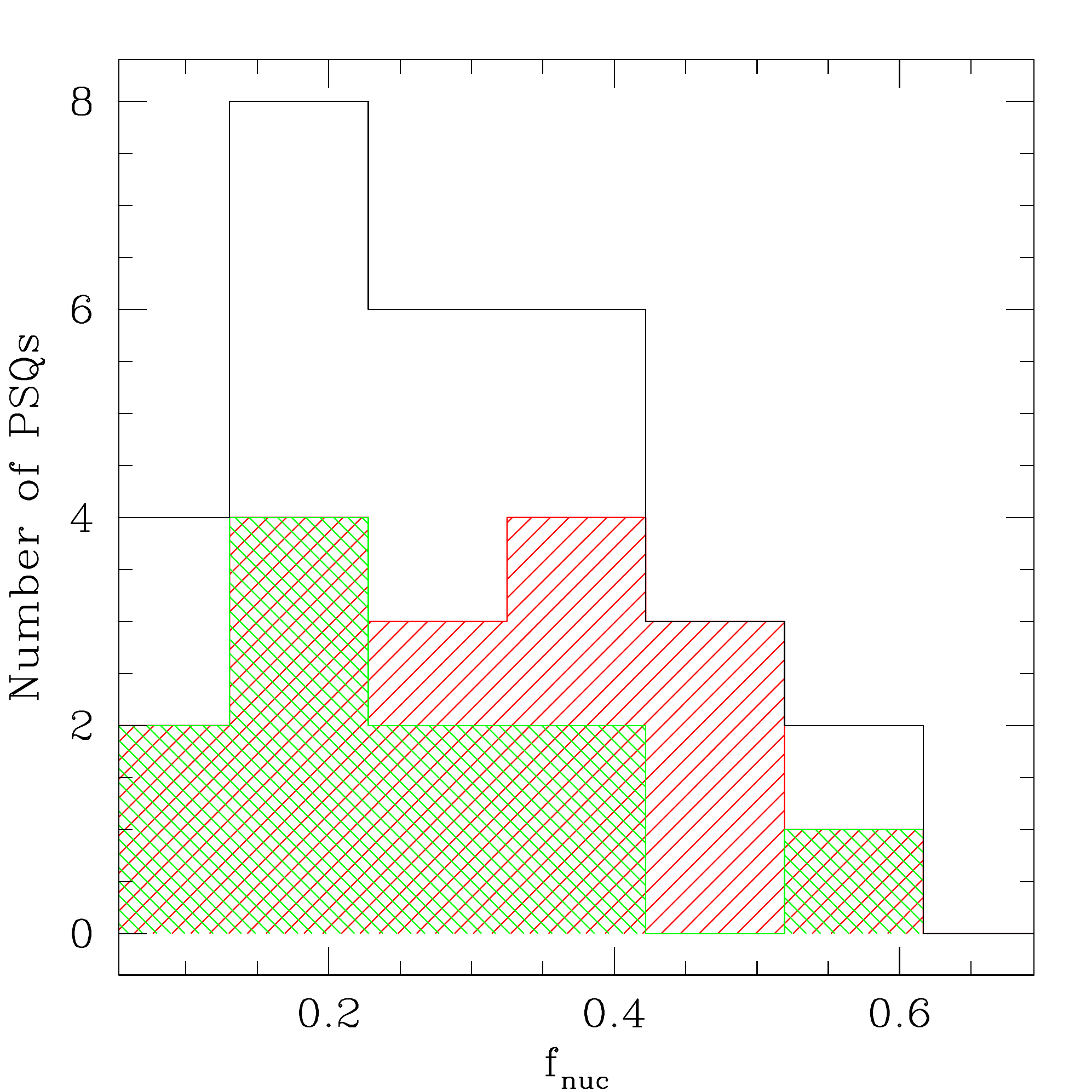}}\\
    \subfloat[][(d) \label{fig:hist_DI_Re}]{\includegraphics[scale=0.3]{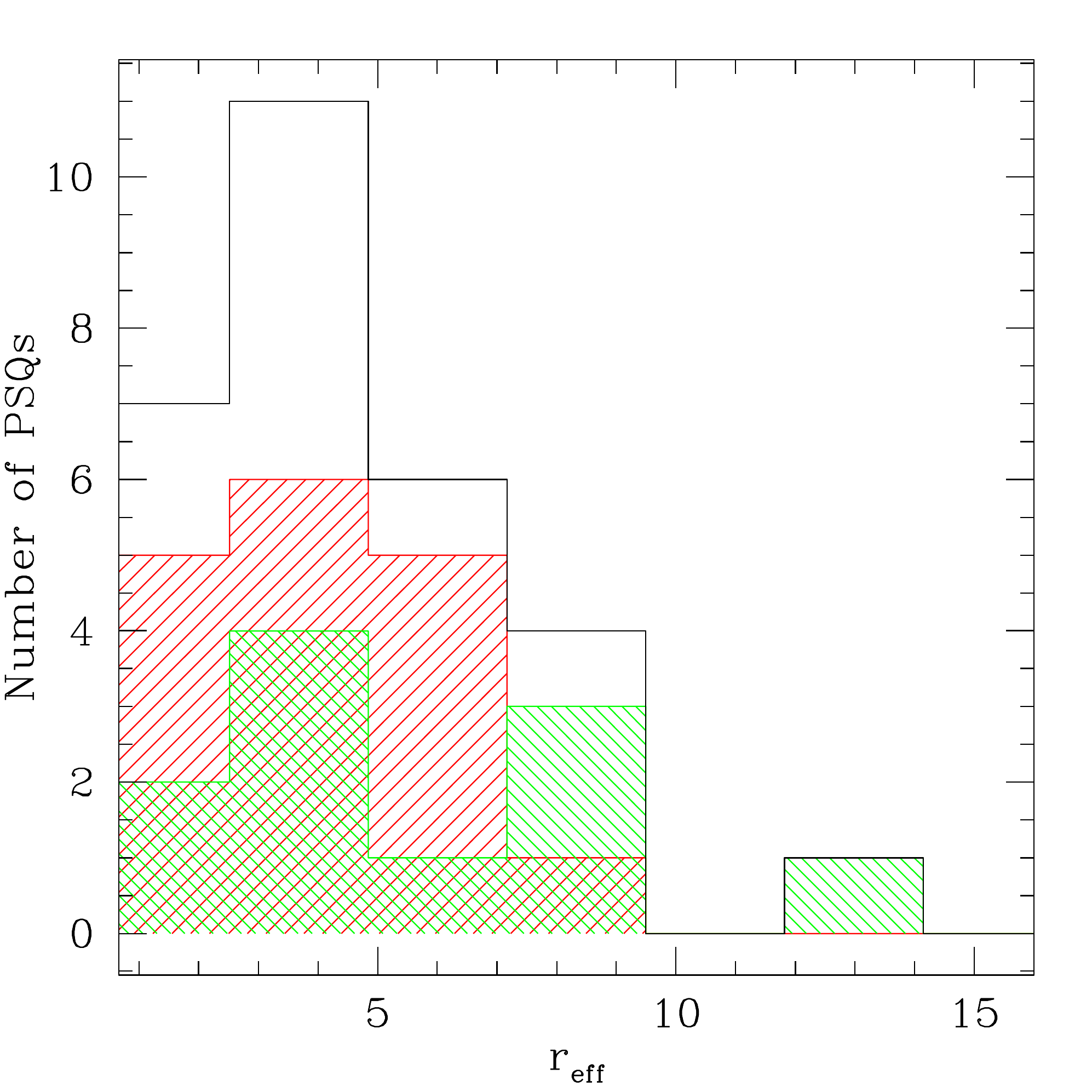}}       
    \subfloat[][(e) \label{fig:hist_DI_n}]{\includegraphics[scale=0.3]{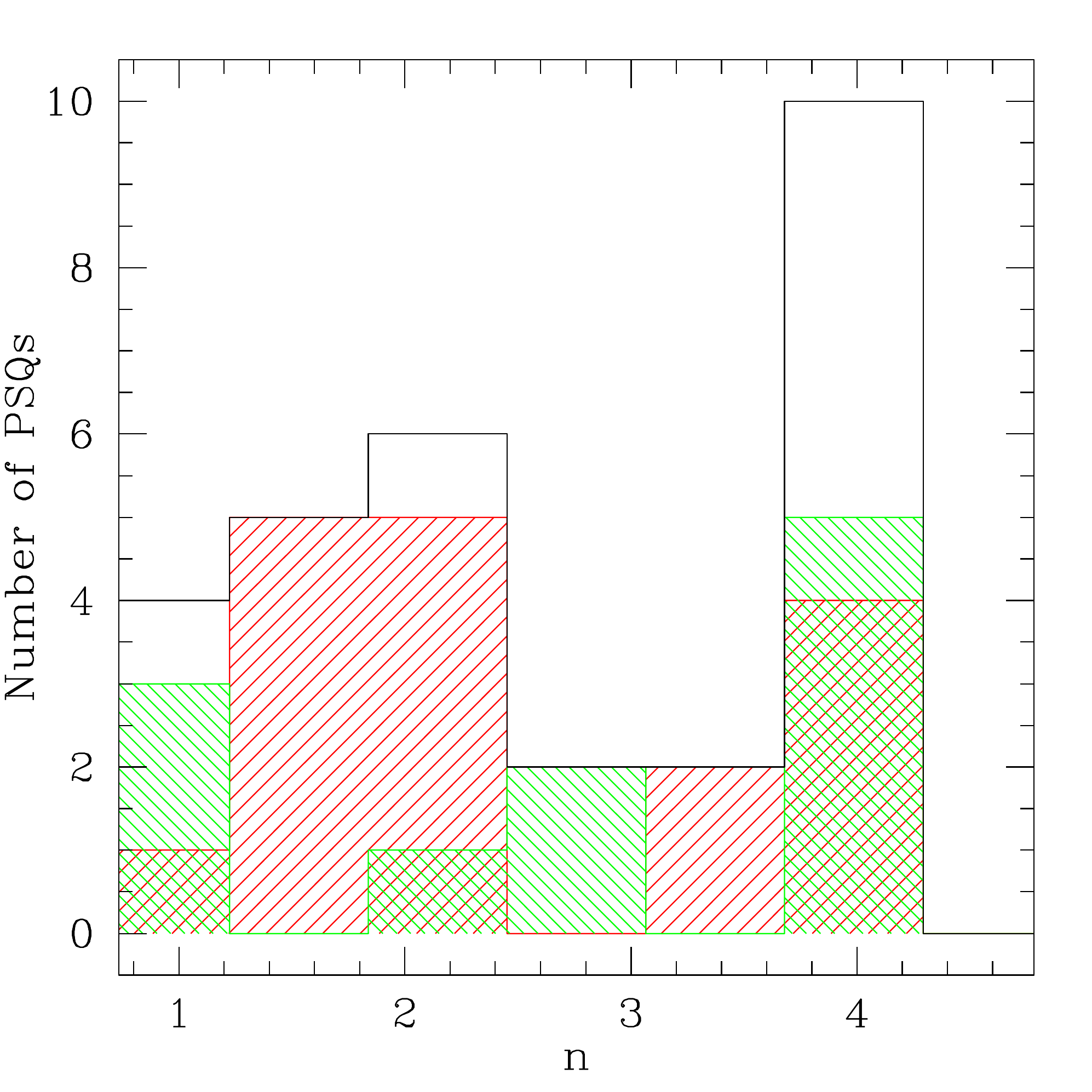}} \\    
    \caption{\small{a-e) The solid black histogram represents the total distribution while the red and green histograms indicate the undisturbed and disturbed/harassed plus highly disturbed/merging host galaxy distributions.\label{fig:DIhist}}}
  \centering
\end{figure}
\clearpage

\subsection{Degree of Disturbance}
\label{sec:Analysis.DI}

   We find evidence of interactions and/or companions in 17 of the 29 PSQs. There are 11 ($\sim$40\%) objects that are visibly disturbed, of which four are highly disturbed. The remaining 12 objects have no signs of interaction, which include undisturbed early-type galaxies, spirals, and barred spirals. It is interesting to note that four of five barred spirals were classified as undisturbed. Studies have suggested that bars can result from instabilities caused by strong interactions \citep{berentzen04, varela04}. Bar instability may also be a secular process. Thus, it is possible that the presence of bars are a possible AGN/starburst fueling mechanism aside from the usual suspects of merger, interaction, and harassment, or we are probing bar formation/evolution just after visible evidence of the perturbing interaction has faded.
   
   We test whether the disturbed (both disturbed and highly disturbed) and undisturbed host galaxies are drawn from the same parent population using the nonparametric Mann-Whitney test.  We give the median host parameter values for undisturbed and disturbed galaxies and the two tailed probability that they are drawn from the same parent population in Table~\ref{tab:spop}. We see bimodal distributions between disturbed and undisturbed host galaxies in host absolute magnitude and nuclear light fraction. Histograms of these parameters are given in Figure~\ref{fig:DIhist}a-e. A total of 5 population tests were performed. The same caveat mentioned in \S~\ref{sec:Analysis.Morph} about dependent parameters applies here. We expect less than 0.25 spurious statistically significant results.
   
   Figure~\ref{fig:hist_DI_CompMag} shows that disturbed galaxies have greater absolute host galaxy magnitudes. This could be due to galaxy-galaxy interaction-induced star formation. However, there seems to be no significant bimodality in the distribution of quasar absolute magnitude with respect to degree of disturbance. The PSF luminosities of the disturbed and undisturbed galaxies are not significantly different. Consistent with the two previous results undisturbed galaxies have greater nuclear light fractions (Figure~\ref{fig:hist_DI_LF}). 

\begin{figure}[!ht]
\label{fig:hist_M+D}
\figurenum{6} 
  \centering
    \subfloat[][(a) \label{fig:hist_di}]{\includegraphics[scale=0.4]{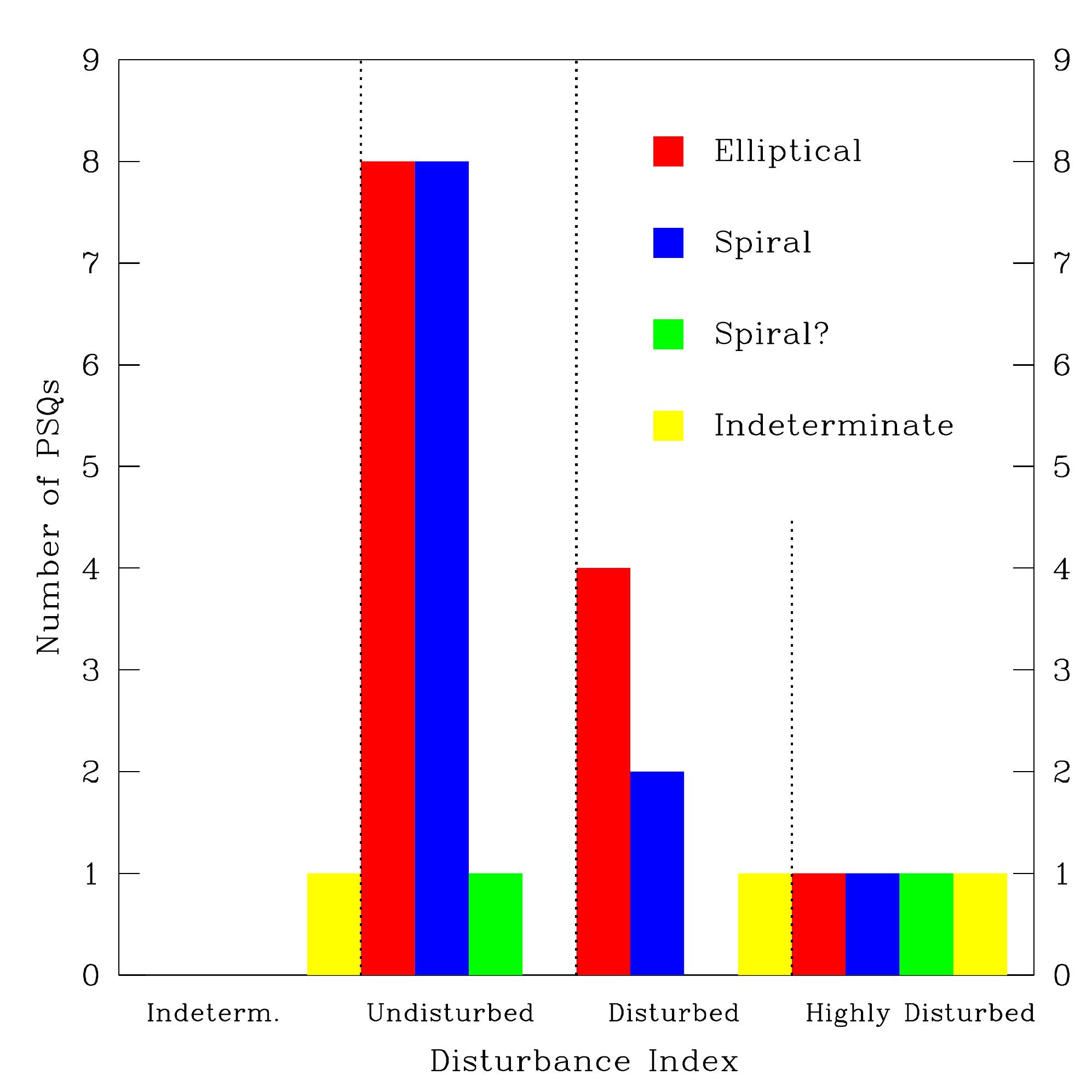}}
    \subfloat[][(b)\label{fig:hist_morph}]{\includegraphics[scale=0.4]{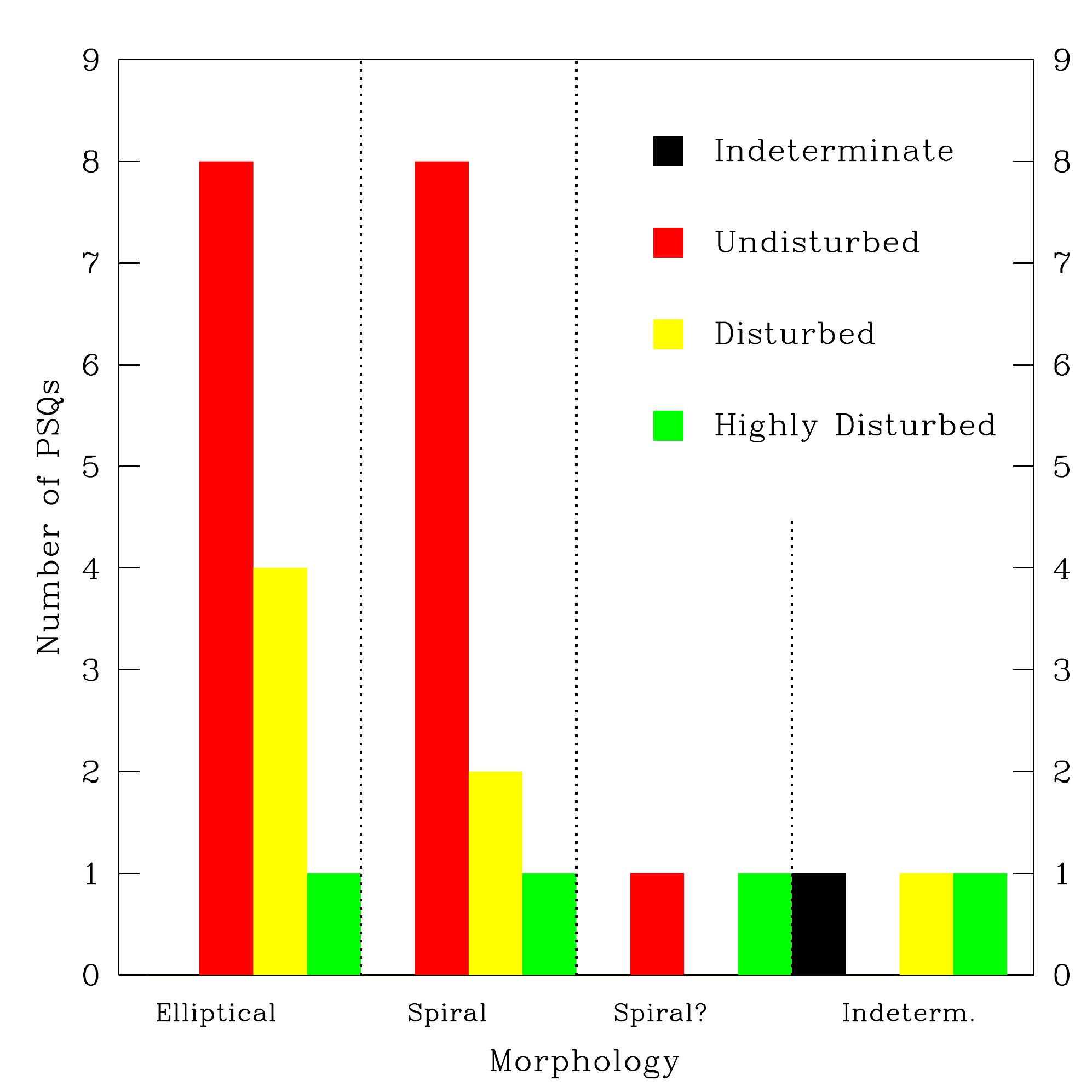}}
    \caption{a) The red, blue, green, and yellow bars indicate the early-type, spiral, `probable' spiral, and indeterminate morphology distributions, respectively, binned by disturbance index and separated by the dotted line. b) The green, yellow, red, and black bars indicate the highly disturbed, disturbed/harassed, undisturbed, and indeterminate host galaxy distributions, respectively, binned by morphology and separated by the dotted line. \label{fig:hist_M+D}}
  \centering
\end{figure}

   A useful tool in analyzing relations between morphology and degree of disturbance are the distributions of Figure~\ref{fig:hist_M+D}. Binned by degree of disturbance (Figure~\ref{fig:hist_di}), the distribution of early-type and spiral hosts are approximately equivalent. Similar numbers of early-type and spiral types are seen in each bin. Viewed another way, binned by morphology (Table~\ref{fig:hist_morph}), we come to the same conclusion.

\clearpage

\subsection{Host Parameter Correlations}
\label{sec:Analysis.Param}

\begin{figure}[tbhp]
\centering
\figurenum{7} 
\includegraphics[width=4in]{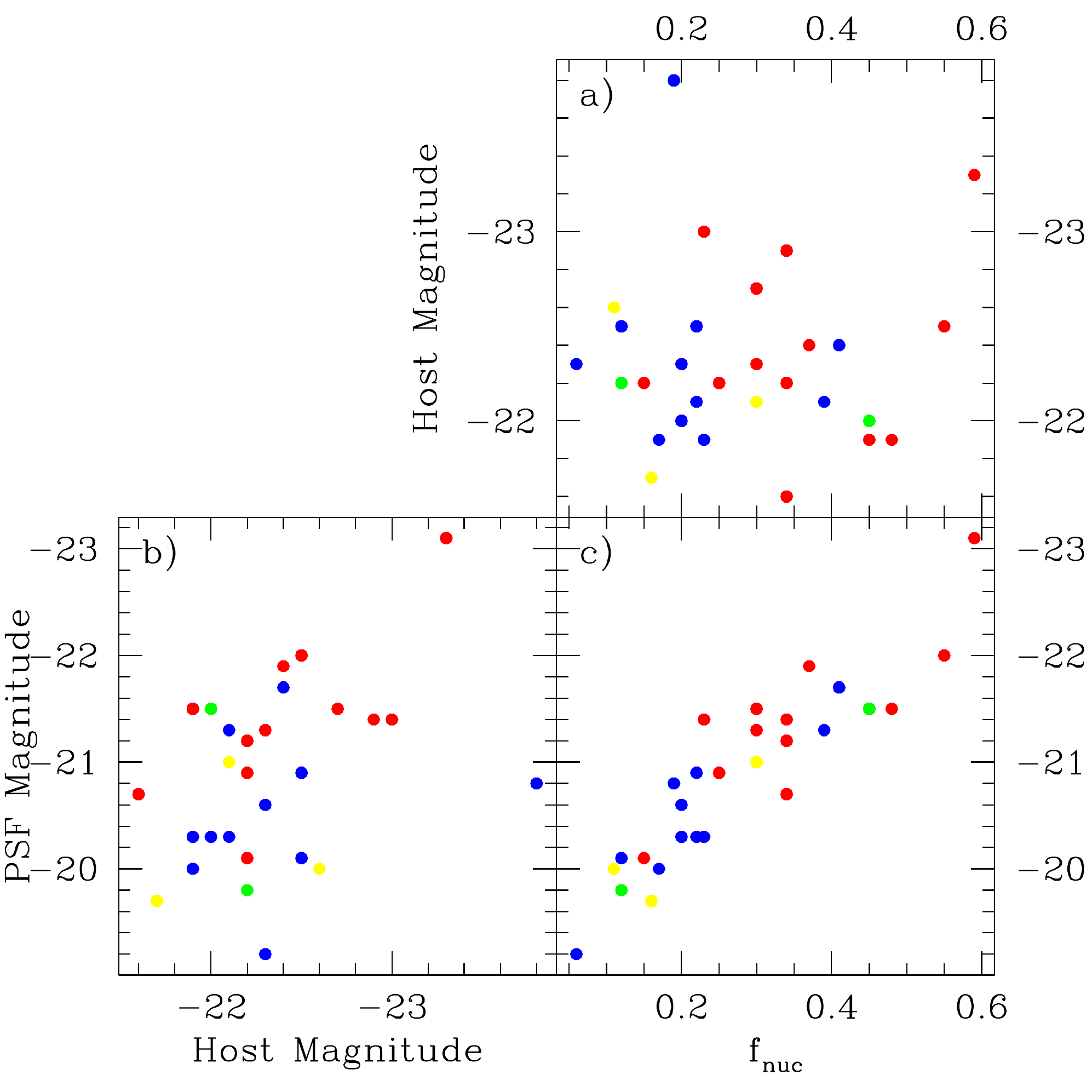}
\caption{The parameters; PSF magnitude, nuclear light fraction, and host magnitude plotted against one another. The points indicate morphology where red, blue, green, and yellow points are the early-type, spiral, `probable' spiral, and indeterminate morphology distributions, respectively. \label{fig:plot_param}}
\end{figure}

   We calculate Spearman-rank correlation coefficient matrices between fitted host galaxy parameters for the: i) S\'{e}rsic model run of all PSQ hosts, ii) PSQ hosts that are best characterized by one component (S\'{e}rsic), and, iii) PSQ hosts that are best characterized by two components (Bulge-plus-Disk). We looked for correlations above the 2$\sigma$ level. We expect less than 3.9 spurious significant correlations of the 78 tests we performed. As mentioned in \S~\ref{sec:Analysis.Morph} and \S~\ref{sec:Analysis.DI} the expected correlations between PSF absolute magnitude and nuclear light fraction are confirmed for each correlation matrix (S\'{e}rsic, adopted single component, and adopted two component). We plot the PSF magnitude, nuclear  light fraction, and host magnitude against each other for the S\'{e}rsic model run in Figure~\ref{fig:plot_param}. We note that there are no significant correlations for the pair's host magnitude versus nuclear light fraction (Figure~\ref{fig:plot_param}a) and PSF magnitude versus host magnitude (Figure~\ref{fig:plot_param}b). However, there is a significant correlation (coefficient of $-$0.811) between PSF magnitude and nuclear light fraction (Figure~\ref{fig:plot_param}c). 
   
   The single component matrix yielded two significant correlations (effective radius-S\'{e}rsic index at p-value 0.011, PSF magnitude-host magnitude at p-value 0.011), while the double component matrix resulted in two significant correlations (bulge magnitude-bulge effective radius at p-value 0.025, disk magnitude-disk $b/a$ at p-value 0.001). These correlations may represent physical (non-spurious) correlations between host parameters. We hope to determine whether this is the case in future studies (see \S~\ref{sec:Summ}).

\section{Discussion}
\label{sec:Disc}

        The morphologies and levels of disturbance of PSQs vary. This sample is not homogeneous; about half are ellipticals, half spirals, half disturbed and half undisturbed. This is perhaps surprising since we used a uniform automated algorithm to select the sample based on the same spectroscopic criteria. This variance could be related to many factors, such as, quasar lifetime, quasar and/or starburst ignition time, fuel availability/consumption, galaxy interaction type. Multiple combinations of quasar and starburst properties could lead to the heterogeneity of PSQs. \citet{schawinski09} argue that to detect both AGN and post-starburst populations a delay between peak AGN and starburst activity on the order of 100 Myrs is needed, thus AGN begin quenching star formation prior to our ability to detect them. In order to detect an object as a PSQ, the post-starburst population and AGN luminosity must be of comparable strength, hence there is a window for detection which is dependent upon the lifetime of each. If the starburst and quasar coeval, PSQs may be probing interaction/triggering timescales within $\sim$100 Myrs. However, if there is a lag between the two, or one triggers the other, we may be probing interaction/triggering timescales of up to 1 Gyr. The objects of this sample may represent snapshots in time along an evolutionary sequence which suggests why their morphologies are so different.
        
           While theory suggests that galaxy formation in the early universe was hierarchical, antihierarchical quenching of star formation in large galaxies is consistent with observations of the local universe \citep{cowie96}. The transition between the two regimes lies at $z\sim$ 1-2, somewhat after the peak of AGN activity. Thus, for the better part of the past 10 billion years, the characteristic stellar mass of forming galaxies has been decreasing (cosmic downsizing) and black hole growth has migrated from the largest SMBH in the early universe to small ones now \citep[AGN cosmic downsizing;][]{heckman04}. Today most of the big systems have merged, as a result the prodigious amount of fuel necessary to power quasars is not available. Thus, we see weaker AGN, powered by smaller engines, in smaller galaxies, and less dramatic interactions or secular evolution (e.g., minor mergers, harassment, bars, instabilities).  
        

   Galaxies may evolve via different modes according to their morphology and epoch. The early-type hosts of this study are possibly the less massive versions of giants at $z \sim 2$ formed via hierarchical merging whereas the spiral hosts may have retained their shape after a minor interaction, harassment or secular processes and be more consistent with downsizing. For example, the evolutionary schema for the elliptical mode induced by a major merger may begin with (1) a ULIRG phase upon first approach (tidal effects inducing intense star formation and a powerful quasar enshrouded by dust) during which coalescence occurs, (2) the quasar swamps the host galaxy light in the QSO phase after the dust is blown out by feedback from the quasar, (3) then a brief PSQ (quasar+post-starburst) phase ensues as the quasar fades to a luminosity comparable to the post-starburst populations, (4) as the quasar continues to fade the post-starburst populations are more clearly visible (E+A), (5) finally even the post-starburst  and merger signatures have faded to reveal the elliptical host \citep{hopkins08}. 
   
   The evolutionary schema for the spiral mode induced by a minor interaction, harassment or secular processes may begin with (1) a starburst phase upon inducement followed by, (2) coalescence/relaxation and an E+A phase, (3) then, once fuel to feed the AGN reaches the nucleus, a PSQ phase ensues, (4) and as the post-starburst population fades an AGN/Seyfert phase follows and an (5) undisturbed spiral is observed once the AGN and possible disturbance signatures have vanished. Both elliptical and spiral hosts are equally present in the PSQ sample, thus we suggest that both of these modes of evolution are important. 
              
\begin{figure}[tbhp]
\centering
\figurenum{8} 
\includegraphics[width=4in]{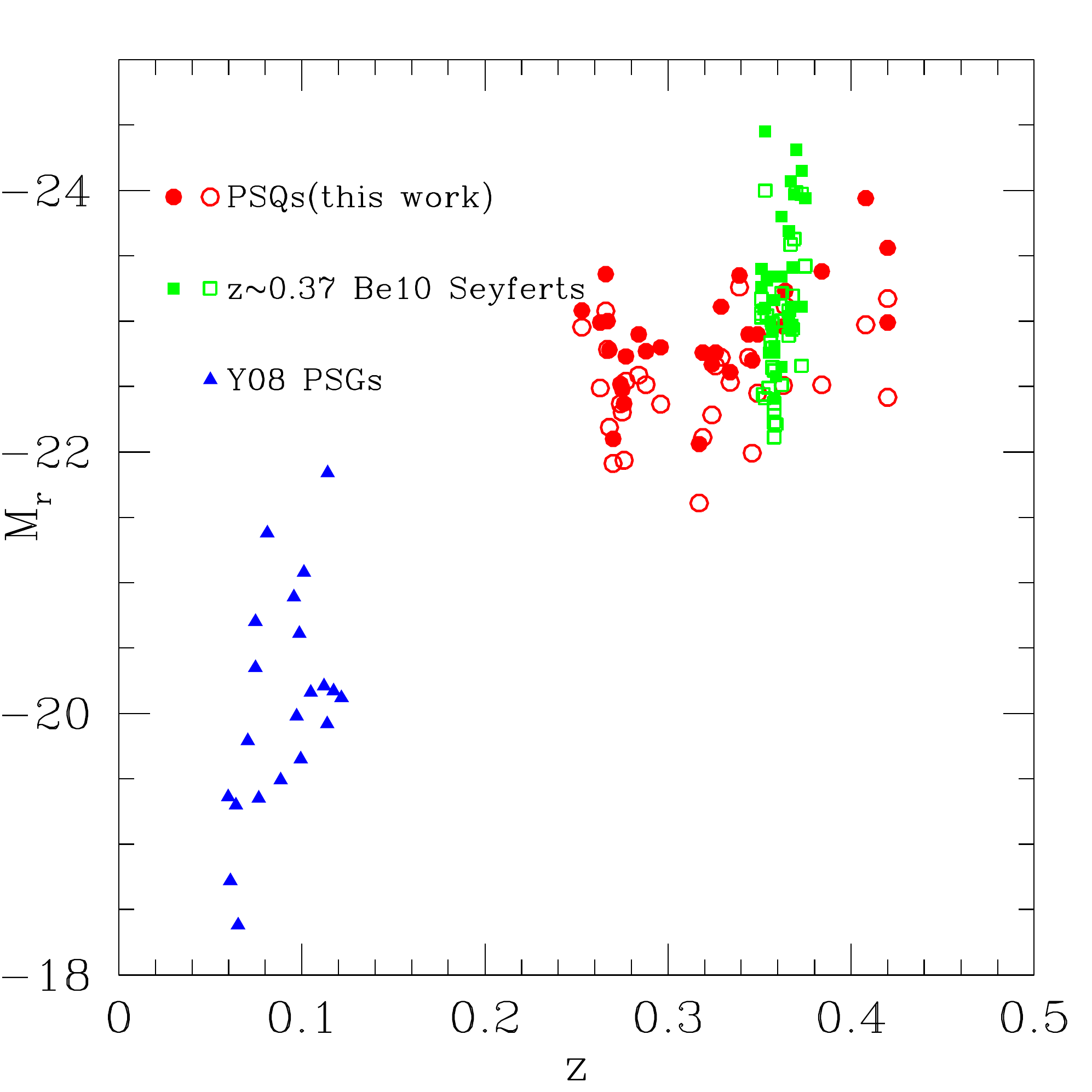}
\caption{A comparison of Galactic dereddened K-corrected $r$ absolute magnitudes of our sample (red circles), that of \citet{bennert10} (green squares), and \citet{yang08} (blue triangles). Filled symbols represent the total magnitude while the open symbols use $f_{nuc}$ to calculate the host galaxy magnitude.\label{fig:comp_plot}}
\end{figure}

   By comparing our sample with others we may be able to gain insight into how these candidate transitioning object fit into evolutionary classes. We give a brief comparison of recent samples of candidate transition objects in Table~\ref{tab:comp} as well as a plot of the samples Galactic dereddened K-corrected $r$ absolute magnitudes as a function of redshift in Figure~\ref{fig:comp_plot}. In the following paragraphs we discuss in more detail comparisons between PSQ and other possible transition types.
                     	
   A recent study by \citet{yang08} of post-starburst, or E+A galaxies, presents the detailed morphologies of 21 E+A galaxies using high-resolution \emph{HST} ACS and WFPC2 images. The selection criterion were similar to our PSQs with the additional constraints of increasing $H_{total}$ to be greater than 5.5 \AA, little [O$_{II}$] emission (EW[O$_{II}$] $< 2.5$ \AA) with lower redshift ($0.07 < z < 0.18$). The study reveals that 11  of 21 (55\%) objects have dramatic tidal features indicative of mergers. The E+As are found to be similar to early-types; having high bulge-to-total light ratios (median $B/T = 0.59$) and high S\'{e}rsic indices ($n \ge 4$). The majority of the objects appear to be early-types while at least two objects are grand design spiral galaxies. Eleven objects are disturbed and the remaining objects show no visual signs of interaction or harassment at this resolution. In comparison to the PSQ sample, the level of disturbance and higher number of early-types suggests that these objects may be in closer alignment with the elliptical mode of evolution in their last throes of morphological disturbance. 
	
 	A PSQ with a more powerful/massive central engine would observationally appear as a QSO, swamping out the light from its own host galaxy. To address the question of whether the majority of QSOs are the result of recent mergers, \citet{canalizo07}, \citet{bennert08} and \citet[\emph{in preparation}]{hancock11} obtained deep (five orbit) \emph{HST} ACS images for 18 QSO host galaxies that were classified morphologically as ellipticals. These objects are relatively nearby and luminous quasars ($M_V < -23.5$, $0.1 < z < 0.25$). Of the initial pilot study of five objects, the majority (4/5) revealed striking signs of tidal interactions such as ripples, tidal tails, and warped disks that were not detected in previous shallower studies. For the PSQ sample, the total exposure time of 720s corresponds to about one quarter of an orbital period. It is possible that observing PSQs at the depth of the QSO study may show more evidence of interaction via the detection of low-surface brightness fine structure hence, increasing the disturbance fraction. The high luminosities and early type host morphology indicate that their sample is likely among the elliptical mode of evolution. This careful study may reveal the answer to what astronomers have been pondering for decades: Can most quasar activity be attributed to transitioning hosts via merger/interaction activity?
   
	We can also compare PSQs to other AGNs of similar redshift and luminosity. \citet{bennert10} obtained \emph{HST} single orbit imaging to determine morphology, nuclear luminosity, and structural parameters of the spheroidal component for a sample of 34 Seyfert galaxies at $z = 0.36$. The majority of the galaxies are intermediate or late-type spirals (29/34). Furthermore, 10/34  galaxies show disturbance/interaction and at least five are merging with companion galaxies and three are morphologically disturbed. The fraction of disturbed hosts is less for the Be10 sample in comparison with PSQs (10/34 and 17/29, respectively ). Again we must note that observing PSQs at the depth of the Be10 study might lead to more evidence of interaction, increasing the disturbed fraction. In comparison to the PSQ sample, the high fraction of spirals and low disturbance fraction suggests that the Be10 galaxies are likely among the spiral mode of evolution. 
	
	We note that the most luminous PSQs with disturbed early-type host galaxies appear to be consistent with merger products. This is not inconsistent with recent studies which show that AGN with $z < 1$ are not found to be overly disturbed in comparison to a control sample of inactive galaxies \citep{cisternas11}. These AGN are less luminous than PSQs and represent a more typical population of AGN at lower redshifts. However, the small volume size from which the \citet{cisternas11} sample is selected represents a small probability that there will be any luminous AGN/quasars in the field. The PSQ sample is more extreme, both in luminosity and post-starburst features. However, we do find that our less luminous objects appear to be Seyfert galaxies that have not required major mergers to trigger the AGN. 
	
	Comparison of PSQs with other types (e.g., nearby luminous quasars of \citet{bennert08}, E+A galaxies of \citet{yang08}, and $z \sim 0.36$ Seyfert 1s of \citet{bennert10} provides more insight into the roles that transitioning galaxies can play in galaxy evolution. At least in objects that are considered transitioning, we see at low redshift, luminous quasars and galaxies can be disturbed merger products \citep{yang08, bennert08}. However, a sample of AGN at similar redshift are more consistent with secularly evolving spiral galaxies \citep{bennert10}. Since PSQs are a heterogeneous population with a variety of morphologies, both dynamical interactions and secular evolution appear to play important roles in the evolution of active galaxies while $z < 1$.

\begin{deluxetable}{llccclc}
\label{tab:comp}
\tabletypesize{\tiny}
\tablecolumns{7}
\tablewidth{0pc}
\tablecaption{Comparison of Sample Types \label{tab:comp}}
\tablehead{
\colhead{Sample} & \colhead{Reference} & \colhead{Number of} & \colhead{$z$} & \colhead{$\langle M_r\rangle$} & \colhead{Morphology} & \colhead{Number} \\
\colhead{} & \colhead{} & \colhead{Objects} & \colhead{Range} & \colhead{} & \colhead{} & \colhead{Disturbed} 
}
\startdata
PSQ & this work & 29 & 0.25:0.45 & $-$22.9 & 13 spiral, 13 early-type, 3 Indeterm. & 17 \\
E+A & Y08 & 21 & 0.07:0.18 & $-$20.07 & Consistent with early-type, with 2 GD spirals & 11 \\
Low-z QSOs & Be08 & 5 & 0.1:0.25 & $M_V < -23.5$ & 5 early-type & 4 \\
$z \sim 0.37$ Seyferts & Be10 & 34 & $z \sim 0.37$ & $-$23.3 & 29 spiral & 20 ($\ge$5 merging + $\ge$3 disturbed)
\enddata
\end{deluxetable}

\section{Summary and Outlook}
\label{sec:Summ}

   We studied in detail the morphologies and host galaxy parameters of 29 PSQs using high resolution  \emph{HST}/ACS-F606W images. We test the idea that PSQs are a phase in the life of galaxies triggered by external events (e.g., mergers, tidal interactions) or whether they are a more heterogenous population in which multiple mechanisms, external and/or internal, can contribute to the class (e.g., mergers, harassment, bar instabilities). The degree of disturbance in these objects is suggestive of dual quasar and starburst activity due to galaxy-galaxy interactions of a variety of types. However, we must note that not all of these objects are disturbed, thus, secular processes of the downsizing epoch may also play a role in dual quasar-starburst activity. We give a summary of the results.
   
\begin{enumerate}
\item 13 of 29 PSQs reside in spiral hosts. 13 of 29 PSQs reside in early-type hosts. Three of the targets are of indeterminate morphology. This implicates that at least two mechanisms are responsible for the triggering the PSQ phase.
\item Early-type galaxies have greater PSF/AGN luminosities, although early-type and spiral hosts have similar host galaxy luminosities. Thus, assuming similar Eddington ratios, early-types have greater black hole masses.
\item The majority of PSQs (17/29) have some sort of interaction or disturbance. Statistically, early-type and spiral hosts are similarly disturbed. Deeper imaging could reveal even more disturbances.
\item Four of the five barred spirals are undisturbed, which may be explained if, bars are a starburst/AGN fueling mechanism in their own right, or bars are a result of strong interactions that happened too long ago for us to detect. 
\item The most luminous PSQs with disturbed early-type host galaxies appear to be consistent with merger products. The less luminous objects appear to be constant with Seyfert galaxies, many of which are barred spirals. 
\item Comparison of PSQs with other types of galaxies shows that both dynamical interactions and secular evolution play important roles in the evolution of active galaxies for $z < 1$. 
\end{enumerate}
   	
   In order to address the outstanding questions involving PSQs and their role in evolutionary scenarios of galaxies we will look at additional properties of post-starburst and AGN via population synthesis modeling in an upcoming paper (Paper II). We have obtained new high signal-to-noise optical spectroscopy of all but one object of the PSQ sample. The analysis of the spectra will focus on the determination of the ages and masses of the host stellar populations (utilizing stellar population synthesis models) and the black hole masses and their Eddington fractions (by making virial mass estimates). By design, the host galaxies of this sample have massive ($M_{burst} \sim 10^{10} M_{\odot}$) and moderate-aged stellar populations (hundreds of Myrs). Preliminary measurements indicate that PSQs have Eddington fractions ($L/L_{Edd}$) on the order of $10\%$ with $M_{BH}\sim 10^7 M_{\odot}$. If the sample represents snapshots along an evolutionary sequence we might expect to see some trends within morphology type in starburst mass, stellar population age, supermassive black hole mass and Eddington fractions. However, these calculations require high signal-to-noise optical spectroscopy to deconvolve the host and AGN components and are beyond the scope of this paper. With this additional information we will revisit issues discussed in this paper. 

We thank Chien Peng for his advice on using GALFIT. We thank the anonymous referee for carefully reading the manuscript and for useful suggestions. We acknowledge support from NASA through the LTSA grant NNG05GE84G. Z. Shang acknowledges support from the national Natural Science Foundation of China through grant 10633040 and support by Chinese 973 Program 2007CB815405. S. L. Cales was supported in part by NASA Headquarters under the NASA Earth and Space Science Fellowship Program - Grant NNX08AX07H.

\bibliographystyle{/Applications/TeX/BibLib/apj}
\bibliography{/Applications/TeX/BibLib/AllCat}

\begin{thebibliography}{68}
\expandafter\ifx\csname natexlab\endcsname\relax\def\natexlab#1{#1}\fi

\bibitem[{{Abazajian} {et~al.}(2005){Abazajian}, {Adelman-McCarthy},
  {Ag{\"u}eros}, {Allam}, {Anderson}, {Anderson}, {Annis}, {Bahcall}, {Baldry},
  {Bastian}, {Berlind}, {Bernardi}, {Blanton}, {Bochanski}, {Boroski},
  {Brewington}, {Briggs}, {Brinkmann}, {Brunner}, {Budav{\'a}ri}, {Carey},
  {Castander}, {Connolly}, {Covey}, {Csabai}, {Dalcanton}, {Doi}, {Dong},
  {Eisenstein}, {Evans}, {Fan}, {Finkbeiner}, {Friedman}, {Frieman},
  {Fukugita}, {Gillespie}, {Glazebrook}, {Gray}, {Grebel}, {Gunn}, {Gurbani},
  {Hall}, {Hamabe}, {Harbeck}, {Harris}, {Harris}, {Harvanek}, {Hawley},
  {Hayes}, {Heckman}, {Hendry}, {Hennessy}, {Hindsley}, {Hogan}, {Hogg},
  {Holmgren}, {Holtzman}, {Ichikawa}, {Ichikawa}, {Ivezi{\'c}}, {Jester},
  {Johnston}, {Jorgensen}, {Juri{\'c}}, {Kent}, {Kleinman}, {Knapp}, {Kniazev},
  {Kron}, {Krzesinski}, {Lamb}, {Lampeitl}, {Lee}, {Lin}, {Long}, {Loveday},
  {Lupton}, {Mannery}, {Margon}, {Mart{\'{\i}}nez-Delgado}, {Matsubara},
  {McGehee}, {McKay}, {Meiksin}, {M{\'e}nard}, {Munn}, {Nash}, {Neilsen},
  {Newberg}, {Newman}, {Nichol}, {Nicinski}, {Nieto-Santisteban}, {Nitta},
  {Okamura}, {O'Mullane}, {Owen}, {Padmanabhan}, {Pauls}, {Peoples}, {Pier},
  {Pope}, {Pourbaix}, {Quinn}, {Raddick}, {Richards}, {Richmond}, {Rix},
  {Rockosi}, {Schlegel}, {Schneider}, {Schroeder}, {Scranton}, {Sekiguchi},
  {Sheldon}, {Shimasaku}, {Silvestri}, {Smith}, {Smol{\v c}i{\'c}}, {Snedden},
  {Stebbins}, {Stoughton}, {Strauss}, {SubbaRao}, {Szalay}, {Szapudi},
  {Szkody}, {Szokoly}, {Tegmark}, {Teodoro}, {Thakar}, {Tremonti}, {Tucker},
  {Uomoto}, {Vanden Berk}, {Vandenberg}, {Vogeley}, {Voges}, {Vogt},
  {Walkowicz}, {Wang}, {Weinberg}, {West}, {White}, {Wilhite}, {Xu}, {Yanny},
  {Yasuda}, {Yip}, {Yocum}, {York}, {Zehavi}, {Zibetti}, \&
  {Zucker}}]{abazajian05}
{Abazajian}, K., {et~al.} 2005, \aj, 129, 1755

\bibitem[{{Bell} {et~al.}(2004){Bell}, {Wolf}, {Meisenheimer}, {Rix}, {Borch},
  {Dye}, {Kleinheinrich}, {Wisotzki}, \& {McIntosh}}]{bell04}
{Bell}, E.~F., {et~al.} 2004, \apj, 608, 752

\bibitem[{{Bennert} {et~al.}(2008){Bennert}, {Canalizo}, {Jungwiert},
  {Stockton}, {Schweizer}, {Peng}, \& {Lacy}}]{bennert08}
{Bennert}, N., {Canalizo}, G., {Jungwiert}, B., {Stockton}, A., {Schweizer},
  F., {Peng}, C.~Y., \& {Lacy}, M. 2008, \apj, 677, 846

\bibitem[{{Bennert} {et~al.}(2010){Bennert}, {Treu}, {Woo}, {Malkan}, {Le
  Bris}, {Auger}, {Gallagher}, \& {Blandford}}]{bennert10}
{Bennert}, V.~N., {Treu}, T., {Woo}, J., {Malkan}, M.~A., {Le Bris}, A.,
  {Auger}, M.~W., {Gallagher}, S., \& {Blandford}, R.~D. 2010, \apj, 708, 1507

\bibitem[{{Berentzen} {et~al.}(2004){Berentzen}, {Athanassoula}, {Heller}, \&
  {Fricke}}]{berentzen04}
{Berentzen}, I., {Athanassoula}, E., {Heller}, C.~H., \& {Fricke}, K.~J. 2004,
  \mnras, 347, 220

\bibitem[{{Blanton} \& {Moustakas}(2009)}]{blanton09}
{Blanton}, M.~R., \& {Moustakas}, J. 2009, \araa, 47, 159

\bibitem[{Brotherton {et~al.}(2011)Brotherton, Stoll, Paul, Diamond-Stanic,
  Shang, Cales, Ganguly, Canalizo, \& Vanden~Berk}]{brotherton11}
Brotherton, M., {et~al.} 2011, submitted

\bibitem[{{Brotherton} {et~al.}(2002){Brotherton}, {Grabelsky}, {Canalizo},
  {van Breugel}, {Filippenko}, {Croom}, {Boyle}, \& {Shanks}}]{brotherton02}
{Brotherton}, M.~S., {Grabelsky}, M., {Canalizo}, G., {van Breugel}, W.,
  {Filippenko}, A.~V., {Croom}, S., {Boyle}, B., \& {Shanks}, T. 2002, \pasp,
  114, 593

\bibitem[{{Brotherton} {et~al.}(1999){Brotherton}, {van Breugel}, {Stanford},
  {Smith}, {Boyle}, {Miller}, {Shanks}, {Croom}, \&
  {Filippenko}}]{brotherton99}
{Brotherton}, M.~S., {et~al.} 1999, \apjl, 520, L87

\bibitem[{{Brown} {et~al.}(2009){Brown}, {Moustakas}, {Caldwell}, {Palamara},
  {Cool}, {Dey}, {Hickox}, {Jannuzi}, {Murray}, \& {Zaritsky}}]{brown09}
{Brown}, M.~J.~I., {et~al.} 2009, \apj, 703, 150

\bibitem[{{Byrd} \& {Howard}(1992)}]{byrdhoward92}
{Byrd}, G.~G., \& {Howard}, S. 1992, \aj, 103, 1089

\bibitem[{Cales {et~al.}(2011)Cales, Brotherton, Stoll, Paul, Diamond-Stanic,
  Shang, Ganguly, Canalizo, \& Vanden~Berk}]{cales11}
Cales, S., {et~al.} 2011, in preparation

\bibitem[{{Canalizo} {et~al.}(2007){Canalizo}, {Bennert}, {Jungwiert},
  {Stockton}, {Schweizer}, {Lacy}, \& {Peng}}]{canalizo07}
{Canalizo}, G., {Bennert}, N., {Jungwiert}, B., {Stockton}, A., {Schweizer},
  F., {Lacy}, M., \& {Peng}, C. 2007, \apj, 669, 801

\bibitem[{{Canalizo} \& {Stockton}(2001)}]{canalizostockton01}
{Canalizo}, G., \& {Stockton}, A. 2001, \apj, 555, 719

\bibitem[{{Canalizo} {et~al.}(2000){Canalizo}, {Stockton}, {Brotherton}, \&
  {van Breugel}}]{canalizo00}
{Canalizo}, G., {Stockton}, A., {Brotherton}, M.~S., \& {van Breugel}, W. 2000,
  \aj, 119, 59

\bibitem[{{Cardelli} {et~al.}(1989){Cardelli}, {Clayton}, \&
  {Mathis}}]{cardelliclaytonmathis89}
{Cardelli}, J.~A., {Clayton}, G.~C., \& {Mathis}, J.~S. 1989, \apj, 345, 245

\bibitem[{{Cisternas} {et~al.}(2011){Cisternas}, {Jahnke}, {Inskip},
  {Kartaltepe}, {Koekemoer}, {Lisker}, {Robaina}, {Scodeggio}, {Sheth},
  {Trump}, {Andrae}, {Miyaji}, {Lusso}, {Brusa}, {Capak}, {Cappelluti},
  {Civano}, {Ilbert}, {Impey}, {Leauthaud}, {Lilly}, {Salvato}, {Scoville}, \&
  {Taniguchi}}]{cisternas11}
{Cisternas}, M., {et~al.} 2011, \apj, 726, 57

\bibitem[{{Cowie} {et~al.}(1996){Cowie}, {Songaila}, {Hu}, \&
  {Cohen}}]{cowie96}
{Cowie}, L.~L., {Songaila}, A., {Hu}, E.~M., \& {Cohen}, J.~G. 1996, \aj, 112,
  839

\bibitem[{{Di Matteo} {et~al.}(2005){Di Matteo}, {Springel}, \&
  {Hernquist}}]{dimatteo05}
{Di Matteo}, T., {Springel}, V., \& {Hernquist}, L. 2005, \nat, 433, 604

\bibitem[{{Dressler} \& {Gunn}(1983)}]{dresslergunn83}
{Dressler}, A., \& {Gunn}, J.~E. 1983, \apj, 270, 7

\bibitem[{{Elmegreen}(1990)}]{elmegreen90}
{Elmegreen}, B.~G. 1990, New York Academy Sciences Annals, 596, 40

\bibitem[{{Falkenberg} {et~al.}(2009){Falkenberg}, {Kotulla}, \&
  {Fritze}}]{falkenberg09}
{Falkenberg}, M.~A., {Kotulla}, R., \& {Fritze}, U. 2009, \mnras, 397, 1940

\bibitem[{{Ferrarese} \& {Merritt}(2000)}]{ferraresemerritt00}
{Ferrarese}, L., \& {Merritt}, D. 2000, \apjl, 539, L9

\bibitem[{{Gebhardt} {et~al.}(2000{\natexlab{a}}){Gebhardt}, {Bender}, {Bower},
  {Dressler}, {Faber}, {Filippenko}, {Green}, {Grillmair}, {Ho}, {Kormendy},
  {Lauer}, {Magorrian}, {Pinkney}, {Richstone}, \& {Tremaine}}]{gebhardt00b}
{Gebhardt}, K., {et~al.} 2000{\natexlab{a}}, \apjl, 539, L13

\bibitem[{{Gebhardt} {et~al.}(2000{\natexlab{b}}){Gebhardt}, {Kormendy}, {Ho},
  {Bender}, {Bower}, {Dressler}, {Faber}, {Filippenko}, {Green}, {Grillmair},
  {Lauer}, {Magorrian}, {Pinkney}, {Richstone}, \& {Tremaine}}]{gebhardt00a}
---. 2000{\natexlab{b}}, \apjl, 543, L5

\bibitem[{{Granato} {et~al.}(2004){Granato}, {De Zotti}, {Silva}, {Bressan}, \&
  {Danese}}]{granato04}
{Granato}, G.~L., {De Zotti}, G., {Silva}, L., {Bressan}, A., \& {Danese}, L.
  2004, \apj, 600, 580

\bibitem[{Hancock(2011)}]{hancock11}
Hancock, M. 2011, in preparation

\bibitem[{{Hasinger}(2008)}]{hasinger08}
{Hasinger}, G. 2008, \aap, 490, 905

\bibitem[{{Heckman} {et~al.}(2004){Heckman}, {Kauffmann}, {Brinchmann},
  {Charlot}, {Tremonti}, \& {White}}]{heckman04}
{Heckman}, T.~M., {Kauffmann}, G., {Brinchmann}, J., {Charlot}, S., {Tremonti},
  C., \& {White}, S.~D.~M. 2004, \apj, 613, 109

\bibitem[{{Hopkins} \& {Hernquist}(2009)}]{hopkinshernquist09}
{Hopkins}, P.~F., \& {Hernquist}, L. 2009, \apj, 694, 599

\bibitem[{{Hopkins} {et~al.}(2008){Hopkins}, {Hernquist}, {Cox}, \& {Kere{\v
  s}}}]{hopkins08}
{Hopkins}, P.~F., {Hernquist}, L., {Cox}, T.~J., \& {Kere{\v s}}, D. 2008,
  \apjs, 175, 356

\bibitem[{{Hopkins} {et~al.}(2006){Hopkins}, {Hernquist}, {Cox}, {Robertson},
  \& {Springel}}]{hopkins06}
{Hopkins}, P.~F., {Hernquist}, L., {Cox}, T.~J., {Robertson}, B., \&
  {Springel}, V. 2006, \apjs, 163, 50

\bibitem[{{Jahnke} {et~al.}(2004){Jahnke}, {S{\'a}nchez}, {Wisotzki}, {Barden},
  {Beckwith}, {Bell}, {Borch}, {Caldwell}, {H{\"a}ussler}, {Heymans}, {Jogee},
  {McIntosh}, {Meisenheimer}, {Peng}, {Rix}, {Somerville}, \&
  {Wolf}}]{jahnke04}
{Jahnke}, K., {et~al.} 2004, \apj, 614, 568

\bibitem[{{Kauffmann} \& {Haehnelt}(2000)}]{kauffmannhaehnelt00}
{Kauffmann}, G., \& {Haehnelt}, M. 2000, \mnras, 311, 576

\bibitem[{{Kauffmann} {et~al.}(2003){Kauffmann}, {Heckman}, {Tremonti},
  {Brinchmann}, {Charlot}, {White}, {Ridgway}, {Brinkmann}, {Fukugita}, {Hall},
  {Ivezi{\'c}}, {Richards}, \& {Schneider}}]{kauffmann03}
{Kauffmann}, G., {et~al.} 2003, \mnras, 346, 1055

\bibitem[{{Kinney} {et~al.}(1996){Kinney}, {Calzetti}, {Bohlin}, {McQuade},
  {Storchi-Bergmann}, \& {Schmitt}}]{kinney96}
{Kinney}, A.~L., {Calzetti}, D., {Bohlin}, R.~C., {McQuade}, K.,
  {Storchi-Bergmann}, T., \& {Schmitt}, H.~R. 1996, \apj, 467, 38

\bibitem[{Krist \& Hook(2004)}]{kristhook04}
Krist, J., \& Hook, R. 2004, The Tiny Tim User's Guide, Version 6.3,
  \texttt{http://www.stsci.edu/software/tinytim/tinytim.pdf}

\bibitem[{{Magorrian} {et~al.}(1998){Magorrian}, {Tremaine}, {Richstone},
  {Bender}, {Bower}, {Dressler}, {Faber}, {Gebhardt}, {Green}, {Grillmair},
  {Kormendy}, \& {Lauer}}]{magorrian98}
{Magorrian}, J., {et~al.} 1998, \aj, 115, 2285

\bibitem[{{Martin} {et~al.}(2007){Martin}, {Wyder}, {Schiminovich}, {Barlow},
  {Forster}, {Friedman}, {Morrissey}, {Neff}, {Seibert}, {Small}, {Welsh},
  {Bianchi}, {Donas}, {Heckman}, {Lee}, {Madore}, {Milliard}, {Rich}, {Szalay},
  \& {Yi}}]{martin07}
{Martin}, D.~C., {et~al.} 2007, \apjs, 173, 342

\bibitem[{{Merritt} \& {Ferrarese}(2001)}]{merrittferrarese01}
{Merritt}, D., \& {Ferrarese}, L. 2001, \mnras, 320, L30

\bibitem[{{Peng} {et~al.}(2002){Peng}, {Ho}, {Impey}, \& {Rix}}]{peng02}
{Peng}, C.~Y., {Ho}, L.~C., {Impey}, C.~D., \& {Rix}, H.-W. 2002, \aj, 124, 266

\bibitem[{{Peng} {et~al.}(2006{\natexlab{a}}){Peng}, {Impey}, {Ho}, {Barton},
  \& {Rix}}]{peng06a}
{Peng}, C.~Y., {Impey}, C.~D., {Ho}, L.~C., {Barton}, E.~J., \& {Rix}, H.
  2006{\natexlab{a}}, \apj, 640, 114

\bibitem[{{Peng} {et~al.}(2006{\natexlab{b}}){Peng}, {Impey}, {Rix},
  {Kochanek}, {Keeton}, {Falco}, {Leh{\'a}r}, \& {McLeod}}]{peng06b}
{Peng}, C.~Y., {Impey}, C.~D., {Rix}, H., {Kochanek}, C.~S., {Keeton}, C.~R.,
  {Falco}, E.~E., {Leh{\'a}r}, J., \& {McLeod}, B.~A. 2006{\natexlab{b}}, \apj,
  649, 616

\bibitem[{{Sanders} \& {Mirabel}(1996)}]{sandersmirabel96}
{Sanders}, D.~B., \& {Mirabel}, I.~F. 1996, \araa, 34, 749

\bibitem[{{Sanders} {et~al.}(1988){Sanders}, {Soifer}, {Elias}, {Neugebauer},
  \& {Matthews}}]{sanders88}
{Sanders}, D.~B., {Soifer}, B.~T., {Elias}, J.~H., {Neugebauer}, G., \&
  {Matthews}, K. 1988, \apjl, 328, L35

\bibitem[{{Schawinski} {et~al.}(2009){Schawinski}, {Virani}, {Simmons}, {Urry},
  {Treister}, {Kaviraj}, \& {Kushkuley}}]{schawinski09}
{Schawinski}, K., {Virani}, S., {Simmons}, B., {Urry}, C.~M., {Treister}, E.,
  {Kaviraj}, S., \& {Kushkuley}, B. 2009, \apjl, 692, L19

\bibitem[{{Schlegel} {et~al.}(1998){Schlegel}, {Finkbeiner}, \&
  {Davis}}]{schlegel98}
{Schlegel}, D.~J., {Finkbeiner}, D.~P., \& {Davis}, M. 1998, \apj, 500, 525

\bibitem[{{Shen} {et~al.}(2003){Shen}, {Mo}, {White}, {Blanton}, {Kauffmann},
  {Voges}, {Brinkmann}, \& {Csabai}}]{shen03}
{Shen}, S., {Mo}, H.~J., {White}, S.~D.~M., {Blanton}, M.~R., {Kauffmann}, G.,
  {Voges}, W., {Brinkmann}, J., \& {Csabai}, I. 2003, \mnras, 343, 978

\bibitem[{{Shields} {et~al.}(2003){Shields}, {Gebhardt}, {Salviander}, {Wills},
  {Xie}, {Brotherton}, {Yuan}, \& {Dietrich}}]{shields03}
{Shields}, G.~A., {Gebhardt}, K., {Salviander}, S., {Wills}, B.~J., {Xie}, B.,
  {Brotherton}, M.~S., {Yuan}, J., \& {Dietrich}, M. 2003, \apj, 583, 124

\bibitem[{{Spergel} {et~al.}(2007){Spergel}, {Bean}, {Dor{\'e}}, {Nolta},
  {Bennett}, {Dunkley}, {Hinshaw}, {Jarosik}, {Komatsu}, {Page}, {Peiris},
  {Verde}, {Halpern}, {Hill}, {Kogut}, {Limon}, {Meyer}, {Odegard}, {Tucker},
  {Weiland}, {Wollack}, \& {Wright}}]{spergel07}
{Spergel}, D.~N., {et~al.} 2007, \apjs, 170, 377

\bibitem[{{Springel} {et~al.}(2005){Springel}, {Di Matteo}, \&
  {Hernquist}}]{springel05}
{Springel}, V., {Di Matteo}, T., \& {Hernquist}, L. 2005, \apjl, 620, L79

\bibitem[{Stoll {et~al.}(2011)Stoll, Brotherton, Cales, Paul, Diamond-Stanic,
  Shang, Ganguly, Canalizo, \& Vanden~Berk}]{stoll11}
Stoll, R., {et~al.} 2011, in preparation

\bibitem[{{Toomre}(1977)}]{toomre77}
{Toomre}, A. 1977, in Evolution of Galaxies and Stellar Populations, ed. B.~M.
  {Tinsley} \& R.~B. {Larson}, 401

\bibitem[{{Tremaine} {et~al.}(2002){Tremaine}, {Gebhardt}, {Bender}, {Bower},
  {Dressler}, {Faber}, {Filippenko}, {Green}, {Grillmair}, {Ho}, {Kormendy},
  {Lauer}, {Magorrian}, {Pinkney}, \& {Richstone}}]{tremaine02}
{Tremaine}, S., {et~al.} 2002, \apj, 574, 740

\bibitem[{{Tremonti} {et~al.}(2007){Tremonti}, {Moustakas}, \&
  {Diamond-Stanic}}]{tremonti07}
{Tremonti}, C.~A., {Moustakas}, J., \& {Diamond-Stanic}, A.~M. 2007, \apjl,
  663, L77

\bibitem[{{Treu} {et~al.}(2004){Treu}, {Malkan}, \& {Blandford}}]{treu04}
{Treu}, T., {Malkan}, M.~A., \& {Blandford}, R.~D. 2004, \apjl, 615, L97

\bibitem[{{Treu} {et~al.}(2007){Treu}, {Woo}, {Malkan}, \&
  {Blandford}}]{treu07}
{Treu}, T., {Woo}, J., {Malkan}, M.~A., \& {Blandford}, R.~D. 2007, \apj, 667,
  117

\bibitem[{{Vanden Berk} {et~al.}(2001){Vanden Berk}, {Richards}, {Bauer},
  {Strauss}, {Schneider}, {Heckman}, {York}, {Hall}, {Fan}, {Knapp},
  {Anderson}, {Annis}, {Bahcall}, {Bernardi}, {Briggs}, {Brinkmann}, {Brunner},
  {Burles}, {Carey}, {Castander}, {Connolly}, {Crocker}, {Csabai}, {Doi},
  {Finkbeiner}, {Friedman}, {Frieman}, {Fukugita}, {Gunn}, {Hennessy},
  {Ivezi{\'c}}, {Kent}, {Kunszt}, {Lamb}, {Leger}, {Long}, {Loveday}, {Lupton},
  {Meiksin}, {Merelli}, {Munn}, {Newberg}, {Newcomb}, {Nichol}, {Owen}, {Pier},
  {Pope}, {Rockosi}, {Schlegel}, {Siegmund}, {Smee}, {Snir}, {Stoughton},
  {Stubbs}, {SubbaRao}, {Szalay}, {Szokoly}, {Tremonti}, {Uomoto}, {Waddell},
  {Yanny}, \& {Zheng}}]{vandenberk01}
{Vanden Berk}, D.~E., {et~al.} 2001, \aj, 122, 549

\bibitem[{{Vanden Berk} {et~al.}(2006){Vanden Berk}, {Shen}, {Yip},
  {Schneider}, {Connolly}, {Burton}, {Jester}, {Hall}, {Szalay}, \&
  {Brinkmann}}]{vandenberk06}
---. 2006, \aj, 131, 84

\bibitem[{{Varela} {et~al.}(2004){Varela}, {Moles}, {M{\'a}rquez}, {Galletta},
  {Masegosa}, \& {Bettoni}}]{varela04}
{Varela}, J., {Moles}, M., {M{\'a}rquez}, I., {Galletta}, G., {Masegosa}, J.,
  \& {Bettoni}, D. 2004, \aap, 420, 873

\bibitem[{{Veilleux}(2006)}]{veilleux06}
{Veilleux}, S. 2006, New Astronomy Review, 50, 701

\bibitem[{{Veron-Cetty} \& {Veron}(2006)}]{veroncettyveron06}
{Veron-Cetty}, M., \& {Veron}, P. 2006, VizieR Online Data Catalog, 7248, 0

\bibitem[{{Wild} {et~al.}(2009){Wild}, {Walcher}, {Johansson}, {Tresse},
  {Charlot}, {Pollo}, {Le F{\`e}vre}, \& {de Ravel}}]{wild09}
{Wild}, V., {Walcher}, C.~J., {Johansson}, P.~H., {Tresse}, L., {Charlot}, S.,
  {Pollo}, A., {Le F{\`e}vre}, O., \& {de Ravel}, L. 2009, \mnras, 395, 144

\bibitem[{{Woo} {et~al.}(2006){Woo}, {Treu}, {Malkan}, \& {Blandford}}]{woo06}
{Woo}, J., {Treu}, T., {Malkan}, M.~A., \& {Blandford}, R.~D. 2006, \apj, 645,
  900

\bibitem[{{Woo} {et~al.}(2008){Woo}, {Treu}, {Malkan}, \& {Blandford}}]{woo08}
---. 2008, \apj, 681, 925

\bibitem[{{Yan} {et~al.}(2006){Yan}, {Newman}, {Faber}, {Konidaris}, {Koo}, \&
  {Davis}}]{yan06}
{Yan}, R., {Newman}, J.~A., {Faber}, S.~M., {Konidaris}, N., {Koo}, D., \&
  {Davis}, M. 2006, \apj, 648, 281

\bibitem[{{Yang} {et~al.}(2008){Yang}, {Zabludoff}, {Zaritsky}, \&
  {Mihos}}]{yang08}
{Yang}, Y., {Zabludoff}, A.~I., {Zaritsky}, D., \& {Mihos}, J.~C. 2008, \apj,
  688, 945

\bibitem[{{Zabludoff} {et~al.}(1996){Zabludoff}, {Zaritsky}, {Lin}, {Tucker},
  {Hashimoto}, {Shectman}, {Oemler}, \& {Kirshner}}]{zabludoff96}
{Zabludoff}, A.~I., {Zaritsky}, D., {Lin}, H., {Tucker}, D., {Hashimoto}, Y.,
  {Shectman}, S.~A., {Oemler}, A., \& {Kirshner}, R.~P. 1996, \apj, 466, 104

\end{thebibliography}
\clearpage

\section*{Appendix A}
\section*{SDSS Spectra}
\label{sec:AppA}

\begin{figure}[h]
   \centering
   \figurenum{9} 
      \subfloat[][Fig. 9$a$]{\includegraphics[width=4.8in]{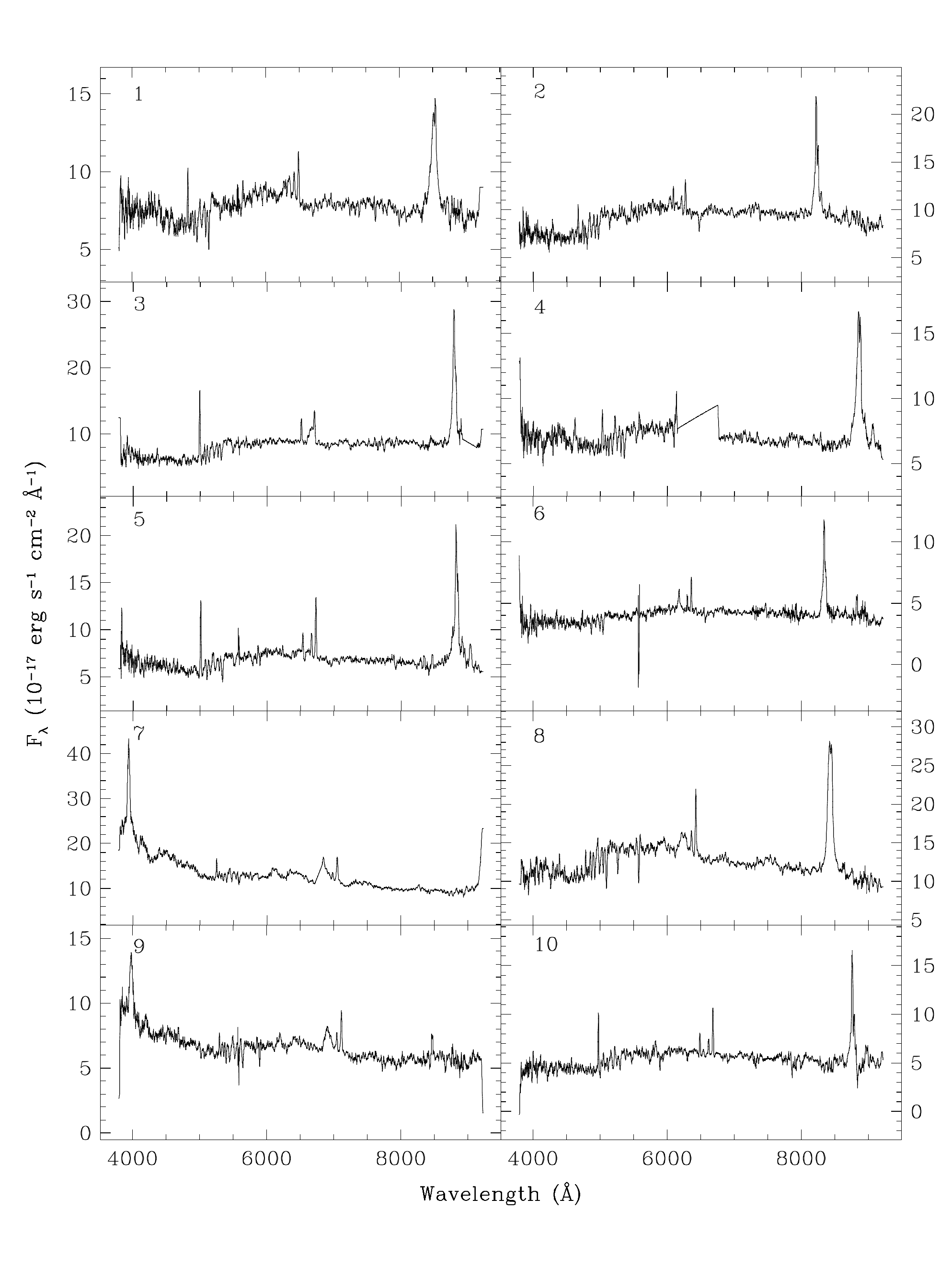}}
      \caption{SDSS DR7 spectra of the PSQ sample numbered as in Table~\ref{tab:sample}. \label{fig:Spectra}}
\end{figure}
\begin{figure}
   \ContinuedFloat
   \centering
      \subfloat[][Fig. 9$b$]{\includegraphics[width=4.8in]{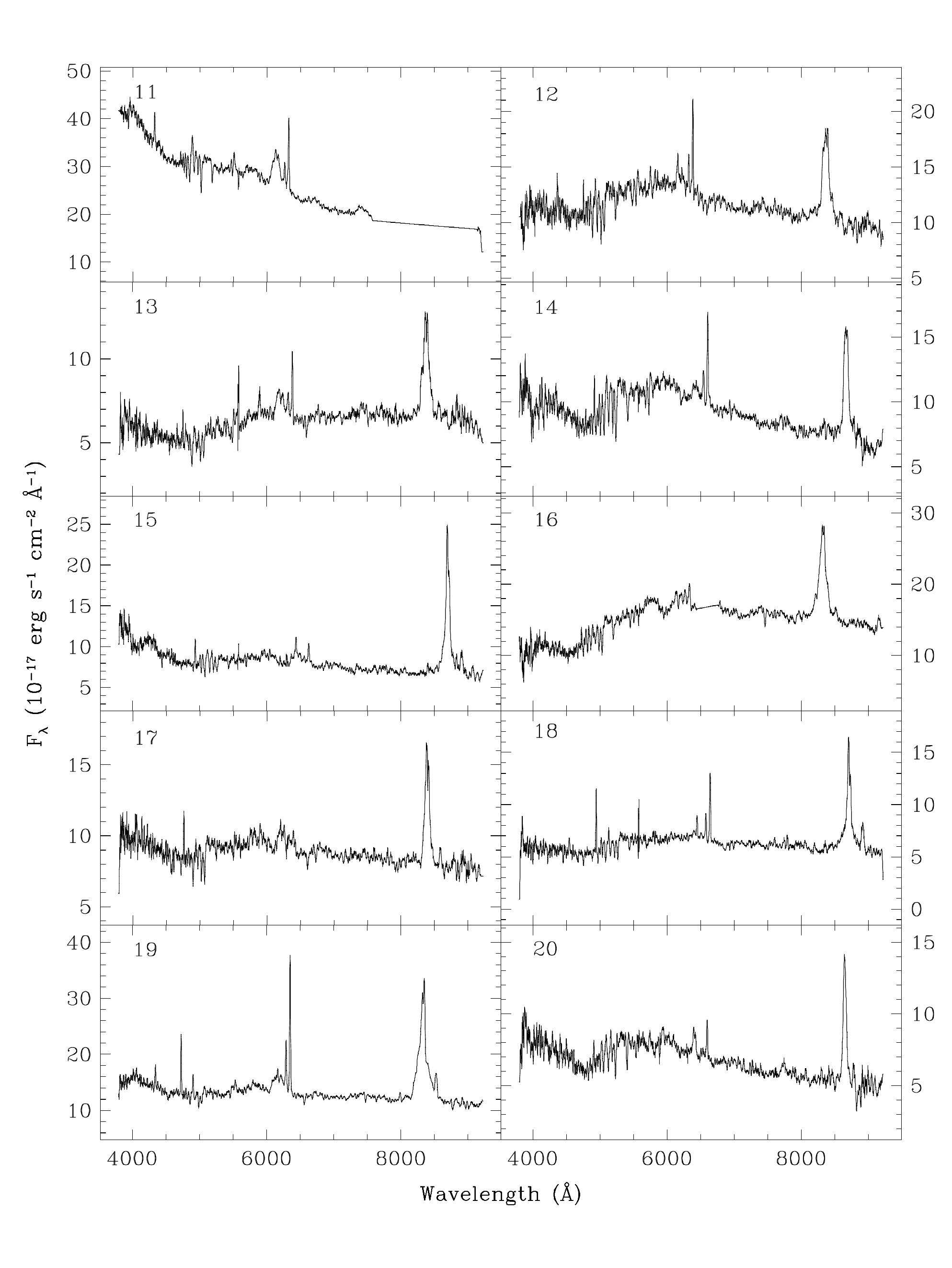}}
\end{figure}
\begin{figure}
   \ContinuedFloat
   \centering
      \subfloat[][Fig. 9$c$]{\includegraphics[width=4.8in]{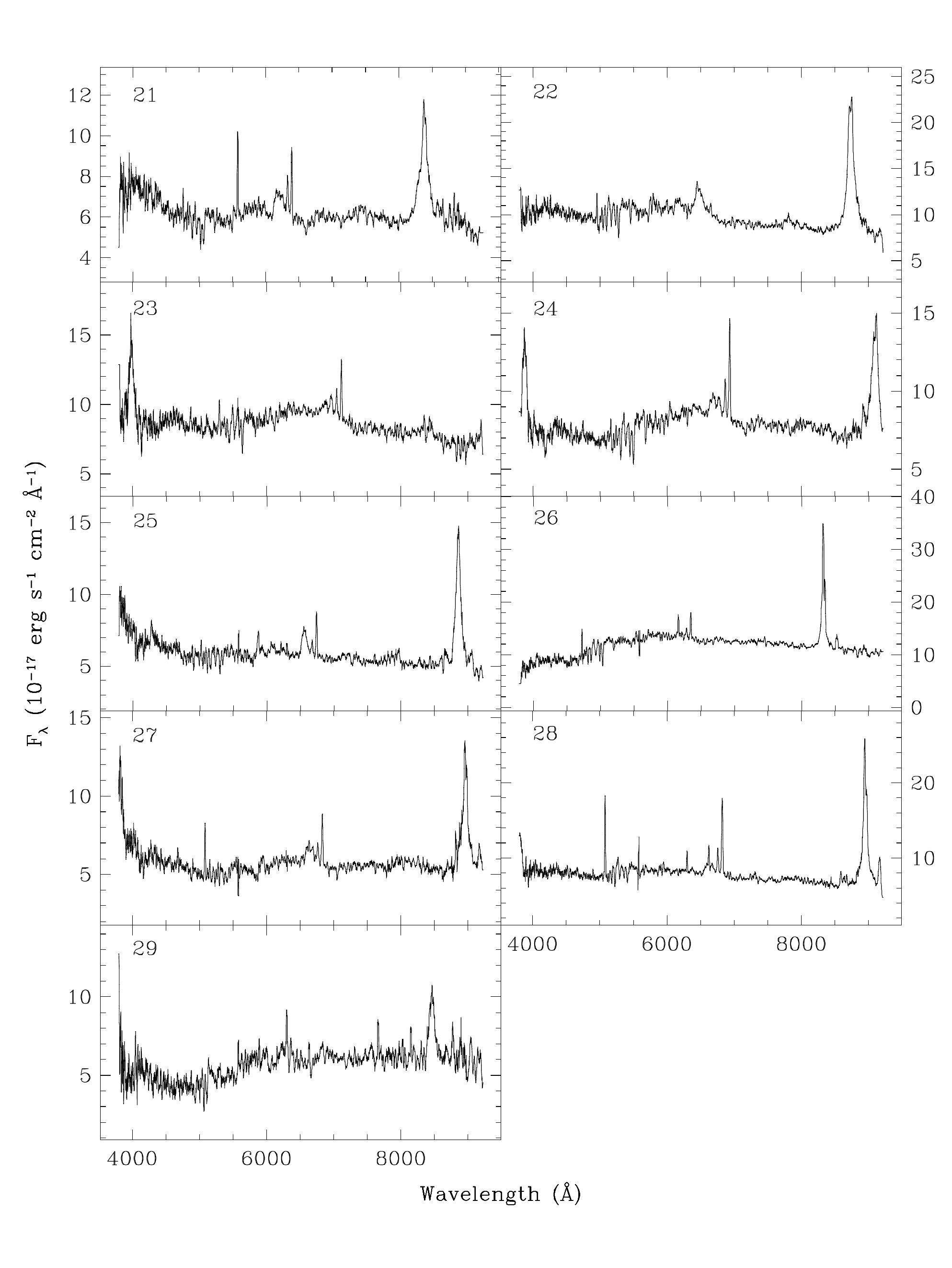}}
\end{figure}
\clearpage

\section*{Appendix B}
\section*{Notes on Individual Objects}
\label{sec:AppB}

   Object 1 (\emph{SDSS J003043.59-103517.6}) $-$ The host is of barred spiral type viewed face-on with two symmetric spiral arms clearly visible. Three faint possible companions are located within about 10\arcs (WNW, NE, and SSE). There is a companion galaxy 14\farcs7 from the central nucleus (PA $=$ 146\arcdeg).

   Object 2 (\emph{SDSS J005739.19+010044.9}) $-$  This host is a slightly asymmetric face-on spiral. 

   Object 3 (\emph{SDSS J020258.94$-$002807.5}) $-$  Classification of the host is indeterminate. The object is clearly disturbed. The arc in the SE quadrant of the object, 0\farcs8 from the nucleus, may be asymmetric star formation or a possible double nucleus.
   
   Object 4 (\emph{SDSS J021447.00$-$003250.6}) $-$  This object appears to be a fairly smooth early-type.
   
   Object 5 (\emph{SDSS J023700.30$-$010130.5}) $-$  This ring galaxy reveals spiral-like structure in the nucleus. A bright source near the ring visible 2\farcs6 to the SSW (PA $=$ 211\arcdeg) of the nucleus is a foreground M star \citep[\emph{in preparation}]{stoll11}.
   
   Object 6 (\emph{SDSS J040210.90$-$054630.3}) $-$  The target here appears to have undergone recent interaction. It could be that this is a spiral with a bright star-forming tidal tail in the east. However, what looks like a tidal tail could (and may well) be the remnant of an interacting companion galaxy. An arc of star formation appears in the NW quadrant of the object at a line of sight semi-major axis distance of 1\farcs2.
   
   Object 7 (\emph{SDSS J074621.06$+$335040.7}) $-$ The target appears to be a smooth early-type.
   
   Object 8 (\emph{SDSS J075045.00$+$212546.3}) $-$ The host appears to be an early-type with possible interation/merger signatures visible in the form of shells (one such shell is apparent in the NE quadrant of the target at a line of sight radius of 3\farcs7).
   
   Object 9 (\emph{SDSS J075521.30$+$295039.2}) $-$ The arms of this spiral are flocculent and asymmetric. 
   
   Object 10 (\emph{SDSS J075549.56$+$321704.1}) $-$ Although at first glance this object might appear to be an early-type, a closer look reveals a faint bar of length 0\farcs9 oriented at PA $\sim 135$\arcdeg. A companion of this object (14\farcs6 from target, PA $=$ 140\arcdeg) is visually disturbed with a jagged `S' appearance and long tidal tail extending toward the north.
        
   Object 11 (\emph{SDSS J081018.67$+$250921.2}) $-$ A faint hint of a tidal tail/fuzz emanating from the central SE quadrant towards the south (3\farcs5, PA $=$ 136\arcdeg) adds fine structure to this otherwise smooth early-type host.

   Object 12 (\emph{SDSS J105816.81$+$102414.5}) $-$ This host appears to be a smooth early-type.
   
   Object 13 (\emph{SDSS J115159.59$+$673604.8}) $-$ Two symmetric, yet faint spiral arms connect via the bar (4\farcs9 across, PA $\sim 135$\arcdeg) of this face-on barred spiral. SDSS photometrically identifies a neighboring galaxy 15\farcs6 from the target at PA $=$ 104\arcdeg.
      
   Object 14 (\emph{SDSS J115355.58$+$582442.3}) $-$ The object appears to be a smooth early-type elongated at PA $\sim 70$\arcdeg. SDSS photometrically identifies a neighboring galaxy 16\farcs3 from the target at PA $=$ 325\arcdeg.

   Object 15 (\emph{SDSS J123043.41$+$614821.8}) $-$ Classification of the host is indeterminate. The host shows faint flocculent spiral structure without revealing anything so organized as arms. SDSS photometrically identifies a neighboring galaxy 5\farcs4 from the target at PA $=$ 164\arcdeg.
     
   Object 16 (\emph{SDSS J124833.52$+$563507.4}) $-$ This host is a highly disturbed early-type. Flare-like shells are apparent in the NW and SE quadrants of the object (radii $=$ 4\farcs4 and 6\farcs1, respectively) as well as a visible dust-lane within a line of sight distance 0\farcs5 from the quasar in a NNW-SSE fashion. SDSS photometrically identifies a neighboring galaxy (which appears to be a spiral) 16\farcs7 from the target at PA $=$ 92\arcdeg.
         
   Object 17 (\emph{SDSS J145640.99+524727.2}) $-$ This host is a face-on barred (2\farcs2 across, PA $\sim 20$\arcdeg) spiral with fine structure. SDSS photometrically identifies a fuzzy neighboring galaxy 8\farcs2 from the target at PA $=$ 297\arcdeg.

   Object 18 (\emph{SDSS J145658.15+593202.3}) $-$ Merger of 2-3 apparent spiral galaxies with another apparent spiral companion within 9\farcs7 at PA $=$ 213\arcdeg. The double nuclei are 0\farcs8 apart, which is in turn 4\farcs1 from the galaxy approaching from the west (PA $=$ 90\arcdeg). SDSS photometrically recognizes several other sources within 10\farcs0. The SDSS photometric pipeline deblends the source into two objects. The magnitude reported in Table~\ref{tab:sample} refers only to the double nucleus source, and not the companion to the East.
   
   Object 19 (\emph{SDSS J154534.55+573625.1}) $-$ At first glance one might classify the host as an early-type. But, on closer inspection there appears to be a flattened disk elongated in the ENE-WSW direction. A dust lane envelopes the nucleus with dimensions 1\farcs3 (line of sight) in the semi-major axis and 0\farcs6 (line of sight) in the semi-minor axis.  SDSS photometrically identifies a fuzzy neighboring galaxy 9\farcs3 from the target at PA $=$ 133\arcdeg.
      
   Object 20 (\emph{SDSS J164444.92+423304.5}) $-$ The host appears to be a smooth early-type. SDSS photometrically identifies what appears to be an edge-on spiral along with another spiral at distances, 3\farcs0 (PA $=$ 0\arcdeg) and 4\farcs5 (PA $=$ 107\arcdeg), respectively,  from the target. 
   
   Object 21 (\emph{SDSS J170046.95+622056.4}) $-$ The disk of this spiral has a ring of star formation 3\farcs6 from the central source (or a dust lane at 2\farcs0, semi-major axis line of sight). This ring of star formation may be due to harassment from a small satellite galaxy 2\farcs6 away (line of sight, PA $=$ 197\arcdeg). SDSS photometrically identifies a fuzzy neighboring galaxy 6\farcs3 from the target at PA $=$ 45\arcdeg. 
   
   Object 22 (\emph{SDSS J210200.42+000501.8}) $-$ This early-type has a tidal tail extending 4\farcs1 (PA $=$ 264\arcdeg) in length from the central source possibly wrapping around behind the host (from W to E) and ending with  the tip of the tail on the opposite side of origin at a distance of 4\farcs1 (PA $=$ 80\arcdeg). The SDSS photometric pipeline classifies part of the tidal tail as a separate source, thus the magnitude of the galaxy reported in Table I will be slightly underestimated.  
      
   Object 23 (\emph{SDSS J211343.20-075017.6}) $-$ The smooth early-type target is at a line of sight distance of 7\farcs5 (PA $=$ 132\arcdeg) from its early-type companion.

   Object 24 (\emph{SDSS J211838.12+005640.6}) $-$ The object appears to be a smooth early-type.

   Object 25 (\emph{SDSS J212843.42+002435.6}) $-$ This early-type, which is elongated in the NE-SW direction, has (what appears to be an early-type) companion located 2\farcs6 (PA $=$ 162\arcdeg) from the central source. There also appears to be more tidal fluff extending 2\farcs6 from nucleus in the S (PA $=$ 180\arcdeg). The SDSS photometry includes both the central source and the companion. 
   
   Object 26 (\emph{SDSS J230614.18-010024.4}) $-$ The companion of this flocculent face-on spiral is 10\farcs8 (line of sight, PA $=$ 246\arcdeg) from the nucleus. The host appears to have a single arm (which could be a tidal tail) located in the NE quadrant of the object, as well as a tidal arc 1\farcs7 from the nucleus in the SE.
      
   Object 27 (\emph{SDSS J231055.50-090107.6}) $-$ This pair is reminiscent of M51 (The Whirlpool Galaxy). The southern arm of the large spiral is intersected by a smaller galaxy at a line of sight center to center distance of 4\farcs1 (PA $=$ 212\arcdeg). Although the orientation of the primary spiral galaxy is not favorable there appears to be a bar of 4\farcs5 across at PA $\sim 55$\arcdeg. The SDSS photometric pipeline deblends the source into two objects.  The magnitude reported in Table~\ref{tab:sample} refers only to the large spiral.

   Object 28 (\emph{SDSS J233430.89+140649.7}) $-$ This object appears to be a fairly smooth early-type with a faint tidal tail disturbance in the SE quadrant.
   
   Object 29 (\emph{SDSS J234403.55+154214.0}) $-$ The host is similar to that of \emph{SDSS J230614.18$-$\\010024.4}, a flocculent face-on spiral.

\end{document}